\documentstyle[12pt,epsfig,axodraw]{article}

\advance\textheight by \topskip
\begin{document}
\tolerance=100000
\thispagestyle{empty}
\setcounter{page}{0}

\newcommand{\HPA}[1]{{\it Helv.\ Phys.\ Acta.\ }{\bf #1}}
\newcommand{\AP}[1]{{\it Ann.\ Phys.\ }{\bf #1}}
\newcommand{\be}{\begin{equation}}
\newcommand{\ee}{\end{equation}}
\newcommand{\br}{\begin{eqnarray}}
\newcommand{\er}{\end{eqnarray}}
\newcommand{\ba}{\begin{array}}
\newcommand{\ea}{\end{array}}
\newcommand{\bi}{\begin{itemize}}
\newcommand{\ei}{\end{itemize}}
\newcommand{\bn}{\begin{enumerate}}
\newcommand{\en}{\end{enumerate}}
\newcommand{\bc}{\begin{center}}
\newcommand{\ec}{\end{center}}
\newcommand{\ul}{\underline}
\newcommand{\ol}{\overline}
\def\l{\left\langle}
\def\r{\right\rangle}
\def\as{\alpha_{s}}
\def\ycut{y_{\mbox{\tiny cut}}}
\def\yij{y_{ij}}
\def\epem{\ifmmode{e^+ e^-} \else{$e^+ e^-$} \fi}
\newcommand{\eeww}{$e^+e^-\rightarrow W^+ W^-$}
\newcommand{\qqQQ}{$q_1\bar q_2 Q_3\bar Q_4$}
\newcommand{\eeqqQQ}{$e^+e^-\rightarrow q_1\bar q_2 Q_3\bar Q_4$}
\newcommand{\eewwqqqq}{$e^+e^-\rightarrow W^+ W^-\ar q\bar q Q\bar Q$}
\newcommand{\eeqqgg}{$e^+e^-\rightarrow q\bar q gg$}
\newcommand{\eeqloop}{$e^+e^-\rightarrow q\bar q gg$ via loop of quarks}
\newcommand{\eeqqqq}{$e^+e^-\rightarrow q\bar q Q\bar Q$}
\newcommand{\eewwjjjj}{$e^+e^-\rightarrow W^+ W^-\rightarrow 4~{\rm{jet}}$}
\newcommand{\eeqqggjjjj}{$e^+e^-\rightarrow q\bar 
q gg\rightarrow 4~{\rm{jet}}$}
\newcommand{\eeqloopjjjj}{$e^+e^-\rightarrow q\bar 
q gg\rightarrow 4~{\rm{jet}}$ via loop of quarks}
\newcommand{\eeqqqqjjjj}{$e^+e^-\rightarrow q\bar q Q\bar Q\rightarrow
4~{\rm{jet}}$}
\newcommand{\eejjjj}{$e^+e^-\rightarrow 4~{\rm{jet}}$}
\newcommand{\jjjj}{$4~{\rm{jet}}$}
\newcommand{\qqbar}{$q\bar q$}
\newcommand{\ww}{$W^+W^-$}
\newcommand{\ar}{\rightarrow}
\newcommand{\sm}{${\cal {SM}}$}
\newcommand{\Dir}{\kern -6.4pt\Big{/}}
\newcommand{\Dirin}{\kern -10.4pt\Big{/}\kern 4.4pt}
\newcommand{\DDir}{\kern -8.0pt\Big{/}}
\newcommand{\DGir}{\kern -6.0pt\Big{/}}
\newcommand{\wwqqqq}{$W^+ W^-\ar q\bar q Q\bar Q$}
\newcommand{\qqgg}{$q\bar q gg$}
\newcommand{\qloop}{$q\bar q gg$ via loop of quarks}
\newcommand{\qqqq}{$q\bar q Q\bar Q$}

\def\st{\sigma_{\mbox{\scriptsize t}}}
\def\Ord{\buildrel{\scriptscriptstyle <}\over{\scriptscriptstyle\sim}}
\def\OOrd{\buildrel{\scriptscriptstyle >}\over{\scriptscriptstyle\sim}}
\def\pl #1 #2 #3 {{\it Phys.~Lett.} {\bf#1} (#2) #3}
\def\np #1 #2 #3 {{\it Nucl.~Phys.} {\bf#1} (#2) #3}
\def\zp #1 #2 #3 {{\it Z.~Phys.} {\bf#1} (#2) #3}
\def\jp #1 #2 #3 {{\it J.~Phys.} {\bf#1} (#2) #3}
\def\pr #1 #2 #3 {{\it Phys.~Rev.} {\bf#1} (#2) #3}
\def\prep #1 #2 #3 {{\it Phys.~Rep.} {\bf#1} (#2) #3}
\def\prl #1 #2 #3 {{\it Phys.~Rev.~Lett.} {\bf#1} (#2) #3}
\def\mpl #1 #2 #3 {{\it Mod.~Phys.~Lett.} {\bf#1} (#2) #3}
\def\rmp #1 #2 #3 {{\it Rev. Mod. Phys.} {\bf#1} (#2) #3}
\def\cpc #1 #2 #3 {{\it Comp. Phys. Commun.} {\bf#1} (#2) #3}
\def\sjnp #1 #2 #3 {{\it Sov. J. Nucl. Phys.} {\bf#1} (#2) #3}
\def\xx #1 #2 #3 {{\bf#1}, (#2) #3}
\def\hepph #1 {{\tt hep-ph/#1}}
\def\preprint{{\it preprint}}

\begin{flushright}
{\large RAL-TR-98-035}\\
{\rm July 1998}\\
\end{flushright}

\vspace*{\fill}

\begin{center}
{\Large {\bf Six-jet production at lepton colliders\footnote{E-mail:  
moretti@v2.rl.ac.uk.}}}\\[1.cm]
{\large 
S.~Moretti}\\[0.4 cm]
{\it Rutherford Appleton Laboratory,}\\
{\it Chilton, Didcot, Oxon OX11 0QX, UK.}
\end{center}

\vspace*{\fill}

\begin{abstract}
{\normalsize
\noindent
We study electron-positron annihilations into six jets at the parton level in 
perturbative Quantum Chromo-Dynamics (QCD), via the elementary processes
$e^+e^-\ar q\bar q gggg$, 
$e^+e^-\ar q\bar q q'\bar q' gg$ and
$e^+e^-\ar q\bar q q'\bar q' q''\bar q''$, for massive quarks $q, q'$
and $q''$ and massless gluons $g$. Several numerical results of
phenomenological relevance are given,
at three different collider energies and for a representative selection of 
jet clustering algorithms.
We also present helicity amplitudes and colour factors needed
for the tree-level calculation.}
\end{abstract}

\vskip1.5truecm
\noindent
{{\it PACS}: 12.38.-t, 12.38.Bx, 13.87.-a, 13.65.+i}\\
\noindent
{{\it Keywords}: perturbative QCD, LO computations, jets, lepton colliders.}
\vspace*{\fill}
\newpage

\section{Introduction and motivation}
\label{Sect_intro}

\subsection{Multi-jet rates}
\label{Subsect_multi}

\begin{figure}[htb]
\begin{center}
\begin{picture}(450,110)
\SetScale{.9}
\SetWidth{1.2}
\SetOffset(50,0)

\ArrowLine(20,50)(-10,80)
\ArrowLine(-10,20)(20,50)
\Text(-20,80)[]{$e^+$}
\Text(-20,10)[]{$e^-$}

\Photon(20,50)(80,50){5}{4}
\Text(40,60)[]{$\gamma,Z$}

\ArrowLine(80,50)(140,110)
\ArrowLine(140,-10)(80,50)
\Text(132,105)[]{$q$}
\Text(132,-15)[]{$\bar q$}

\Gluon(130,100)(160,100){3}{3}
\ArrowLine(160,100)(190,120)
\ArrowLine(190,80)(160,100)

\Line(175,110)(180,110)
\Gluon(180,110)(210,110){3}{3}
\Text(180,115)[]{$q'$}
\Text(180,70)[]{$\bar q'$}

\ArrowLine(210,110)(240,130)
\ArrowLine(240,90)(210,110)
\Text(230,120)[]{$q''''$}
\Text(230,78)[]{$\bar q''''$}
\Line(225,100)(230,100)
\Gluon(230,100)(255,100){3}{3}

\Gluon(110,80)(140,80){3}{3}

\Gluon(90,60)(120,60){3}{3}

\Gluon(110,20)(140,20){3}{3}
\Gluon(140,20)(170,40){3}{3}
\Gluon(170,0)(140,20){3}{3}

\ArrowLine(170,0)(200,20)
\ArrowLine(200,-20)(170,0)
\Text(190,20)[]{$q''$}
\Text(190,-20)[]{$\bar q''$}

\Gluon(170,40)(190,60){3}{3}
\Gluon(190,20)(170,40){3}{3}
\Line(170,40)(180,40)
\Gluon(180,40)(210,40){3}{3}

\ArrowLine(210,40)(240,60)
\ArrowLine(240,20)(210,40)
\Text(230,55)[]{$q'''$}
\Text(230,20)[]{$\bar q'''$}

\Text(320,50)[]{\rm etc.}

\end{picture}
\vskip1.0cm
\caption{Schematic
description of multi-parton production in $e^+e^-$ annihilations.}
\label{fig_cascade}
\end{center}
\end{figure}
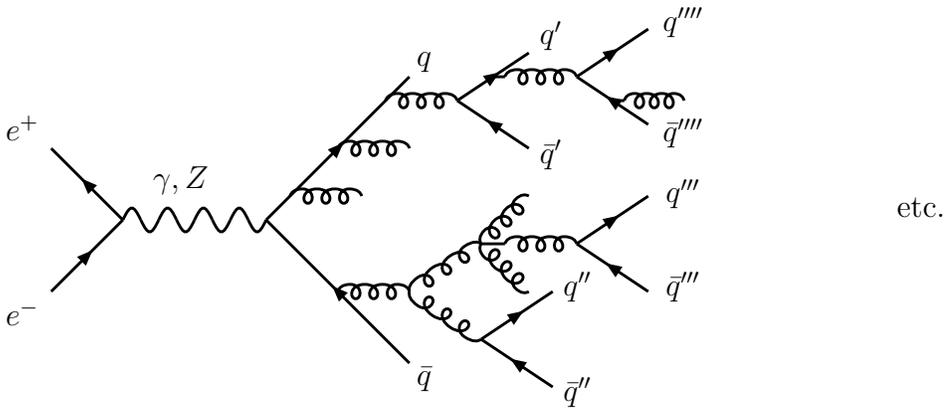

As the centre-of-mass (CM) energy of modern accelerators
grows, the number of final state partons that can be accommodated
in the available phase space becomes larger. 
Many of these will be quarks and gluons, as
QCD forces are the strongest among the fundamental interactions.
In electron-positron annihilations, coloured partons can be produced 
via an initial splitting of a $\gamma$ or 
$Z$ current into quark-antiquark pairs, 
followed by no hard radiation at all or by (multiple) gluon bremsstrahlung, 
with the vector particles eventually yielding further quark-antiquark
pairs or other gluons, as schematically recalled in Fig.~\ref{fig_cascade}. 
\vskip0.75cm
\vskip-0.25cm
Clearly, as this so-called
{\sl parton shower} or {\sl cascade} evolves in time, the 
amount of energy exchanged among the various partons diminishes, 
so that the process
will eventually enter a non-perturbative regime.
At such low energy scales (roughly, of the order of 1 GeV), one is 
unable to describe quantitatively the above process in terms of a
series expansion in powers of the strong coupling constant $\as$ using
Feynman diagram techniques, since $\as$ becomes significantly large
and the series cannot be resummed reliably. Note that 1 GeV is also 
the scale at which the partons are conventionally assumed to hadronise, that 
is, to convert themselves into the observed hadrons through the formation of
colourless bound states (see Ref.~\cite{QCDbook} for an up-to-date
review of hadronisation mechanisms), which in turn make up the jets observed
in the experimental apparatus. 

This fact clearly illustrates an intrinsic limitation of the perturbative 
approach. A further, more practical, restraint that one has to face is due
to our scarce ability for calculating multi-particle amplitudes using
standard perturbative techniques. In fact, with the number of final states
particles increasing, the complexity of the expressions
of the fundamental scattering amplitudes and the large number of terms involved
(arising from the combinatorics of the 
diagrams) that one has to master become in the end an obstacle
difficult to overcome, in spite of the help of modern computers.
For example, if one limits oneself in Fig.~\ref{fig_cascade} 
to the case of gluon radiation 
only (that is to $q\bar q + ng$ final states), one has the 
following impressive escalation of numbers of Feynman  diagrams involved
(considering the $\gamma$ and $Z$ propagators together):
1 for $n=0$, 2 for $n=1$, 8 for $n=2$, 50 for $n=3$, 428 for $n=4$,
4670 for $n=5$, etc\footnote{Under these circumstances, the use of spinor
techniques based on the helicity decomposition of the Feynman amplitudes
becomes mandatory, in order to simplify the calculation
to a certain extent (for a good introduction to helicity methods
and complete references, see \cite{NP17} and \cite{HELbook}).}.
%Also note that the number of those involving
%$g\ar q\bar q$ splittings, although smaller, is of the same
%order of magnitude

An even higher degree of complications is encountered if one goes
beyond the leading order (LO) and tries to incorporate loop corrections
in the calculation. In fact, it should be recalled that the actual number
of jets produced in electron-positron annihilations is defined according
to a phenomenological {\sl jet clustering} algorithm, in which jets are 
constructed out of primary objects (hereafter, labelled by $i,j$ etc.),
the latter being hadrons or calorimeter cells in the real experimental
case and partons in the theoretical calculation.  The term
{\sl exclusive} means that each primary object is assigned to a
unique jet and each final state has a unique jet multiplicity, for a
given value of the jet {\sl resolution parameter} $\ycut$. Algorithms
of this type are generically 
based on a binary clustering, wherein the number of objects 
is reduced one at a time by combining the two nearest (in some sense)
ones. The joining procedure is stopped by testing against some
criterion (the parameter $\ycut$), 
and the final clusters are called jets (we will treat some of 
these algorithms in greater detail
later on, in Subsect.~\ref{Subsect_algorithms}). 

From the point of view of perturbative calculations, it is evident that
the $n$-jet rate  will then be constituted not only of the fraction 
of $n$-parton
events in which  all partons are resolved, but also by the $(n+m)$-parton
contributions (with $m\ge1$) in which
$m$ partons remain unresolved. Now, in the 
$(n+m)$-parton $\rightarrow$ $n$-jet transitions, two partons can well be
combined in singular configurations. This can happen when one has either
two collinear partons within or one soft parton outside the
`jet cones'. Both these contributions are (in general, positively) divergent.
Thanks to the Bloch-Nordsieck \cite{BN} and Kinoshita-Lee-Nauenberg
\cite{KLN} theorems (see also Ref.~\cite{Sterman}), these collinear
and soft singularities are cancelled at the same order in 
$\as$  by the
divergent contributions (generally, negative) provided by the virtual loop 
graphs. If one recombination has occurred then one-loop diagrams must be
computed, when two mergings have taken place two-loop graphs are needed,
and so on. As a matter of fact, the computation of loop diagrams is even more
demanding than that of tree-level ones (both using trace \cite{loop}
and helicity \cite{loop_hel} techniques). 
Not surprisingly then, the number of {\sl exact} calculations of $e^+e^-\ar 
n$-parton matrix elements (ME) available in the literature is rather limited.

As for the tree-level, till a few years ago only up to 
five-parton final states were computed,
via the elementary scatterings: $e^+e^-\ar q\bar q ggg$ and $e^+e^-\ar q\bar q
q'\bar q'g$. This calculation was tackled in Ref.~\cite{five}. The case of 
four partons via $e^+e^-\ar q\bar q gg$ and $e^+e^-\ar q\bar q
q'\bar q'$ has been known since long time \cite{four}.
The case of three partons (i.e., $e^+e^-\ar q\bar qg$) 
was very trivial \cite{QCDbook} and the lowest-order case of two 
(i.e., $e^+e^-\ar q\bar q$) is in fact a textbook example \cite{two}.
Only recently, the task of computing six-parton cross
sections (via
$e^+e^-\ar q\bar q gggg$, 
$e^+e^-\ar q\bar q q'\bar q' gg$ and
$e^+e^-\ar q\bar q q'\bar q' q''\bar q''$) 
has been accomplished, in Ref.~\cite{iosix}, where numerical results
for the massless case were presented. It is the main purpose of the present 
paper to present new phenomenological analyses of six-jet production
at lepton colliders
as well as to give the analytical expressions corresponding to those 
six-parton processes, further 
extending the calculations of Ref.~\cite{iosix} to the case of quarks with
generic mass. Notice that the retention of finite values of
$m_{q,q',q''}$ in the amplitudes turns out to be a serious complication 
in many instances, because of the unavoidable proliferation of terms
and/or the slowness of execution of the numerical routines that it generates.
As a matter of fact, depending on the CM energy, the actual flavour
of the quark being produced and also the
region in phase space under study, in many cases the effect of the masses 
can safely be ignored\footnote{So was the case before the LEP era for the more 
complicated contribution of 
the intermediate $Z$ (and its interference with the photon) in the primary 
$e^+ e^-$ annihilation, provided the energy at which the latter takes place
is well below the mass of the former (at PEP and PETRA, for example).}.

As for loop and radiative diagrams,
the highest-order corrections calculated to date in $e^+e^-\ar n$-jet
annihilations are the next-to-leading order (NLO) ones to the four-jet rate, 
that is,  terms
proportional to $\alpha_s^3$ \cite{cpt,slac},
in the massless approximation. The $\alpha_s^2$ ones date back to 
Ref.~\cite{four} for the massless case 
while the massive one has been completed
  very recently \cite{massivethreenlo}. In fact, considerable 
simplifications can be attained also at loop level when the quark mass 
is neglected. Two- and higher-order-loop corrections are not yet known. 

But let us review the theoretical progress in evaluating 
$e^+e^-\ar n$-parton rates and its interplay with the experiment.
This will also eventually help us to motivate the need of the availability
of the calculations carried out in Ref.~\cite{iosix} and in this paper.

The history of calculations of multi-jet production in $e^+e^-$  
scatterings \cite{NP3}--\cite{NP16} began at the times when jets were 
first observed at PEP and PETRA. At those energies 
the $Z$ contribution was small and could safely be ignored 
\cite{four}, \cite{NP3}--\cite{NP6}, (see also \cite{NP9,NP11}).
Mass effects were already taken into account in some instances. However,
apart from the simplest cases \cite{NP3,NP4},
only the formulae for massless fermions were compact enough
to be published \cite{four,NP5,NP10}.
With the advent of LEP and the advances in spinor techniques
these results were improved in various ways.
In particular, in \cite{NP13} the complete triply differential
cross section for $e^+e^-\rightarrow q\bar q g$ for massive quarks
with both intermediate photon and $Z$ was presented.
In Ref.~\cite{NP14} the decay $Z\rightarrow 4f$ (with $f=q$ or $\ell$: i.e., 
quark or lepton) was studied and from the
expression given there it is possible to reconstruct the full amplitude for
$e^+e^-\rightarrow q\bar q q'\bar q'$. Very simple formulae for 
$e^+e^-\rightarrow q\bar q gg,\, q\bar q q'\bar q'$ were given 
in Ref.~\cite{NP15} for massless quarks. The tree-level amplitudes
for three-, four- and five-parton events including all mass effects
and both the $\gamma$ and $Z$ contributions were given in analytic
form in Refs.~\cite{noiall,voi}.
Finally, one-loop three-parton
and tree-level four-parton terms have been combined and incorporated in 
Monte Carlo simulation programs \cite{EERAD,EVENT,EVENT2}, as has 
been done for the one-loop four-parton and tree-level five-parton amplitudes
\cite{DEBRECEN,MENLO_PARC,EERAD2}, 
for massless quarks in both cases, whereas the massive 
contributions to 
the $\as^2$ three-jet rates with massive partons will soon be. 

A further point should be noted. On the one hand, at 
small values of the resolution parameter $\ycut$, the fixed-order perturbative
calculations of $n$-jet rates become unreliable because of the presence of 
large logarithms of the form $\ln(\ycut)$ (in other words, because of
 the growing of the $(n+m)$-jet contributions, with $m\ge1$). On the other 
hand, the exact higher order calculations can be technically unavailable. 
Under these circumstances, the resummation to all $\as$ orders in perturbation 
theory of terms of the form ${\cal O}({\as^n\ln^m(1/y)})$, e.g.,  with $m=l$
and $m=l-1$ (to leading and next-to-leading accuracy, with $n$ and $m$ 
depending on the jet observable), 
can improve substantially the reliability of the theoretical predictions, 
as was shown to be the case for the two- and tree-jet rates
\cite{y3resum}--\cite{CDFW1}. Recall, however, that such a resummation 
is only possible within the framework of certain jet clustering schemes,
such as the Durham and Cambridge ones (see Subsect.~\ref{Subsect_algorithms}),
for which the jet fractions show the usual Sudakov exponentiation of 
multiple soft-gluon emission.

The availability of three-, four- and five-parton matrix elements 
(tree- and loop-level and supplemented with the mentioned logarithms resummed)
has 
been of crucial importance in studying the underlying structure of QCD.
Throughout the years  
and at different colliders (PEP, PETRA, LEP, SLC), several tests have been 
performed. The strong coupling constant $\alpha_s$ has been determined from jet
rates and from shape variables \cite{alphas} and both the
flavour independence and the running with $\sqrt s$
have been verified (see, e.g., Ref.~\cite{flavour_running}).
Three- \cite{3jd} and four-jet \cite{4jd} differential 
distributions have been long studied  and their behaviour agrees with
QCD predictions calculated to second and third order in $\alpha_s$,
respectively.
The colour factors, which determine the gauge
group responsible for strong interactions, have been measured
\cite{colfac,ALEPHgluino}. Models alternative to QCD
have been ruled out and the coupling of the  
triple gluon vertex has been
verified to be consistent with the QCD theory \cite{Abelian}.
The latest QCD tests are currently being carried out at LEP2. In addition, 
at this energy, one of the key issues is to quantify to a high degree of
accuracy
the amount of background in multi-jet events from QCD to the $W^+W^-$ hadronic
decays, as the corresponding signal (though biased by systematic uncertainties 
due to relatively unknown Bose-Einstein \cite{bose} and Colour
 Interconnection \cite{CR} effects) still represents
to date the experimentally
preferred decay channel, because of its somewhat higher statistics and
because kinematic constraints can tighten the precision of $M_{W^\pm}$
measurements \cite{lep2w,Oxford}. 

In the not too distant future (presumably not long after the year 2005), a new
electron-positron machine will be operative, the Next Linear Collider
(NLC), which is expected to run initially at an energy of 500 GeV, later
increasing up to 2 TeV or so. 

\subsection{Six-jet rates}
\label{Subsect_six}

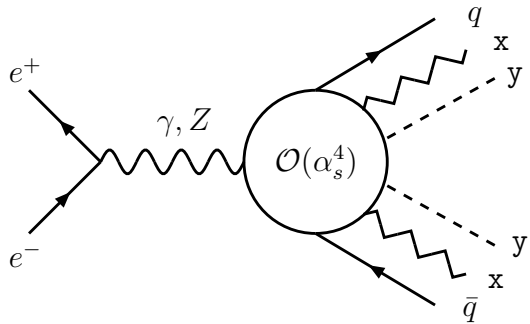
\begin{figure}[htb]
\begin{center}
\begin{picture}(450,110)
\SetScale{.9}
\SetWidth{1.2}
\SetOffset(50,0)

\ArrowLine(120,50)(90,80)
\ArrowLine(90,20)(120,50)
\Text(80,80)[]{$e^+$}
\Text(80,10)[]{$e^-$}

\Photon(120,50)(180,50){5}{4}
\Text(140,60)[]{$\gamma,Z$}

\CArc(210,50)(30,270,90)
\CArc(210,50)(30,90,270)

\Text(190,45)[]{${\cal O}(\alpha_s^4)$}

\ArrowLine(210,80)(260,110)
\ArrowLine(260,-10)(210,20)
\Text(250,100)[]{$q$}
\Text(248,-11)[]{$\bar q$}

\ZigZag(230,71.5)(273,97){4}{3}
\ZigZag(273,3)(230,28){4}{3}
\Text(260,90)[]{${\tt x}$}
\Text(258,0)[] {${\tt x}$}

\DashLine(240,60)(285,85){4}%{3}
\DashLine(285,15)(240,40){4}%{3}
\Text(265,77)[]{${\tt y}$}
\Text(267,13)[]{${\tt y}$}
\end{picture}
\vskip0.5cm
\caption{Schematic
Feynman diagrams contributing in lowest order to $e^+e^-\ar $
6 partons. The jagged and dashed lines and the labels ${\tt x,y}$
refer to either a quark 
or a gluon, see (\ref{qqgggg})--(\ref{qqqqqq}).}
\label{fig_diagrams}
\end{center}
\end{figure}

The phenomenology of six-jet events produced in electron-positron annihilations
(shown schematically in Fig.~\ref{fig_diagrams}) is not as all well known 
as that
of smaller multiplicity jet rates, from both the experimental and 
theoretical point of view. This should not sound surprising, given
the complications arising from, on the one hand,
the poor event rate and large number of tracks, 
and, on the other hand, the mentioned complexity of the 
perturbative QCD calculations.

\vskip-0.25cm
In our opinion, the time has now come to approach the calculation
of $e^+e^-\ar 6$~jet events. We are driven to this belief 
from the following  considerations. 
LEP1 was the era of the $Z$-peak and of its two-jet
(and several higher-order) decays. LEP2 is now the age of 
the $W^+W^-$-resonance 
and of its four-jet (and some higher-order) decays. 
As we will move into the NLC epoch, we will step into
a long series of resonant processes, typically 
ending up with six-jet signatures.
One can mention top physics for a start. As this is one
of the main goals of the NLC, we should expect a lot of experimental
studies concerned with $t\bar t\ar b\bar b W^+W^-\ar \mbox{6~jet}$-events,
since the $W^\pm$'s mainly decay hadronically into two jets.
Then one should not 
forget the new generation of gauge boson resonances, such as $W^+W^-Z$ and
$ZZZ$, and their favourite six-jet decays. 
One could also add highly-virtual photonic processes, like $\gamma W^+W^-$, 
$\gamma ZZ$, $\gamma\gamma Z$ and $\gamma\gamma\gamma$ 
as well as those involving the Higgs particle, via 
$ZH\ar ZW^+W^-$, $ZH\ar ZZZ$ and $ZH\ar ZHH$, 
provided this is light enough to be 
produced at those energy regimes.

A collection of dedicated articles illustrating the phenomenology of
top-antitop and three-boson production at NLC
can be found in Ref.~\cite{ee500}. 
It is beyond the scope of this paper to recall the salient features of
those studies and the interplay of the mentioned resonant events with the
six-jet background produced via QCD. Instead, what we would like
to emphasise here,
is that one should be ready with all the appropriate phenomenological
instruments to challenge the new kind of multi-jet experimental studies
that will have to be carried out at the new generation of $e^+e^-$ machines.

As for the theoretical progress in this respect, 
studies of $e^+e^-\ar$~6-fermion electroweak processes are under way. 
A brief account of
approaches and methods can be found in Ref.~\cite{nico5}. These kinds of
reactions have already been calculated and analysed for the case involving up 
to four quarks in the final state \cite{sandro} (and other similar studies
are in preparation 
\cite{grace}), while in Ref.~\cite{nico} the concern was for
Higgs processes with two quarks produced.
One should probably expect the upgrade of the codes used for those
studies to the case of six-quark production via electroweak interactions
quite soon. However, a large fraction 
of the six-jet cross section  comes
from QCD interactions involving gluon propagators, gluon emissions and 
quark-gluon couplings. The case of six-jet production from $W^+W^-$ decays
was considered in Ref.~\cite{WW6}.

In the present paper, the main focus will be on the ${\cal O}(\as^4)$
QCD processes at tree-level:
\be\label{qqgggg}
e^+e^-\ar q\bar q gggg, 
\ee
\be\label{qqqqgg}
e^+e^-\ar q\bar q q'\bar q' gg
\ee 
\be\label{qqqqqq}
e^+e^-\ar q\bar q q'\bar q' q''\bar q'',
\ee
where $q, q'$ and $q''$ represent any possible flavours of quarks
(massless and/or massive) and $g$ is the QCD gauge boson.

As a by-product of our analysis, we will study the case of 
electromagnetic (EM) radiation.
Since the QED emission of photons off quark lines is dynamically
similar to that of gluons (apart from the different strength of the coupling 
and the colour factor associated with it), we will modify our MEs
with external gluons (\ref{qqgggg})--(\ref{qqqqgg})
appropriately, so as to allow for the case of multiple hard photon emission
from the final state (Final State Radiation, FSR), in
$e^+e^-\ar q\bar q ggg\gamma$,  
$e^+e^-\ar q\bar q gg\gamma\gamma$,     
$e^+e^-\ar q\bar q g\gamma\gamma\gamma$,
$e^+e^-\ar q\bar q \gamma\gamma\gamma\gamma$ (via two-quark) and
$e^+e^-\ar q\bar q q'\bar q'g\gamma$,
$e^+e^-\ar q\bar q q'\bar q'\gamma\gamma$ (via four-quark) events. 
In order to calculate the photoproduction cross sections,
we further supplement the FSR diagrams with those in which 
photon emission takes place
from the initial state (Initial State Radiation, ISR) along with those in
which photons are radiated from both the incoming and outgoing
fermions. Notice that hard photons 
produced by electrons and positrons can also be included 
by means of a convolution of the non-radiative diagrams 
with the so-called Electron Structure Functions (ESF) \cite{structure}.
Such implementation allows one to include
the exact photon corrections to the $e^+e^-$ annihilation subprocess
resummed up to the third order \cite{alpfaem3} 
in the EM coupling constant $\alpha_{em}$ in the infrared limit 
(in particular, they embody both soft/collinear and virtual photon emission).
We have not exploited here this approach, as we have used our tree-level
diagrams. In doing so, however, additional cuts are necessary
in order to screen our results from the collinear singularities
due to the infrared emission from the initial state. 
We have required $p^\gamma_T > 5$ GeV
and ${|\cos\theta_\gamma | < 0.85}$. Such cuts are similar to the photon
selection
criteria adopted by the LEP  Collaborations.

We will apply our results to the case of LEP1, LEP2 and, particularly,
NLC energies. 
A short description
of the computational techniques adopted will be given in Sect.~\ref{Sect_ME}.
Results are in Sect.~\ref{Sect_results}. 
A brief summary is given in Sect.~\ref{Sect_summa}. The explicit
helicity amplitude formulae are in Appendix I, whereas the colour factors
are presented in Appendix II.

\section{Matrix Elements}
\label{Sect_ME}

In order to master the large number of Feynman diagrams (of the order
of several hundred) entering in 
processes (\ref{qqgggg})--(\ref{qqqqqq}) at tree-level, 
we have used spinor methods \cite{HELbook}.
In fact, if one adopts the technique of deriving the expression of the squared 
amplitude $|A|^2=|\sum_k a_k|$ by using the textbook method of the traces
($a_k$ being the amplitude associated to diagram $k$), one results in
 a prohibitive number of terms to be evaluated. Typically, if $n$ is
the number of diagrams, then $\frac{n(n+1)}{2}$ traces
have to be computed. Recalling the numbers of 
Sect.~\ref{Sect_intro}, the task is clearly prohibitive for six-parton events.
 The calculation 
becomes  simpler if one uses so-called
`helicity amplitude' (or `spinor') techniques. In such an approach, given
a set of helicities assigned to the external particles, one computes the
contribution of every single diagram $k$ as a complex number $a_k$,
sums over all $k$'s and takes the modulus squared. This way, in the above 
example, one only needs to compute $n$ (complex) terms.
To obtain unpolarised cross sections one simply sums the modulus squared for
the various helicity combinations.
Thus, a further advantage is intrinsic to the spinor method, 
that one can naturally 
compute also polarised MEs (e.g., for the SLC and NLC), 
without the need of inserting into the 
expression of the amplitudes the helicity projection operators \cite{Itzi}.

\vfill\clearpage

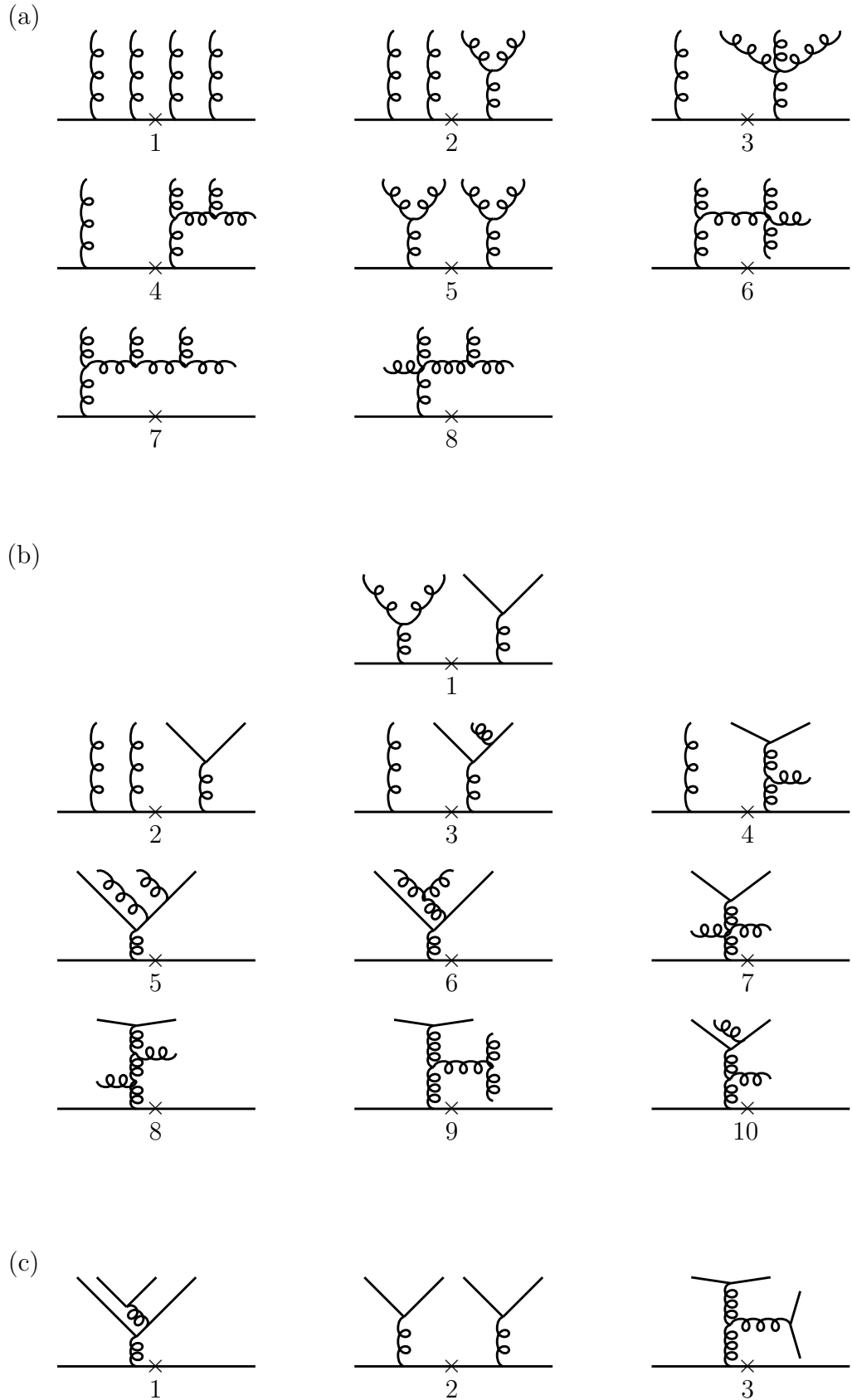
\begin{figure}[!t]
\begin{center}
\vskip-2.0cm
\begin{picture}(500,100)
\SetScale{.9}
\SetWidth{1.2}
\SetOffset(0,0)

\Text(30,110)[]{(a)}

\Text(90,63.25)[]{$\times$}
\Text(90,53)[]{$1$}
\Line( 50,70)(150,70)
\Gluon(70,70)(70,115){3}{3}
\Gluon(90,70)(90,115){3}{3}
\Gluon(110,70)(110,115){3}{3}
\Gluon(130,70)(130,115){3}{3}

\Text(225,63.25)[]{$\times$}
\Text(225,53)[]{$2$}
\Line(200,70)(300,70)
\Gluon(220,70)(220,115){3}{3}
\Gluon(240,70)(240,115){3}{3}
\Gluon(270,70)(270,95){3}{2}
\Gluon(270,95)(255,115){3}{2}
\Gluon(285,115)(270,95){3}{2}

\Text(360,63.25)[]{$\times$}
\Text(360,53)[]{$3$}
\Line(350,70)(450,70)
\Gluon(365,70)(365,115){3}{3}
\Gluon(415,70)(415,95){3}{2}
\Gluon(415,95)(415,115){3}{2}
\Gluon(415,95)(385,115){3}{3}
\Gluon(445,115)(415,95){3}{3}

\Text(90,-4.5)[]{$\times$}
\Text(90,-14.75)[]{$4$}
\Line( 50,-5)(150,-5)
\Gluon(65,-5)(65,40){3}{3}

\Gluon(110,-5)(110,20){3}{2}
\Gluon(110,20)(110,40){3}{2}
\Gluon(110,20)(130,20){3}{2}
\Gluon(130,20)(130,40){3}{2}
\Gluon(130,20)(150,20){3}{2}

\Text(225,-4.5)[]{$\times$}
\Text(225,-14.75)[]{$5$}
\Line(200,-5)(300,-5)
\Gluon(230,-5)(230,20){3}{2}
\Gluon(230,20)(215,40){3}{2}
\Gluon(245,40)(230,20){3}{2}
\Gluon(270,-5)(270,20){3}{2}
\Gluon(270,20)(255,40){3}{2}
\Gluon(285,40)(270,20){3}{2}

\Text(360,-4.5)[]{$\times$}
\Text(360,-14.75)[]{$6$}
\Line(350,-5)(450,-5)
\Gluon(375,-5)(375,20){3}{2}
\Gluon(375,20)(375,40){3}{2}
\Gluon(375,20)(410,20){3}{3}
\Gluon(410,20)(410,40){3}{2}
\Gluon(430,20)(410,20){3}{2}
\Gluon(410,0)(410,20){3}{2}

\Text(90,-72)[]{$\times$}
\Text(90,-82.25)[]{$7$}
\Line(50,-80)(150,-80)
\Gluon(65,-80)(65,-55){3}{2}
\Gluon(65,-55)(65,-35){3}{2}
\Gluon(65,-55)(90,-55){3}{2}
\Gluon(90,-55)(90,-35){3}{2}
\Gluon(90,-55)(115,-55){3}{2}
\Gluon(115,-55)(115,-35){3}{2}
\Gluon(115,-55)(140,-55){3}{2}

\Text(225,-72)[]{$\times$}
\Text(225,-82.25)[]{$8$}
\Line(200,-80)(300,-80)
\Gluon(235,-80)(235,-55){3}{2}
\Gluon(235,-55)(235,-35){3}{2}
\Gluon(235,-55)(215,-55){3}{2}
\Gluon(235,-55)(260,-55){3}{3}
\Gluon(260,-55)(260,-35){3}{2}
\Gluon(260,-55)(280,-55){3}{2}

\end{picture}
%\vskip3.0cm
%\caption{Feynman topologies contributing in lowest order to $e^+e^-\ar 
%q\bar q gggg$. The symbol $\times$ refers to the insertion of the $e^+e^-\ar
%\gamma,Z$ current. Permutations are not shown.} 
%\label{fig_2q4g}
\end{center}
\begin{center}
\vskip4.0cm
\begin{picture}(500,170)
\SetScale{.9}
\SetWidth{1.2}
\SetOffset(0,0)

\Text(30,180)[]{(b)}

\Text(225,131)[]{$\times$}
\Text(225,120.75)[]{$1$}
\Line(200,145)(300,145)
\Gluon(225,145)(225,165){3}{2}
\Gluon(225,165)(205,190){3}{2}
\Gluon(245,190)(225,165){3}{2}
\Gluon(275,145)(275,170){3}{2}
\Line(255,190)(275,170)
\Line(295,190)(275,170)

\Text(90,63.25)[]{$\times$}
\Text(90,53)[]{$2$}
\Line( 50,70)(150,70)
\Gluon(70,70)(70,115){3}{3}
\Gluon(90,70)(90,115){3}{3}
\Gluon(125,70)(125,95){3}{2}
\Line(105,115)(125,95)
\Line(145,115)(125,95)

\Text(225,63.25)[]{$\times$}
\Text(225,53)[]{$3$}
\Line(200,70)(300,70)
\Gluon(220,70)(220,115){3}{3}
\Gluon(260,70)(260,95){3}{2}
\Line(240,115)(260,95)
\Line(280,115)(260,95)
\Gluon(270,105)(260,115){3}{2}

\Text(360,63.25)[]{$\times$}
\Text(360,53)[]{$4$}
\Line(350,70)(450,70)
\Gluon(370,70)(370,115){3}{3}
\Gluon(410,70)(410,87.5){3}{2}
\Gluon(410,87.5)(410,105){3}{2}
\Line(390,115)(410,105)
\Line(430,115)(410,105)
\Gluon(430,87.5)(410,87.5){3}{2}

\Text(90,-4.5)[]{$\times$}
\Text(90,-14.75)[]{$5$}
\Line( 50,-5)(150,-5)
\Gluon(90,-5)(90,10){3}{2}
\Line(60,40)(90,10)
\Line(120,40)(90,10)
\Gluon(70,40)(95,15){3}{3}
\Gluon(90,40)(105,25){3}{2}

\Text(225,-4.5)[]{$\times$}
\Text(225,-14.75)[]{$6$}
\Line(200,-5)(300,-5)
\Gluon(240,-5)(240,10){3}{2}
\Line(210,40)(240,10)
\Line(270,40)(240,10)
\Gluon(235,25)(245,15){3}{2}
\Gluon(220,40)(235,25){3}{2}
\Gluon(235,25)(250,40){3}{2}

\Text(360,-4.5)[]{$\times$}
\Text(360,-14.75)[]{$7$}
\Line(350,-5)(450,-5)
\Gluon(390,-5)(390,10){3}{2}
\Gluon(390,10)(390,25){3}{2}
\Gluon(390,10)(410,10){3}{2}
\Gluon(390,10)(370,10){3}{2}
\Line(370,40)(390,25)
\Line(410,40)(390,25)

\Text(90,-72)[]{$\times$}
\Text(90,-82.25)[]{$8$}
\Line(50,-80)(150,-80)
\Gluon(90,-80)(90,-66){3}{2}
\Gluon(90,-66)(70,-66){3}{2}
\Gluon(90,-66)(90,-52){3}{2}
\Gluon(110,-52)(90,-52){3}{2}
\Gluon(90,-52)(90,-38){3}{2}
\Line(70,-35)(90,-38)
\Line(110,-35)(90,-38)

\Text(225,-72)[]{$\times$}
\Text(225,-82.25)[]{$9$}
\Line(200,-80)(300,-80)
\Gluon(240,-80)(240,-59){3}{3}
\Gluon(240,-59)(240,-38){3}{3}
\Gluon(240,-59)(270,-59){3}{3}
\Gluon(270,-59)(270,-42){3}{2}
\Gluon(270,-76)(270,-59){3}{2}
\Line(220,-35)(240,-38)
\Line(260,-35)(240,-38)

\Text(360,-72)[]{$\times$}
\Text(359,-82.25)[]{$10$}
\Line(350,-80)(450,-80)
\Gluon(390,-80)(390,-65){3}{2}
\Gluon(390,-65)(390,-50){3}{2}
\Gluon(390,-65)(410,-65){3}{2}
\Line(370,-35)(390,-50)
\Line(410,-35)(390,-50)
\Gluon(397,-45)(382,-35){3}{2}

\end{picture}
%\vskip2.5cm
%\caption{Feynman topologies contributing in lowest order to $e^+e^-\ar 
%q\bar q q'\bar q' gg$. The symbol $\times$ refers to the insertion of 
%the $e^+e^-\ar \gamma,Z$ current. Permutations are not shown.} 
%\label{fig_4q2g}
\end{center}
\begin{center}
\vskip4.25cm
\begin{picture}(500,100)
\SetScale{.9}
\SetWidth{1.2}
\SetOffset(0,0)

\Text(30,110)[]{(c)}

\Text(90,63.25)[]{$\times$}
\Text(90,53)[]{$1$}
\Line(50,70)(150,70)
\Gluon(90,70)(90,85){3}{2}
\Line(60,115)(90,85)
\Line(120,115)(90,85)
\Gluon(85,100)(95,90){3}{2}
\Line(70,115)(85,100)
\Line(85,100)(100,115)

\Text(225,63.25)[]{$\times$}
\Text(225,53)[]{$2$}
\Line(200,70)(300,70)
\Gluon(275,70)(275,95){3}{2}
\Line(255,115)(275,95)
\Line(295,115)(275,95)
\Gluon(225,70)(225,95){3}{2}
\Line(205,115)(225,95)
\Line(245,115)(225,95)

\Text(360,63.25)[]{$\times$}
\Text(360,53)[]{$3$}
\Line(350,70)(450,70)
\Gluon(390,70)(390,91){3}{3}
\Gluon(390,91)(390,112){3}{3}
\Gluon(390,91)(420,91){3}{3}
\Line(420,91)(425,108)
\Line(425,74)(420,91)
\Line(370,115)(390,112)
\Line(410,115)(390,112)

\end{picture}
\vskip-0.5cm
\caption{Tree-level Feynman topologies contributing to:
(a) $e^+e^-\ar q\bar q gggg$, 
(b) $e^+e^-\ar q\bar q q'\bar q' gg$ and
(c) $e^+e^-\ar q\bar q q'\bar q' q''\bar q''$. 
The symbol $\times$ refers to the insertion of 
the $e^+e^-\ar \gamma,Z$ current. Permutations are not shown.} 
%\label{fig_6q}
\label{fig_6partons}
\end{center}
\end{figure}

\vfill\clearpage

\subsection{The spinor methods}
\label{Subsect_helicity}

The use of helicity amplitude techniques in high energy physics
dates back to Refs.~\cite{jw,bj}. Since then, 
many different approaches have been
developed \cite{PL3}--\cite{PL12} and
a review of the vast literature on this subject is beyond
the intentions of this paper. For this, 
we remind the reader of Ref.~\cite{NP17}. 
Of those available in literature, we employed here {\sl two} methods.
 We have done so in order to be able to check adequately the correctness of our
calculations, as usual tests of gauge invariance might fail to reveal possible 
{\sl bugs} in the numerical codes. Note that the two
 methods can be used both with massless and massive particles.

The first one is based on the formalism of Ref.
 \cite{HZ} and exploits the {\tt HELAS} subroutines \cite{HELAS}. It is 
the same already  employed to produce the numerical results presented in 
\cite{iosix}. Here, one adopts
 a specific representation for the Dirac matrices. It is then
possible to obtain explicit expressions for the spinors, for the matrices at
the vertices and for the fermion propagators, and thus 
the components of the polarisation vectors of the gauge bosons
can be written directly in terms of their four-momenta.
The complex number corresponding
to a given diagram is obtained by multiplying these matrices. Therefore,
 one most conveniently programs the 
amplitudes directly as deduced from the Feynman rules 
and no manipulations on these need be performed. Masses
in this case turn out to be an entry in a propagator matrix.
It would be possible to consider some algebraic simplifications of the 
formulae, e.g., 
by separating the two independent chirality components of the {\sl massless}
spinors and cancelling by hand the terms proportional to the quark masses.
However, the method of Ref.~\cite{HZ}
is particularly well suited for numerical calculations
and such algebraic treatment would not significantly improve the speed of 
the programs. Therefore, we have not exploited here any chiral decomposition
in case of massless spinors.
Under these circumstances, the analytical expressions in the formalism of
Ref.~\cite{HZ} would simply be
a rewriting of the Feynman rules with the mentioned matrices expressed 
explicitly
in terms of the external momenta and helicities, an exercise that we do not 
perform here.

A second approach was based on the method developed in    
Refs.~\cite{KS,mana,method,ioPRD}.  
Since we will express the helicity amplitudes in this formalism, we
will devote same space to its technicalities. (This is also helpful
in order to introduce a notation that will be used in the Appendices and
to render the formulae appearing there more transparent to the reader.)

\begin{table}[!t]
\begin{center}
\begin{tabular}{|c|c|}
\hline
$\lambda_1\lambda_3$ &
$X(p_1,\lambda_1;p_2;p_3,\lambda_3;c_R,c_L)$ \\ \hline\hline
$++$  & $(\mu_1\eta_2 + \mu_2\eta_1)
(c_R\mu_2\eta_3 + c_L\mu_3\eta_2) + c_R S(+,p_1,p_2) S(-,p_2,p_3)$ \\ \hline
$+-$ & $c_L(\mu_1\eta_2 + \mu_2\eta_1)   S(+,p_2,p_3)
      + c_L (c_L\mu_2\eta_3 + c_R\mu_3\eta_2 )S(+,p_1,p_2)$ \\ \hline
\end{tabular}
\caption{The $X$ functions for the two independent helicity combinations
in terms of the functions $S$, $\eta$ and $\mu$ defined in the text.
The remaining $X$ functions can be obtained by flipping the sign
of the helicities
and exchanging $+$ with $-$ in the $S$ functions and $R$ with $L$ in the
chiral coefficients.}
\label{tab_X}
\end{center}
\end{table}

\begin{table}[!ht]
\begin{center}
\begin{tabular}{|c|c|}
\hline
$\lambda_1\lambda_2\lambda_3\lambda_4$ &
$Z(p_1,\lambda_1;p_2,\lambda_2;p_3,\lambda_3;p_4,
\lambda_4;c_R,c_L;c'_R,c'_L)$\\\hline \hline
$++++$ & $-2[S(+,p_3,p_1)S(-,p_4,p_2)c'_Rc_R
            -\mu_1\mu_2\eta_3\eta_4c'_Rc_L
            -\eta_1\eta_2\mu_3\mu_4c'_Lc_R]$  \\ \hline
$+++-$ & $-2\eta_2c_R[S(+,p_4,p_1)\mu_3c'_L-S(+,p_3,p_1)\mu_4c'_R]$ \\ \hline
$++-+$ & $-2\eta_1c_R[S(-,p_2,p_3)\mu_4c'_L-S(-,p_2,p_4)\mu_3c'_R]$ \\ \hline
$+-++$ & $-2\eta_4c'_R[S(+,p_3,p_1)\mu_2c_R-S(+,p_3,p_2)\mu_1c_L]$ \\ \hline
$++--$ & $-2[S(+,p_1,p_4)S(-,p_2,p_3)c'_Lc_R
            -\mu_1\mu_2\eta_3\eta_4c'_Lc_L
            -\eta_1\eta_2\mu_3\mu_4c'_Rc_R]$  \\ \hline
$+-+-$ & $0$ \\ \hline
$+--+$ & $-2[\mu_1\mu_4\eta_2\eta_3c'_Lc_L
+\mu_2\mu_3\eta_1\eta_4c'_Rc_R
-\mu_2\mu_4\eta_1\eta_3c'_Lc_R
-\mu_1\mu_3\eta_2\eta_4c'_Rc_L]$ \\ \hline
$+---$ & $-2\eta_3c'_L[S(+,p_2,p_4)\mu_1c_L-S(+,p_1,p_4)\mu_2c_R]$ \\ \hline
\end{tabular}
\caption{The $Z$ functions for all independent helicity combinations
in terms of the functions $S$, $\eta$ and $\mu$ defined in the text.
The remaining $Z$ functions can be obtained by flipping the sign 
of the helicities
and exchanging $+$ with $-$ in the $S$ functions and $R$ with $L$ in the
chiral coefficients.}
\label{tab_Z}
\end{center}
\end{table}

In this method, all spinors for any physical momentum are defined in
terms of a basic spinor of an  auxiliary light-like 
momentum. By decomposing the
internal momenta in terms of the external ones, using Dirac algebra
and rewriting the polarisation of external vectors by means of a spinor
current, all amplitudes can eventually be 
reduced to an algebraic combination of spinors 
products $\bar u(p_i)u(p_j)$ (with $i,j,...$ labelling the external 
particles). 

\vskip.15in

{  (i)} {\it Spinors.}
External fermions\footnote{Unless stated otherwise, we shall use the term
``fermion'' and the symbol ``$u$'' for both particles and antiparticles.}
 of mass $m$ and momentum $p^\mu$
are described by spinors
corresponding to states of definite helicity $\lambda$,
$u(p,\lambda)$\footnote{Here, $p$($\lambda$) represents a
generic (anti)spinor four-momentum(helicity).}, verifying the Dirac equations
\be
p\Dir u(p,\lambda)=\pm m u(p,\lambda),\qquad
\bar u(p,\lambda)p\Dir =\pm m \bar u(p,\lambda),
\ee
and the spin sum relation
\be\label{dirac}
\sum_{\lambda=\pm} u(p,\lambda)\bar u(p,\lambda)=
p\Dir\pm m,
\ee
where the sign $+(-)$ refers (here and in the following)
to a particle(antiparticle).
One can choose two arbitrary
vectors $k_0$  and $k_1$ such that
\be 
k_0\cdot k_0=0, \quad\quad k_1\cdot k_1=-1, \quad\quad k_0\cdot k_1=0,
\ee
and express the
spinors $u(p,\lambda)$ in terms of chiral ones
$w(k_0,\lambda)$ as
\be
u(p,\lambda)=w(p,\lambda)+\mu w(k_0,-\lambda),
\ee
where
\be
w(p,\lambda)=p\Dir w(k_0,-\lambda)/\eta,
\ee
and
\be
 \mu=\pm {m\over{\eta}}, \quad\quad\quad \eta=\sqrt{2|p\cdot k_0|}.
\ee
The spinors $w(k_0,\lambda)$ satisfy
\be
w(k_0,\lambda)\bar w(k_0,\lambda)=
{{1+\lambda\gamma_5}\over{2}}{k\Dir}_0,
\ee
and therefore
\be
\sum_{\lambda=\pm}w(k_0,\lambda)\bar w(k_0,\lambda)=
{k\Dir}_0.
\ee
The phase between chiral states is fixed by

\be
w(k_0,\lambda)=\lambda {k\Dir}_1 w(k_0,-\lambda).
\ee
The freedom in choosing $k_0$ and $k_1$ provides a powerful tool
for checking the correctness of any calculation.
A convenient, though not unique choice, is the following:
$k_0=(1,0,0,-1)$ and $k_1=(0,1,0,0)$ \cite{method}.
In such a case the {\sl massless} spinors in the two methods
\cite{KS} and \cite{HZ} coincide exactly, so that it is possible 
to compare in greater detail the two corresponding numerical codes.
In particular, the results obtained with the two formalisms must agree
for every single diagram and every polarisation of external particles. 

\vskip.15in

{(ii)} {\it Polarisation vectors for massless gauge 
bosons\footnote{Polarisation
 vectors for massive gauge bosons can be introduced on a similar
footing, however, since we do not need them for our calculation, we refer
the reader to the mentioned bibliography for this particular case.}.}
External spin 1 massless gauge bosons of momentum $p^\mu$ are described
by polarisation vectors
corresponding to states of definite helicity $\lambda$,
$\varepsilon^{\mu}(p,\lambda)$, fulfilling the identities
\be
\varepsilon(p,\lambda)\cdot p=0,
\quad\quad\quad
\varepsilon(p,\lambda)\cdot \varepsilon(p,\lambda)=0,
\ee
\be\label{spin}
\varepsilon^\mu(p,-\lambda)=\varepsilon^{\mu *}(p,\lambda),
\quad\quad\quad
\varepsilon(p,\lambda)\cdot \varepsilon(p,-\lambda)=-1,
\ee
and the spin sum relation (e.g., in the axial gauge)
\be\label{spinsum}
\sum_{\lambda=\pm}
\varepsilon^\mu(p,\lambda)
\varepsilon^{\nu *}(p,\lambda)=-g^{\mu\nu}+
{{q^\mu p^\nu+q^\nu p^\mu}\over{p\cdot q}},
\ee
where $q^\mu$ is any four--vector not proportional to $p^{\mu}$.
Any object
$\varepsilon^\mu(p,\lambda)$ obeying the relations 
(\ref{spin})--(\ref{spinsum})
makes an acceptable choice for the polarisation vectors. For instance
\be\label{polar}
\varepsilon^\mu(p,\lambda)=N
[\bar u(p,\lambda)\gamma^\mu u(q,\lambda)],
\ee
$N$ being the normalisation factor
\be\label{gluon_norm}
N=[4(q\cdot p)]^{-1/2}.
\ee
The existing freedom in choosing $q^\mu$ corresponds to fixing the gauge.
The final results do not depend on the choice of $q^\mu$.

\vskip.15in

{(iii)} {\it The $S$, $X$ and $Z$ functions.}
Using the above definitions one can compute the functions
\begin{equation}
S(\lambda,p_1,p_2)=[\bar u(p_1,\lambda) u(p_2,-\lambda)],
\end{equation}
\begin{equation}
X(p_1,\lambda_1;p_2;p_3,\lambda_3;c_R,c_L)=
[\bar u(p_1,\lambda_1) {p_{2}}\Dirin\Gamma u(p_3,\lambda_3)],
\end{equation}
and
$$
\hskip -1.2in
Z(p_1,\lambda_1;p_2,\lambda_2;p_3,\lambda_3;p_4,\lambda_4;
c_R,c_L;c'_R,c'_L)= 
$$
\begin{equation}
\hskip1.7in
[\bar u(p_1,\lambda_1) \Gamma^{\mu} u(p_2,\lambda_2)]
[\bar u(p_3,\lambda_3) \Gamma'_{\mu} u(p_4,\lambda_4)],
\end{equation}
where
\begin{equation}
\Gamma^{(')\mu}=\gamma^{\mu}\Gamma^{(')},
\end{equation}
and
\begin{equation}\label{vertex}
\Gamma^{(')}=c^{(')}_R P_R + c^{(')}_L P_L,
\end{equation}
with
\begin
{equation}P_R={{1+\gamma_5}\over{2}},\quad\quad\quad
P_L={{1-\gamma_5}\over{2}},
\end{equation}
the chiral projectors. \par
By computing the resulting traces one easily finds ($\epsilon^{0123} = 1$
is the Levi-Civita tensor) \cite{KS,mana}
\begin{equation}
S(+,p_1,p_2)= 2{{(p_1\cdot k_0)(p_2\cdot k_1)
 -(p_1\cdot k_1)(p_2\cdot k_0)
 +i\epsilon_{\mu\nu\rho\sigma}
  k^\mu_0k^\nu_1p^\rho_1p^\sigma_2}\over{\eta_1\eta_2}},
\end{equation}
for the $S$ functions
and the expressions listed in Tabs.~\ref{tab_X} and \ref{tab_Z}
for the $X$ and $Z$ functions, respectively. For the $S$ functions, one has
$S(-,p_1,p_2)= S(+,p_2,p_1)^*$,
while the remaining $X$ and $Z$ functions can be obtained as described
in the captions of Tabs.~\ref{tab_X}--\ref{tab_Z}.

%It is instructive to give a visual representation of the $S$, $Y$ and
%$Z$ functions, see 
%Fig.~\ref{fig_functions}.
%
%\begin{figure}[!t]
%\begin{center}
%\begin{picture}(450,250)
%\SetScale{.9}
%\SetWidth{1.2}
%\SetOffset(50,0)
%
%\ArrowLine(70,200)(40,230)
%\ArrowLine(40,170)(70,200)
%\Text(5,210)[]{$p_1,+\lambda$}
%\Text(5,150)[]{$p_2,-\lambda$}
%
%\Text(103,181)[]{$= S(\lambda,p_1,p_2)$}
%
%\ArrowLine(70,50)(40,80)
%\ArrowLine(40,20)(70,50)
%\Text(10,75)[]{$p_1,\lambda_1$}
%\Text(10,15)[]{$p_2,\lambda_2$}
%\Text(44,46)[]{$\Gamma$}
%
%\DashLine(70,50)(130,50){4}
%\Text(187,45)[]{$= Y(p_1,\lambda_1; p_2, \lambda_2; c_R, c_L)$}
%
%\ArrowLine(70,-100)(40,-70)
%\ArrowLine(40,-130)(70,-100)
%\Text(10,-60)[]{$p_1,\lambda_1$}
%\Text(10,-120)[]{$p_2,\lambda_2$}
%\Text(44,-90)[]{$\Gamma^\mu$}
%
%\Photon(70,-100)(130,-100){5}{4}
%
%\ArrowLine(130,-100)(160,-70)
%\ArrowLine(160,-130)(130,-100)
%\Text(168,-60)[]{$p_3,\lambda_3$}
%\Text(168,-120)[]{$p_4,\lambda_4$}
%\Text(138,-90)[]{${\Gamma'}_\mu$}
%
%\Text(264,-90)[]{$= Z(p_1, \lambda_1; p_2, \lambda_2; 
%                      p_3, \lambda_3; p_4, \lambda_4; 
%                      c_R, c_L; {c'}_R, {c'}_L)$}
%
%\end{picture}
%\vskip5.5cm
%\caption{Diagrammatic representation of the $S$, $Y$ and $Z$ functions.
%Here, the wavy line mimics a spin 1 boson (apart from the factor
%$\frac{1}{q^2-M^2}$, with $q=p_1+p_2\equiv 
%p_3+p_4$) whereas the dashed one refers
%to a spin 0 one.}
%\label{fig_functions}
%\end{center}
%\end{figure}

The colour factors have been calculated with two different
methods, as illustrated in Ref.~\cite{color} and by means of the 
techniques reported in Ref.~\cite{Cvitanovic}. Some possible
expressions of both the amplitudes and the colour
matrices will be given in the Appendices. The coded
helicity amplitudes have been checked for gauge invariance and integrated
using {\tt VEGAS} \cite{VEGAS}, in both formalisms. The agreement
at the (squared) amplitude level is always within 12 significant digits in 
{\tt REAL*8} precision.
 
Although the actual number of Feynman diagrams needed to compute processes
(\ref{qqgggg})--(\ref{qqqqqq}) is very large, 
only a much
smaller portion of different topologies\footnote{By `topology' we mean
a collection of Feynman diagrams which differ only in the permutations of
gauge vectors along the fermion lines (neglecting their colour structure).}
describe these. In fact, in the case of 
two-quark-four-gluon diagrams the topologies are those shown in 
Fig.~\ref{fig_6partons}(a), eight in total. 
For four-quark-two-gluons, Fig.~\ref{fig_6partons}(b), one has ten
and for six-quarks, Fig.~\ref{fig_6partons}(c), three topologies.
The same implementation of each of the topologies pictured
in Fig.~\ref{fig_6partons} can then 
be used several times by means of recursive 
permutations of the momenta of the external particles. Furthermore, each
common `sub-topology' to those in  Fig.~\ref{fig_6partons} is saved and then 
reused in the same numerical evaluation.
The actual number of Feynman diagrams corresponding to the 
topologies in Fig.~\ref{fig_6partons},
including all possible permutations of momenta and counting
as either one or three each diagram involving a four-gluon vertex 
(which carries in fact three different colour structures), is given in 
Tab.~\ref{tab_numbers}.
 
\begin{table}[!t]
\begin{center}
\begin{tabular}{|c|c|c|c|c|c|c|c|c|c|}
\hline
\multicolumn{10}{|c|}
{\rule[0cm]{0cm}{0cm}
$e^+e^-\ar q\bar q gggg$} \\ \hline
\rule[0cm]{0cm}{0cm}
a1  & a2  & a3 & a4 & a5 & a6 & a7 & a8 & $~$ & $~$ \\ \hline
120 & 144 & 24[72] & 72 & 36 & 8[24]  & 12 & 12[36] & $~$ & $~$ \\ \hline
\multicolumn{10}{|c|}
{\rule[0cm]{0cm}{0cm}
Total 428[516]} \\ \hline\hline

\multicolumn{10}{|c|}
{\rule[0cm]{0cm}{0cm}
$e^+e^-\ar q\bar q q'\bar q' gg$ $(q\neq q')$} \\ \hline
\rule[0cm]{0cm}{0cm}
b1 & b2 & b3 & b4 & b5 & b6 & b7 & b8 & b9 & b10 \\ \hline
12 & 48 & 48 & 24 & 24 & 8  & 4[12]  & 8  & 4  & 16 \\ \hline
\multicolumn{10}{|c|}
{\rule[0cm]{0cm}{0cm}
Total 196[204]} \\ \hline\hline

\multicolumn{10}{|c|}
{\rule[0cm]{0cm}{0cm}
$e^+e^-\ar q\bar q q'\bar q' q''\bar q''$ $(q\neq q'\neq q'')$} \\ \hline
\rule[0cm]{0cm}{0cm}
c1 & c2 & c3 & $~$ & $~$ & $~$ & $~$ & $~$ & $~$ & $~$ \\ \hline
24 & 18 & 6  & $~$ & $~$ & $~$ & $~$ & $~$ & $~$ & $~$ \\ \hline
\multicolumn{10}{|c|}
{\rule[0cm]{0cm}{0cm}
Total 48} \\ \hline
\end{tabular}
\caption{Number of Feynman diagrams corresponding to the topologies
in Fig.~\ref{fig_diagrams}. If two quark flavours are identical, numbers
are multiplied by two, if three are, by six. In square brackets are the same
numbers if the four-gluon diagrams are counted thrice, according to their
three colour structures.}
\label{tab_numbers}
\end{center}
\end{table}

\subsection{Jet clustering algorithms}
\label{Subsect_algorithms}

As jet clustering algorithms we have adopted the Jade (J) \cite{jade}, 
Durham (D) \cite{durham}
and Cambridge (C) \cite{cambridge} schemes\footnote{In this Section as well
as in the following ones, in order to 
simplify the notation, we shall use $y$ to represent $\ycut$, the 
jet resolution parameter discussed in Sect.~\ref{Sect_intro}. 
In addition, we acknowledge here our abuse in referring to the various
jet `finders'
both as algorithms and as schemes, since the last term was originally
intended to identify the composition law of four-momenta when pairing two 
clusters. This is in fact a well admitted
habit which we believe will not generate confusion in our discussion.}.
The first one uses the expression
\be
\label{measureJ}
y_{ij}^J = {{2E_i E_j(1-\cos\theta_{ij})}
\over{s}},
\ee
for the `jet measure', whereas the other two are based on the same 
test variables
\be
\label{measureD}
y_{ij}^{D,C} = {{2\min (E^2_i, E^2_j)(1-\cos\theta_{ij})}
\over{s}},
\ee
but differ in the clustering procedure. 
In eqs.~(\ref{measureJ})--(\ref{measureD}), 
$E_i$ and $E_j$ represent the energies of any pair of partons $i$ 
and $j$, whereas $\theta_{ij}$ is their relative angle. A six-jet sample
is selected by imposing the constraint $y_{ij}\ge y$ on all $n$ possible
parton combinations $ij$ (with $n=15$ for the Jade and Durham and 
$5\leq n\leq 15$ for the Cambridge algorithm). The $n=15$ pairs 
of the J and D schemes are simply obtained by pairing
all partons in the final state in all possible different combinations $ij$.
The variable number $5\leq n\leq 15$ for the C scheme is a consequence of the
so-called `soft freezing' step, which is described in 
\cite{cambridge}, to which we refer the reader for further details.

\subsection{Numerical parameters}
\label{Subsect_numerics}

We have adopted $M_Z=91.17$ GeV, $\Gamma_Z=2.516$ GeV,
$\sin^2 \theta_W=0.23$, $\alpha_{em}= 1/128$ and, unless otherwise stated,
 the two-loop expression
for $\alpha_{s}$, with 
$\Lambda_{\tiny{\mbox{QCD}}}^{n_f=4}=0.298$ GeV. 
\begin{table}[!h]
\begin{center}
\begin{tabular}{|c|c|c|c|c|}
\hline
\rule[0cm]{0cm}{0cm}
$~$ & $\gamma$ & $Z$ & $W$ & $g$  \\ \hline\hline
\rule[0cm]{0cm}{0cm}
$c_R$ & $Q^{f}$& $g_R^f/s_W c_W$        &  
$0$ & $1$ \\ \hline
$c_L$ & $Q^{f}$& $g_L^f/s_W c_W$ &  
$1/\sqrt{2}s_W$  &  $1$ \\ \hline
\end{tabular}
\caption{The couplings $c_R$ and $c_L$ of eq.~(\ref{vertex}) for $u$- and
$d$-type (anti)quarks and electrons/positrons
to the gauge bosons $\gamma$, $Z$, $W$ and $g$. One has
(adopting the notations $s_W\equiv \sin\theta_W$ and 
$c_W\equiv \cos\theta_W$)
$g_R^f=-Q^fs^2_W$ and $g_L^f=T_3^f-Q^fs^2_W$ (with $q=u,d$), where
$(Q^u,T^u_3)=(+\frac{2}{3},+\frac{1}{2})$, 
$(Q^d,T^d_3)=(-\frac{1}{3},-\frac{1}{2})$ and
$(Q^e,T^e_3)=(-{1},-\frac{1}{2})$
are the fermion charges and isospins.}
\label{tab_cRcL}
\end{center}
\end{table}
We have kept all quarks massless as a default, to speed up the 
numerical evaluation.
Though, in several instances we have retained  a finite value for the bottom
mass, i.e., $m_b=4.25$ GeV. Electron and positron have mass zero too. 
When $e^+e^-\ar t\bar t\ar b\bar b W^+W^-\ar$ 6-jet events have been 
calculated (at tree-level), 
the following values of $t$- and $W^\pm$-parameters were adopted:
$m_t=175$ GeV, $\Gamma_t=1.54$ GeV, 
$M_W=80.23$ GeV and $\Gamma_W=2.2$ GeV. 
As centre-of-mass (CM) energies representative of LEP1, LEP2 and NLC, we have
used the values $\sqrt s=M_Z,180$ GeV and 500 GeV, respectively.

The vertex couplings (or `chiral coefficients') $c_R$ and $c_L$ of 
eq.~(\ref{vertex})
between quarks/leptons and gauge bosons
entering in the expressions of the $S$, $Y$ and $Z$ functions can be found
in Tab.~\ref{tab_cRcL} (we have included also the $W^\pm q\bar q'$ vertex, 
for completeness, though it will not appear in our formulae).

\section{Results}
\label{Sect_results}

\begin{figure}[htb]
\begin{center}
~\epsfig{file=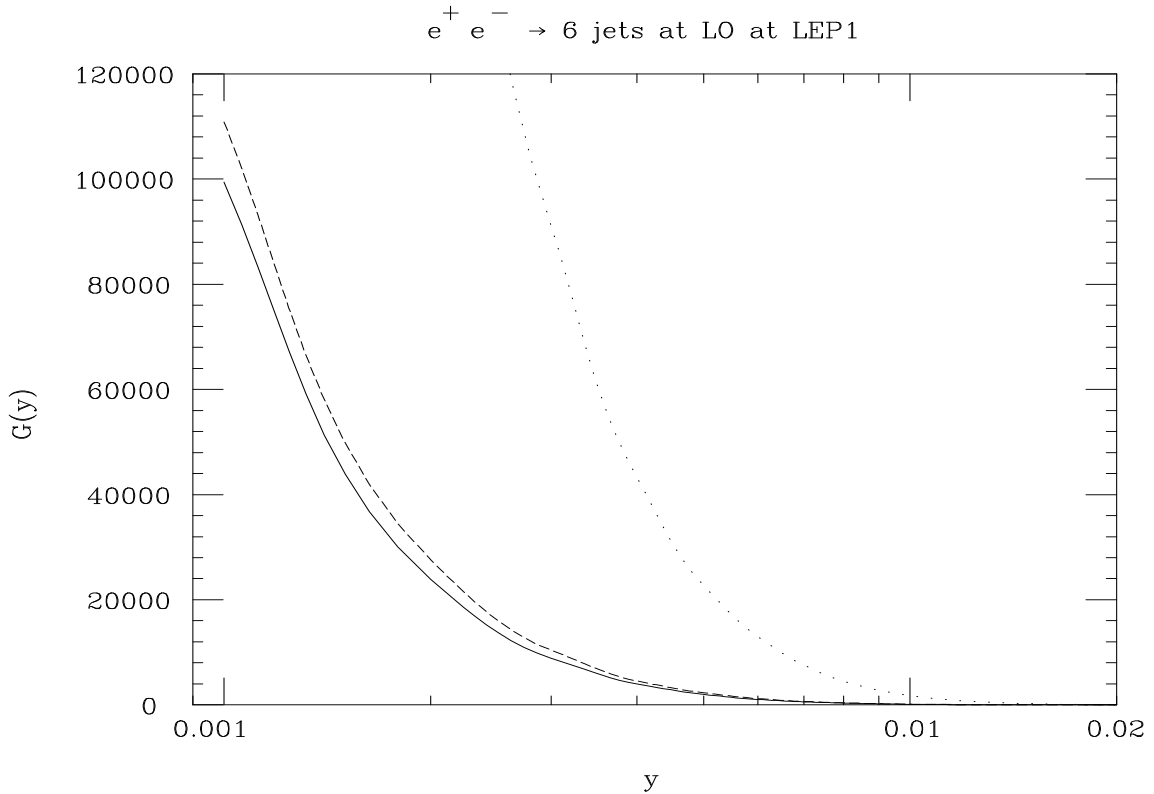,width=12cm,height=18cm,angle=90}
\caption{The parton level $G(y)$ function entering in the six-jet fraction
at LO in the D (continuous line), C (dashed line) and J (dotted line)
schemes, at LEP1.}
\label{fig_newjet6}
\end{center}
\end{figure}

We define the $y$-dependent six-jet fraction by means of the relation
\be
\label{def6}
f_6(y)=\frac{\sigma_6(y)}
                  {\sum_m \sigma_m(y)}
            =\frac{\sigma_6(y)}
                  {\st},
\ee
where $\sigma_6(y)$ is the actual six-parton cross section and $\st$
identifies the {\sl total} hadronic rate
$\st=\sigma_{0}(1+\alpha_s/\pi+...)$, $\sigma_{0}$ being the 
Born cross section. In perturbative QCD one can rewrite eq.~(\ref{def6})
in terms of a series in $\alpha_s$, beginning with its fourth power, as
\be
\label{f6}
f_6(y) =     \left( \frac{\alpha_s}{2\pi} \right)^4  G(y) +..., 
\ee
where $G(y)$ is the lowest-order (or leading-order, LO)
`coefficient function' of the six-jet rate.
We show this quantity in Fig.~\ref{fig_newjet6}, for the case of  
the Durham, Cambridge and Jade schemes, at LEP1\footnote{Here and in the 
following, unless otherwise stated, 
the summations over the three reactions (\ref{qqgggg})--(\ref{qqqqqq}) and 
over all possible massless
combinations of quark flavours in each of these have been performed.}.
Our rates for the Jade and Durham schemes nicely match the `leading-order'
predictions (both in $\alpha_s$ and $\log(1/y)$) given in 
Ref.~\cite{garth} in the region of $y\approx1\times10^{-4}$ for the Durham
scheme and when $y\approx5\times10^{-4}$ for the Jade one.
Here and in the following, though, we will take as a lower limit of
our $y$-range the value of $0.001$, where fixed-order perturbation theory
should be much more reliable.

A  salient feature of Fig.~\ref{fig_newjet6} is that the Jade rate
is much above the Durham and Cambridge ones. At $y=0.001$ the ratio between
the former and the latter two is about 8.0. (Notice that the corresponding
number for the three-, four- and five-jet rates 
is 2.3, 3.6 and 5.7, respectively.) This effect is due to the fact
that $y_{ij}^J$ is always greater than or equal to $y_{ij}^{D,C}$, as
can be easily understood from eqs.~(\ref{measureJ})--(\ref{measureD}). 
Therefore, for a specified $y$ and a given configuration of energies
and angles $E_i,E_j$, $\cos\theta_{ij}$, more events will 
pass the cut $y_{ij}^{J,D,C}>y$ in the J case than in the D and C ones.

We do not push any further similar comparisons involving the J scheme.
We have retained it in Fig.~\ref{fig_newjet6} because it has been used
extensively at LEP1 in the past years. However, such algorithm has been 
proven to be more sensitive to hadronisation effects than the D and C
algorithms, especially at low $y$ values, where the six-jet fraction
onsets \cite{cambridge,BKSS}. Furthermore, the Jade rates 
do not show the usual Sudakov exponentiation of
multiple soft-gluon emission \cite{BS1}, despite having
an expansion of the form $\as^n\ln^m(1/y)$ at small values of the resolution
parameter, contrary to the case of the Durham and Cambridge schemes
(see \cite{cambridge} and references therein),
and generally have a scale dependence of the fixed-order
perturbative rates 
(e.g., the three- \cite{BKSS} and four-jet \cite{slac2} fractions)
larger than that of the D and C schemes.
Thus, given the mentioned flaws,  in the following
we will give total and differential rates only for the last two 
cases\footnote{Apart from one example, when we will need to compute
a cross section using the J scheme for purpose of comparison against
results published in literature.}.

As for these, one should notice that the 
larger rate for the C scheme as compared to the D one is
a direct consequence of the `soft freezing' procedure of resolved jets
described in  Ref.~\cite{cambridge}. The step of eliminating
from the sequence of clustering 
the less energetic one in a resolved pair of particles (i.e., 
with $y_{ij}^{C}>y$), implemented in the C algorithm,
tends to enhance the final jet multiplicity
of the original D scheme. In fact, the procedure prevents the attraction 
of the remaining particles (at large angle) into new unresolved pairs,
whose momenta would then be merged together, producing a lower number of 
final jets. For example, at $y=(0.001)[0.005]\{0.010\}$, relative differences
between the two algorithms are of the order of $(10)[14]\{16\}\%$.

The jet fraction $f_6(y)$ and cross section $\sigma_6(y)$  
corresponding to the rates in Fig.~\ref{fig_newjet6} 
are given in Tab.~\ref{tab_jet6}, for a representative selection
of $y$'s. The value of $\alpha_s$ adopted is 0.120, whereas that
used for $\st$ is 39.86 nb. 
Given the total luminosity collected at LEP1 in the 1989-1995 years (more than
$10^7$ hadronic events have been recorded by the four Collaborations), the 
six-jet event rate is comfortably measurable. For example, for a luminosity
of, say, 100 pb$^{-1}$ per experiment, perturbative QCD predicts at LO
some 58,000 original six-parton events recognised as six-jet ones, for
$y=0.001$ in the C scheme (a number that decreases to approximately 52,000
if the D algorithm is adopted instead).

\begin{table}[t]
\begin{center}
\begin{tabular}{|c|c|c|}
\hline
\rule[0cm]{0cm}{0cm}
$y$ & $f_6(y)$ & $\sigma_6(y)$ (pb) \\ \hline\hline
\rule[0cm]{0cm}{0cm}
$0.001$ & $(1.31)[1.47]\times10^{-2}$  & $(523.91)[584.19]$ \\
$0.002$ & $(3.16)[3.65]\times10^{-3}$  & $(125.95)[145.44]$ \\
$0.003$ & $(1.17)[1.37]\times10^{-3}$  & $(46.81)[54.72]$ \\
$0.004$ & $(5.29)[6.01]\times10^{-4}$  & $(21.09)[23.96]$ \\
$0.005$ & $(2.63)[3.05]\times10^{-4}$  & $(10.49)[12.17]$ \\
$0.006$ & $(1.38)[1.59]\times10^{-4}$  & $(5.49)[6.35]$ \\
$0.007$ & $(7.90)[8.85]\times10^{-5}$  & $(3.15)[3.53]$ \\
$0.008$ & $(4.80)[5.20]\times10^{-5}$  & $(1.91)[2.07]$ \\
$0.009$ & $(2.77)[3.08]\times10^{-5}$  & $(1.10)[1.23]$ \\
$0.010$ & $(1.64)[1.93]\times10^{-5}$  & $(0.65)[0.77]$ \\ \hline\hline
\multicolumn{3}{|c|}
{\rule[0cm]{0cm}{0cm}
LEP1} \\ \hline
\end{tabular}
\caption{Jet fraction and cross section rates for $e^+e^-\ar$~6~jets 
at LEP1, for several values of $y$ in the Durham and Cambridge schemes,
in round and squared brackets, respectively. The numerical errors do not
affect the significant digits shown.}
\label{tab_jet6}
\end{center}
\end{table}

The rates given in Figs.~\ref{fig_newjet6}
and Tab.~\ref{tab_jet6} are LO results. 
Next-to-leading order (NLO) corrections proportional to ${\cal O}(\alpha_s^5)$
are expected to be large. 
In fact, the size of higher-order (HO) 
contributions generally increases with the
power of $\alpha_s$ and with the number
of particles in the final state or, in other terms, with 
their possible permutations (i.e., with the number of different
possible attachments of both the additional 
real and virtual particles to those appearing in lowest order). 
As already mentioned,
the highest-order corrections calculated to date in $e^+e^-\ar n$-jet
annihilations are the NLO ones to the four-jet rate, that is,  terms
proportional to $\alpha_s^3$ \cite{cpt,slac}. The total 
four-jet cross section at NLO was found to be larger
than that obtained at LO in $\alpha_s^2$ by a K-factor of 
1.5 or more, depending
on the value of $y$ implemented during the clustering procedure and
the algorithm used as well. 
Therefore, we should expect that the rates given in Tab.~\ref{tab_jet6}
underestimate the six-jet rates by {\sl at least} a similar factor.
Indeed, assuming a K~$\approx2$ correction, our results are in a
satisfactory agreement with both data and phenomenological 
Monte Carlo (MC) programs \cite{jetfractionsLEP1}.

In this respect the MCs \cite{Jetset}--\cite{Ariadne} 
largely exploited in experimental analyses apparently
perform better than the fixed-order
 perturbative calculations, especially at low $y$-values, where
the latter need to be supported by the additional contribution of
perturbation series involving powers of 
$\log(1/y)$ 
resummed to all orders (for the case of the D and C scheme, see 
Refs.~\cite{CDOTW,BS2,cambridge}) in order to fit the same data.
However, as such MCs generally implement
only the infrared (i.e., soft and collinear) dynamics of 
quarks and gluons in the standard `parton-shower (PS) + ${\cal O}(\alpha_s)$
Matrix Element (ME)' modelling, in many cases their description of the 
large $y$-behaviour and/or that of the interactions of secondary
`branching products' is (or should be expected to be) no longer adequate. 
In fact, this has been shown to be the case, e.g., for some typical
angular quantities of four-jet events \cite{ALEPHgluino,GC}. 
As discussed in Ref.~\cite{spin}, these differences are presumably due to a
poor description  of
the spin correlations among  the various partons
implemented in the MCs:
in particular, between the primary (i.e., from the
$\gamma^*,Z^{(*)}$-splitting)
and secondary (i.e., from the gluon splitting into gluons or quarks)
pairs. In
contrast, once ${\cal O}(\alpha_s^2)$ MEs are inserted and properly matched
to the hadronisation stage, e.g., using the JETSET string fragmentation model
\cite{Pythia} (see also Ref.~\cite{andre}), then the agreement is 
recovered\footnote{In this context, one should however mention that  
ME models with `added-on' fragmentation, i.e.,
`${\cal O}(\alpha_s^2)$ + hadronisation'  cannot be reliably extrapolated
from one energy to another, as the fragmentation tuning is 
energy dependent. To obviate this, a special
`${\cal O}(\alpha_s^2) + \mbox{PS} + \mbox{cluster hadronisation}$' 
version of HERWIG  
is now in preparation \cite{doc}.}.
If one considers that four-jet events represent only the 
next-to-lowest-order QCD interactions in $e^+e^-$ scatterings, then
it is not unreasonable to argue that further complications might well
arise as the final state considered gets more and 
more sophisticated, such as in six-jet events. 

\begin{figure}[htb]
\begin{center}
~\epsfig{file=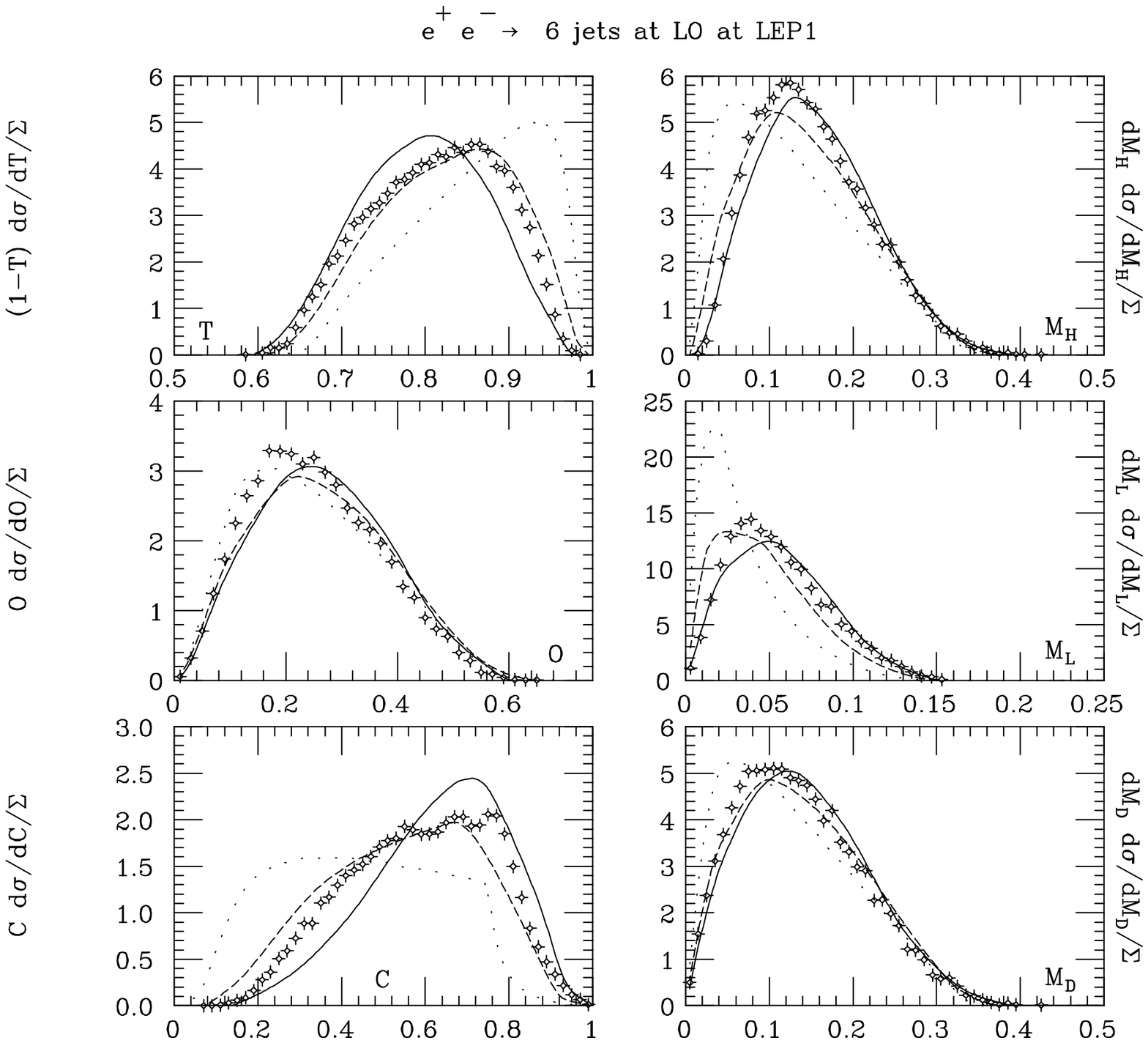,width=16cm,height=18cm,angle=0}
\caption{Differential distributions in the following shape variables:
thrust ($T$), oblateness ($O$), C-parameter ($C$), heavy ($M_H$), light 
($M_L$) and difference ($M_D$) jet mass
for the six- (continuous line), five- (dashed line) and four-jet 
(dotted line) rates at LO, at LEP1.
The `star' points correspond instead to the six-jet rate as obtained by using
the HERWIG Monte Carlo. 
The C scheme was used with resolution parameter $y=0.001$.
Notice that the distributions have been normalised to a common
factor (one) for readability.}
\label{fig_shape1_6j_lep1}
\end{center}
\end{figure}

\begin{figure}[htb]
\begin{center}
~\epsfig{file=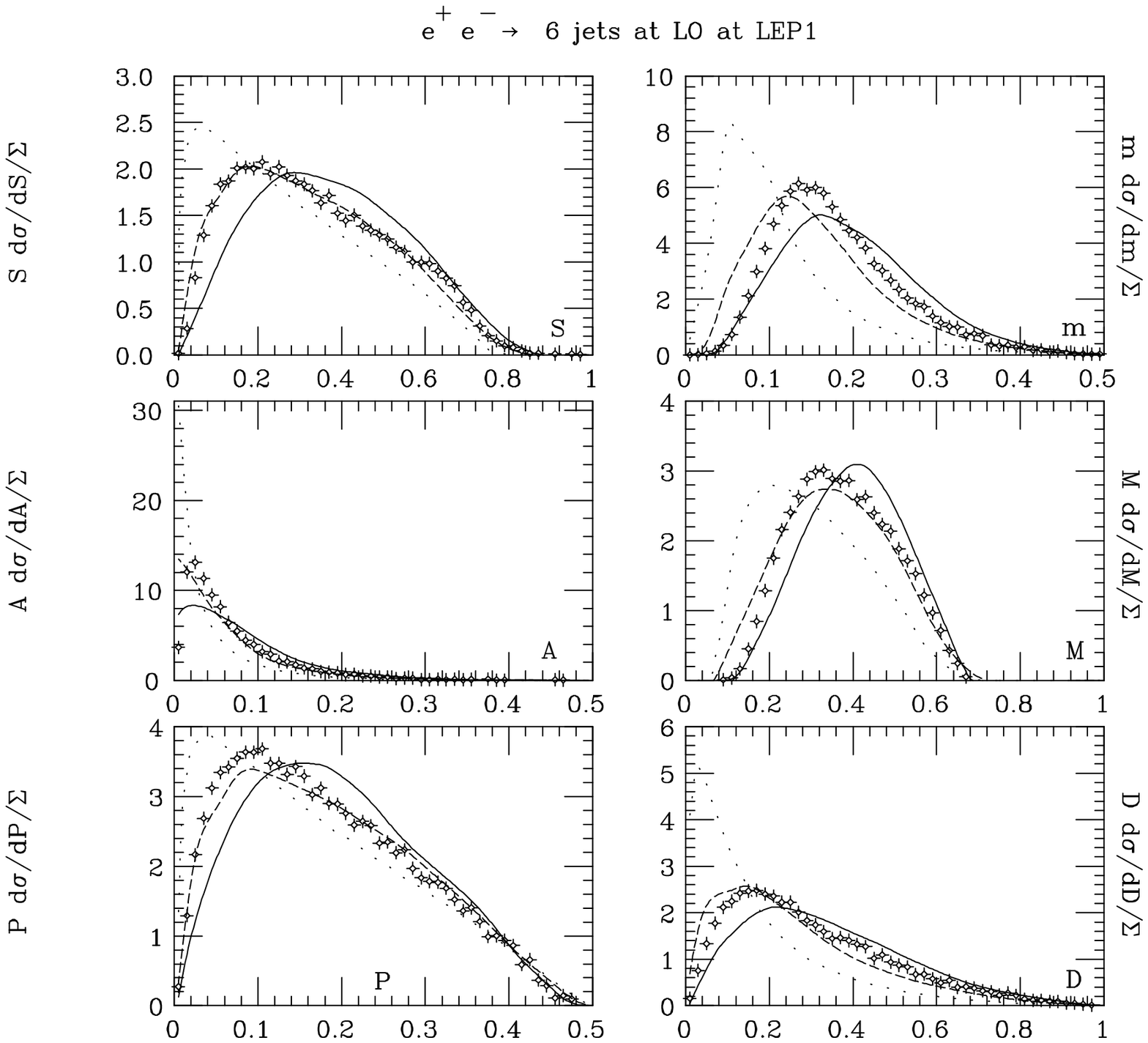,width=16cm,height=18cm,angle=0}
\caption{Differential distributions in the following shape variables:
sphericity ($S$), aplanarity ($S$), planarity ($P$), minor  ($m$) and
major ($M$) value  and D-parameter ($D$)
 for the six- (continuous line), five- (dashed line) and four-jet 
(dotted line) rates at LO, at LEP1.
The `star' points correspond instead to the six-jet rate as obtained by using
the HERWIG Monte Carlo. 
The C scheme was used with resolution parameter $y=0.001$.
Notice that the distributions have been normalised to a common
factor (one) for readability.}
\label{fig_shape2_6j_lep1}
\end{center}
\end{figure}

In order to emphasise this point, we present in 
 Figs.~\ref{fig_shape1_6j_lep1}--\ref{fig_shape2_6j_lep1} 
the differential distributions of the six-jet rate (along
with the five- and four-jet ones)
in some typical shape variables used in QCD studies in electron-positron
annihilations. They are the thrust $T$ \cite{T}, the oblateness $O$ \cite{O}, 
the C-parameter $C$ \cite{C}, the heavy $M_H$, light $M_L$ and difference 
$M_D$ jet masses \cite{M_H}, the sphericity $S$, aplanarity
$A$ and planarity $P$ \cite{S}, the minor $m$ and
major $M$ values \cite{O} and the D-parameter \cite{C}.
(Their allure  in testing the underlying QCD dynamics has been well
illustrated in, e.g., Ref.~\cite{EVENT}.) 
In Figs.~\ref{fig_shape1_6j_lep1}--\ref{fig_shape2_6j_lep1} 
 the algorithm used
to produce the plots was the Cambridge one, with $y=0.001$.

As compared to the same distributions in the case of lower jet multiplicity 
events (such as, e.g., four-jet production), those in 
Figs.~\ref{fig_shape1_6j_lep1}--\ref{fig_shape2_6j_lep1} for the 
six- (more markedly) and five-jet (in some instances) rates are much more 
`central'. That is, the peaks of the distributions are generally further away
from the infrared regions, those  which dominate the phenomenology of
$n$-jet events, for $n=3$ and $4$. 
(The only exception is the C-parameter, which is indeed
expected to play a r{\^o}le apart since there exists a 
singularity at the boundary
of the three-parton phase space $C=2/3$, see Ref.~\cite{CatWeb}, 
affecting the four-jet distribution but not the five- and six-jet ones.)  
This is a clear indication that other dynamics (other than the soft and 
collinear emission of gluons off quark lines) has a perceptible influence on 
differential jet rates, as the multiplicity of the latter increases. 
In other terms,
one should expect the above mentioned spin correlations among the partons 
 to 
modify quite significantly the infrared behaviour dictated by 
the propagators when the energy of the particles decreases,
as happens in the six-parton case and
particularly for the secondary and tertiary branching products of 
the $g\ar q\bar q$, $g\ar gg$ and $g\ar ggg$ splittings (see 
Fig.~\ref{fig_cascade}). 

To further stress this, 
we also plot in Figs.~\ref{fig_shape1_6j_lep1}--\ref{fig_shape2_6j_lep1} 
the same shape variables distributions as obtained at hadron level 
for the six-jet (i.e., $n=6$) rate
by using the HERWIG event generator, with identical choice of jet algorithm
and resolution as above. As a matter of fact, the MC spectra depart
in most cases quite substantially from those obtained by using the exact 
${\cal O}(\as^4)$ MEs, notably, they `point' towards the singularities.
Such behaviour is particular evident for thrust and sphericity, two
variables to which we will come back below.
Under these circumstances, we believe the
availability of exact perturbative calculations of six-jet events to be 
essential in order to  model correctly
the HO parton dynamics for both future tests of QCD (like those 
outlined here) and QCD background studies (as we will
show later on for the case of $t\bar t\ar6$-jet production and decay 
at NLC).

\bigskip

\begin{figure}[htb]
\begin{center}
~\epsfig{file=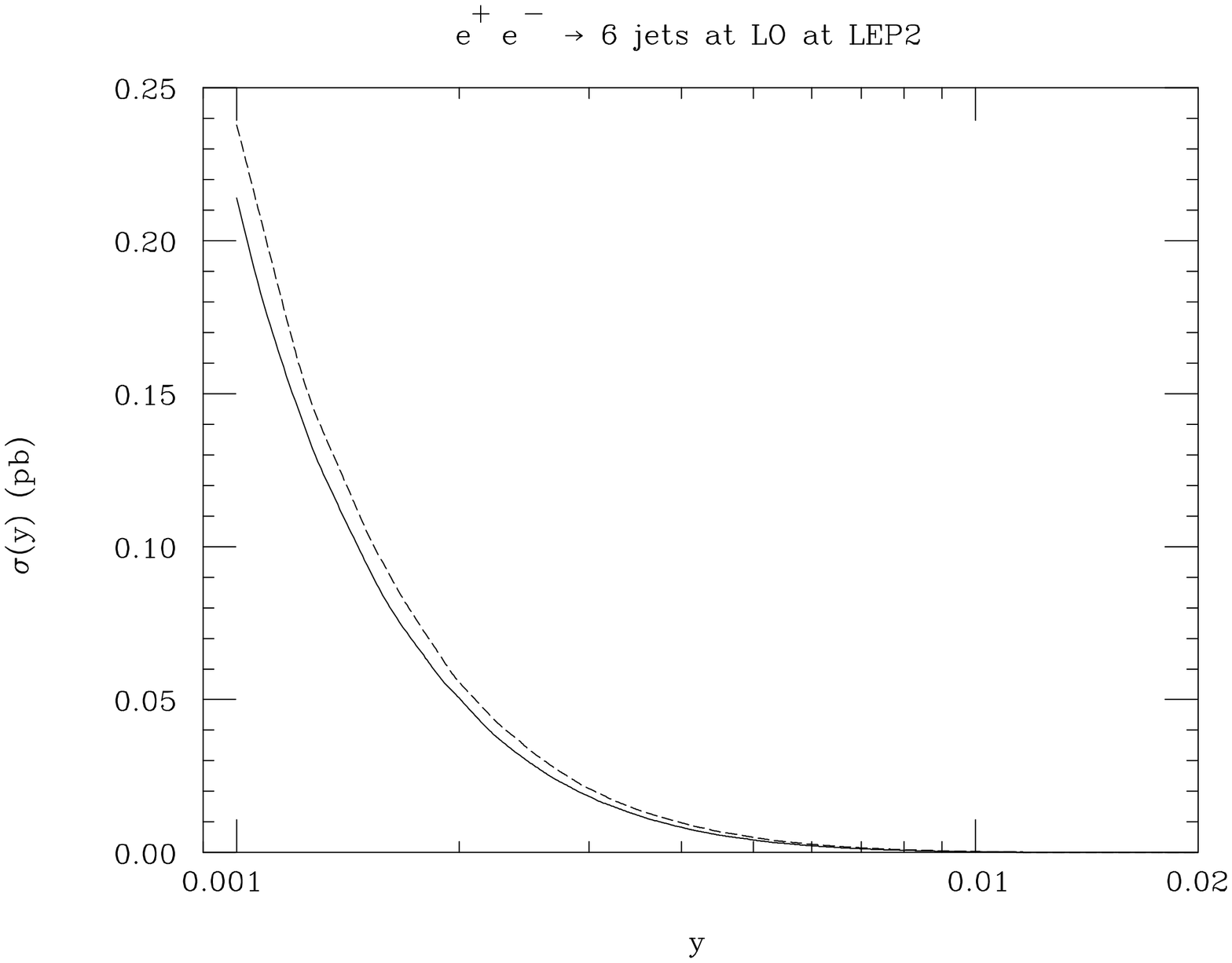,width=12cm,height=16cm,angle=90}
\caption{The total cross section of six-jet events
at LO in the D (continuous line) and C (dashed line) schemes, at LEP2.}
\label{fig_newjet6_lep2}
\end{center}
\end{figure}

The total hadronic cross section falls drastically 
when increasing the CM energy from
the LEP1 to the LEP2 values, by more than three orders of magnitude,
and so does the six-jet rate. Nonetheless, six-jet fractions have been
measured also during the 1995-1996 runs at the CM energies of 130--136,
161 and 172 GeV, when a total luminosity of around 27 pb$^{-1}$ was collected,
and are currently under further investigation (at 183 GeV and higher).
Results can be found for the lower energies in Ref.~\cite{jetfractionsLEP2}.
Within a `typical' K-factor of 2 (which tentatively
quantifies the ratio between the NLO and the LO six-jet rates also
at LEP2)
the values of $f_6(y)$ as reconstructed from our rates  
are always well compatible with those produced by the MCs used
by e.g., the ALEPH Collaboration \cite{6j}. (Notice that this is also true 
at $\sqrt s=161$ GeV, where a six-jet `excess' has apparently been 
observed ~\cite{6j}). For reference, we present
the jet rates at $\sqrt s=180$ GeV in Fig.~\ref{fig_newjet6_lep2} in the
form of the total cross section. The relative differences between
the two D and C algorithms are similar to those already
encountered at LEP1. 

Quite apart from its relevance in QCD jet fractions, the six-jet rate
at LEP2 can in principle represent a source of background to the so
called `colour reconnected' $W^+W^-$ hadronic decays \cite{colour}.
The problem goes as follows. Colour exchange (both at perturbative and
non-perturbative level) can take place
between the decay products of the two $W^\pm$'s, thus destroying the notion
of two separate $W^\pm$ decays and hence casting doubts on our abilities
to reconstruct accurately the mass of the resonances. In perturbative terms,
such effect is due to the presence of interference terms in the total
cross section for $W^+W^-\ar$ multi-parton production and decay, 
which account for the
exchange of real or virtual partons emitted in the decay of
 one of the $W^\pm$'s and absorbed by the partons generated by the other.
 Simple rules of colour conservation show that such colour effects
cannot occur at first order in $\as$, but only in higher orders
(i.e., at least two gluons need to be exchanged).   
At the non-perturbative level, a similar phenomenon takes place
when the colour strings generated among the partons
from the first $W^\pm$ decay overlap with those from the second one,
break down and form new strings which may join quarks from different
shower cascades
(e.g., in the language of the Lund fragmentation model \cite{lund}).
Of the two contributions, the non-perturbative one is expected
 to be dominant
  \cite{tor_val}. The consequence is similar at both levels: namely,
particles are generated that cannot be unambiguously assigned to one $W^\pm$
decay or another.

It is of decisive importance to quantify accurately the above effects,
especially if one wants to achieve a high 
precision measurement of the mass $M_W$ of the $W^\pm$ boson (a
target accuracy of $\pm(40-50)$~MeV or less is expected, 
see, e.g., Ref.~\cite{accuracy}). 
In fact, at present, their dynamics is
 relatively unknown. On the one hand, a complete
${\cal O}(\as^2)$ perturbative 
calculation of $W^+W^-$ production and decay does not
exist to date, as only the tree-level part of $e^+e^-\ar W^+W^-\ar$
6 jets has been computed so far \cite{WW6}. On the other hand, 
in the non-perturbative modelling, many more additional parameters are needed
to describe the reconnection effects than the well tested 
mechanisms of the quark and gluon fragmentation, so that
there are many phenomenological programs now on the market, indeed
predicting rather different sizes for the above colour effects 
\cite{allreconnection}.

\begin{figure}[htb]
\begin{center}
~\epsfig{file=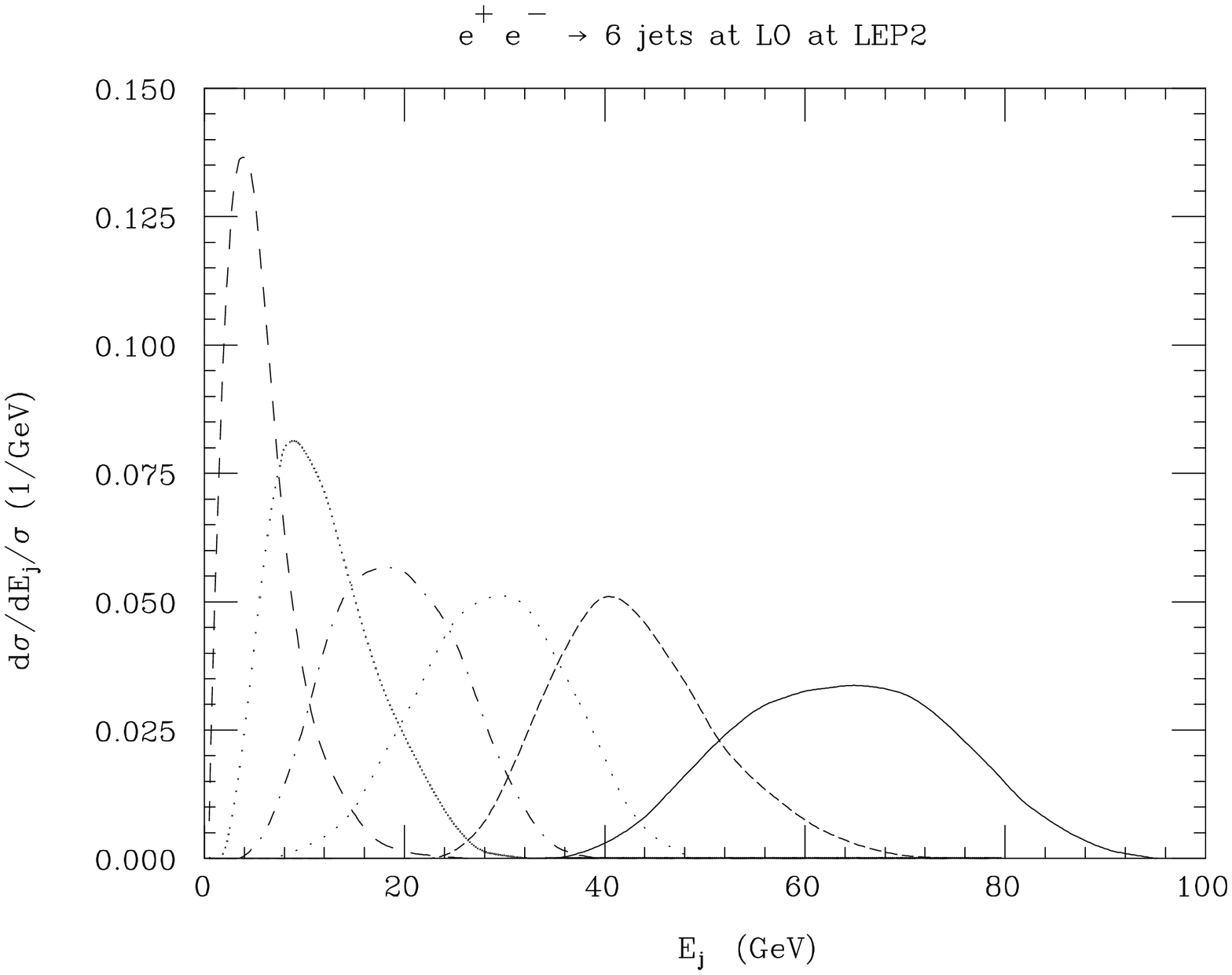,width=12cm,height=16cm,angle=90}
\caption{The energy spectra $E_j$ of six-jet events
at LO in the J scheme for $y=0.001$, at LEP2 (here,
$\protect\sqrt s=175$ GeV). Jets are energy-ordered such that
 $E_1>E_2>...>E_5>E_6$: 
$j=1$ (continuous line),
$j=2$ (short-dashed line),
$j=3$ (dotted line),
$j=4$ (dot-dashed line),
$j=5$ (fine-dotted line) and
$j=6$ (long-dashed line).
Normalisation is to unity.}
\label{fig_jet6_energy_lep2}
\end{center}
\end{figure}

Whereas the treatment of non-perturbative colour rearrangement effects is
beyond the scope of the paper, we come back to the perturbative ones.
In particular, we would like to recall that Ref.~\cite{WW6}
gave an estimate of the size of the latter at LO, in the form of
the part of the $e^+e^-\ar W^+W^-\ar$~6-jet cross section due to the mentioned
interference terms. The effect was found to be quite small: e.g.,  at the level
of $\approx4\times 10^{-3}$ pb at $y=0.001$ for the
Jade algorithm, with $\sqrt s=175$ GeV and using $\alpha_s=0.115$. 
The cumulative rates for processes (\ref{qqgggg})--(\ref{qqqqqq})
for the same jet clustering scheme and the identical combination of $y$, 
$\sqrt s$ and  $\alpha_s$ amount to approximately $2$ pb, well above the colour
reconnected rates in $W^+W^-\ar$~6-jet events. Since in the latter case colour
exchange proceeds via gluons, in 
Ref.~\cite{WW6} the corresponding energy spectra 
were  studied, in order to give some hints about the sort of energy at which
perturbative colour reconnection takes place. The softest gluon energy 
distribution was found to peak at about 8 GeV, whereas that of the
most energetic one lies at approximately 20 GeV. To a good degree of 
approximation, the two gluon energy spectra coincide with those of the
two least energetic particles in $W^+W^-\ar$~6-jet decays.  
In Fig.~\ref{fig_jet6_energy_lep2}, we have studied the energy spectra of
the partons produced via reactions (\ref{qqgggg})--(\ref{qqqqqq}). 
Coincidentally enough, the fifth and the fourth most energetic ones
show their maxima around 8 and 20 GeV, 
respectively\footnote{Further notice that also in events of the type
(\ref{qqgggg})--(\ref{qqqqqq}) the fifth and fourth most energetic partons
are in most cases
gluons, because of the dominance in the total QCD cross section of
two-quarks-four-gluon events, with the two fermion being the most energetic
ones.}. That is, they behave
on the same footing as the gluon jets in $e^+e^-\ar W^+W^-\ar$ 6-jet events,
which are responsible for perturbative colour rearrangement. Though both
the studies in Ref.~\cite{WW6} and those presented here
need to be eventually supported by loop-calculations as well, it is clear
that six-jet events from QCD represent a considerable source of irreducible
background to hadronic $W^+W^-$ events perturbatively 
reconnected.

\begin{figure}[htb]
\begin{center}
~\epsfig{file=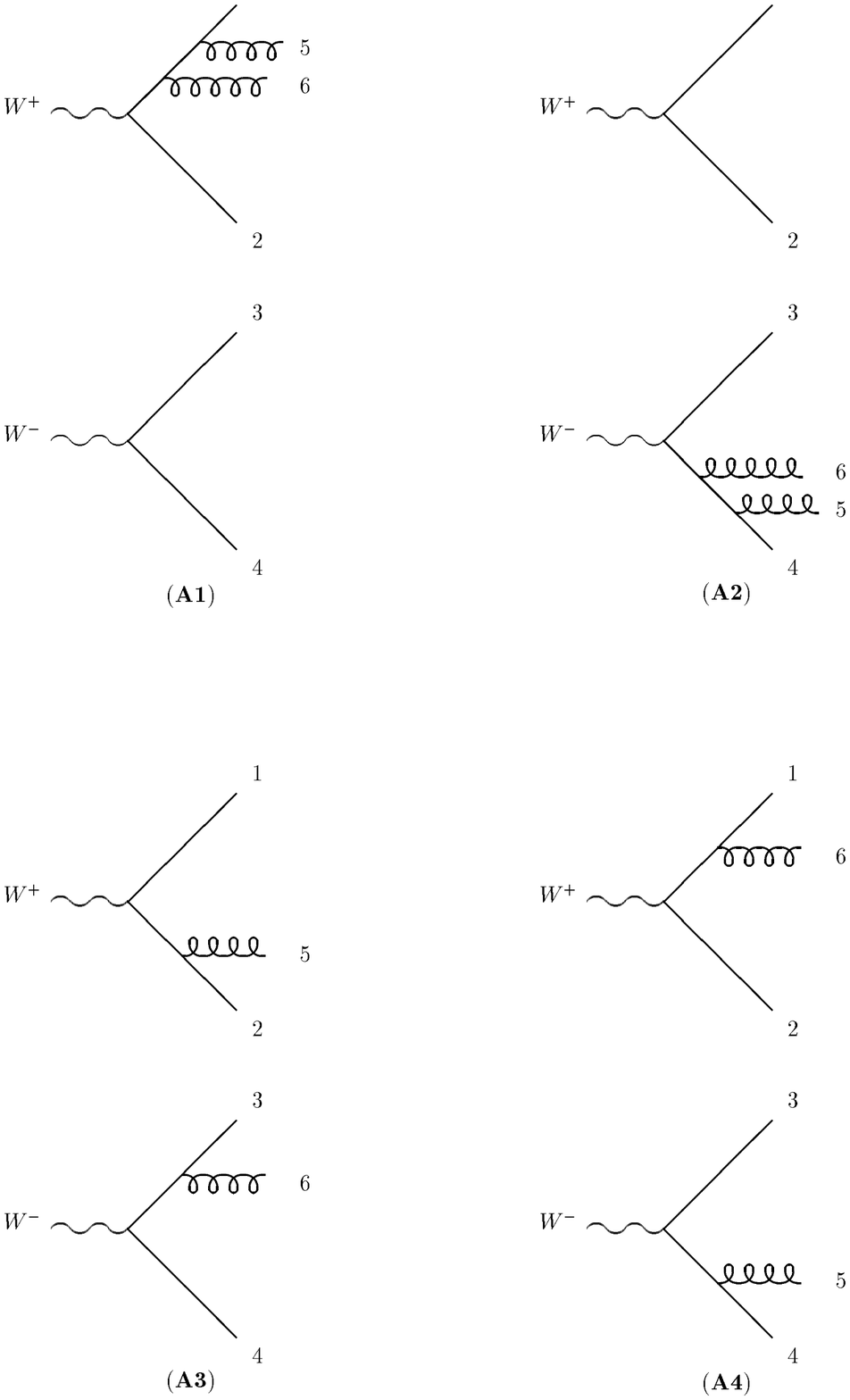,width=16cm,height=18cm,angle=0}
\vskip0.001cm
\caption{Subsets of Feynman diagrams contributing to 
$q\bar q q'\bar q' gg$ production via $W^+W^-$. For simplicity the leptonic
part of the diagrams is not shown. 
Permutations are not shown either.}
\label{fig_reconnection}
\end{center}
\end{figure}

\begin{figure}[tbh]
\begin{center}
~\epsfig{file=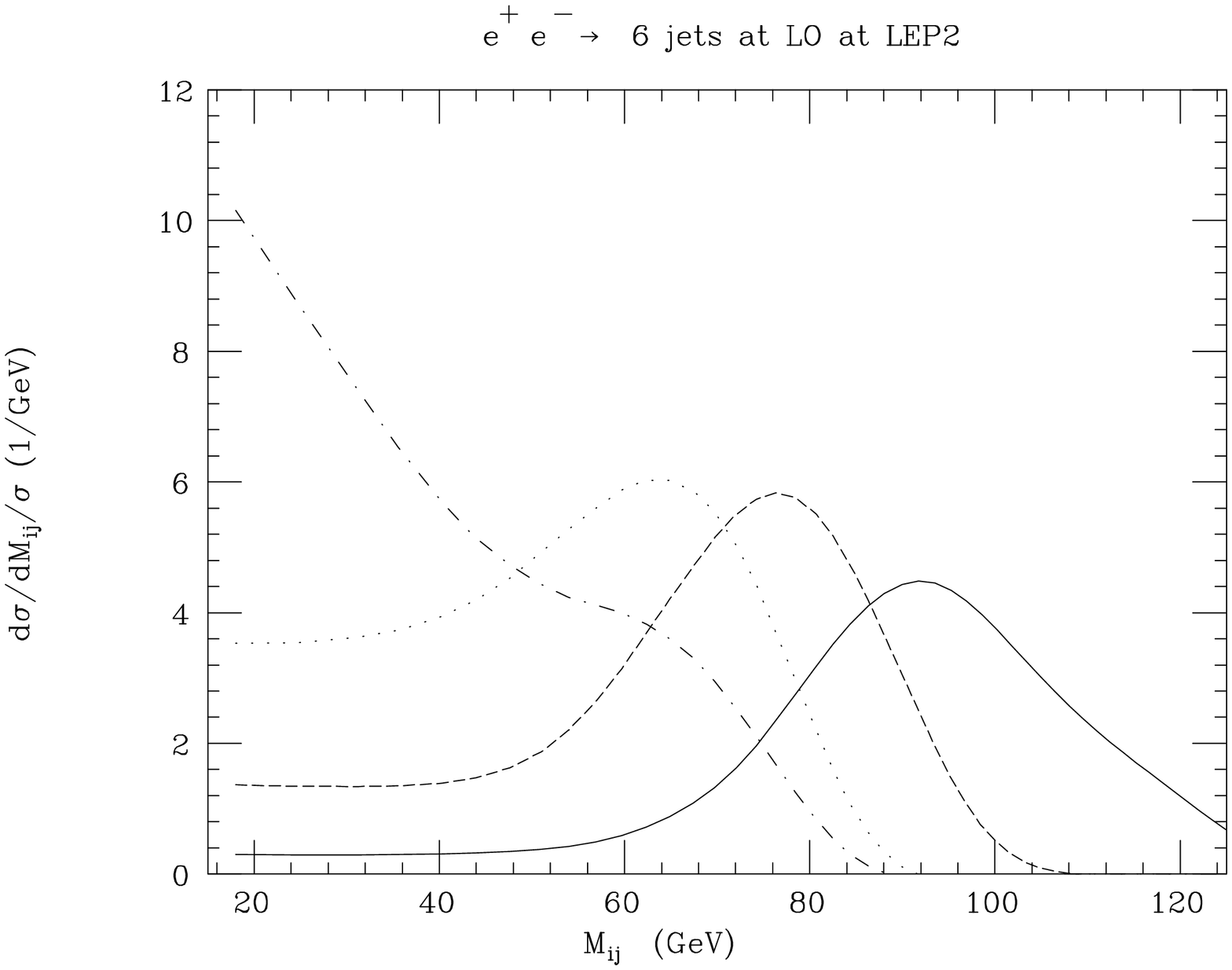,width=12cm,height=16cm,angle=90}
\caption{The distributions in the true invariant mass $M_{ij}$  
of events of the type (\ref{qqgggg})--(\ref{qqqqqq}) at LO in the C scheme
with $y=0.001$, at LEP2, for the following combinations of parton pairs $ij$: 
${12}$ (continuous lines),  
${13}$ (dashed lines),  
${14}$ (dotted lines) and   
${23}$ (dot-dashed lines).}
\label{fig_mass_6j_lep2}
\end{center}
\end{figure}

As a matter of fact, one might consider imposing a cut such
as $M_{jj}\approx M_W$, that is, asking that one di-jet invariant mass 
(see below, eq.~(\ref{mij}))
is close to the value of the $W^\pm$ mass,
as one might naively expect to occur naturally 
in $W^+W^-\ar$ 6-jet events. However, this could
well be counter-productive, for two reasons. 
Firstly, the colour reconnected contribution to $e^+e^-\ar W^+W^-\ar$
6-jet events is defined as the sum of two terms, namely, 
the interference between 
the set of diagrams labelled ({\bf A1}) and ({\bf A2}) 
plus that between  ({\bf A3}) and ({\bf A4}) in
Fig.~\ref{fig_reconnection}\footnote{We
would like to thank the authors of
Ref.~\cite{WW6} for their kind permission of exploiting
here one of the figures of their paper.}.
(Note that the cross term between ({\bf A1})$\oplus$({\bf A2})
 and ({\bf A3})$\oplus$({\bf A4}) is identically
zero). Neither of the two terms has  
a distinctive peak at $M_{jj}\approx M_W$,
in line with the notion of colour exchange taking place 
between the two $W^\pm$ decays.
Secondly, the invariant mass of the di-jet combination formed by
pairing the first and third most energetic jets from six-jet QCD events peaks
not far from the $W^\pm$ mass, i.e., at $\approx76$ GeV (less than 
two $\Gamma_W$ from $M_W$) and is rather broad in the region 
$M_{jj}\approx M_W$, as can be appreciated in Fig.~\ref{fig_mass_6j_lep2}
(here, our default algorithm, resolution parameter and CM energy for LEP2
have been used again, as the shape of the distribution is not significantly
affected by either of those, provided $\sqrt s\gg 2M_W$ for the energy). 
Further notice that other QCD di-jet mass distributions populate
the region around the $M_W$ resonance, such as the combinations $ij=12,14$
and 23 (the other spectra are much softer, as they concentrate at low
invariant masses).
Evidently then, under these circumstances, our understanding of perturbative 
colour interconnection effects in $W^+W^-\ar$ 6-jet decays is bound to our
ability in suppressing the corresponding QCD 
background. This operation can however be attempted only in the
context of a detailed Monte Carlo simulation, including parton shower, 
hadronisation and detector effects, which is beyond our capabilities.

\bigskip

The six-jet event rate that will be collected at
 NLC can be rather large, despite the cross section being more than a factor
$10^4$ smaller than at LEP1 (e.g., at $\sqrt s=500$ GeV).
This is due to the large yearly luminosity expected 
at this machine, around 10 fb$^{-1}$. For such a value and assuming a 
standard evolution of the coupling constant with the increasing energy 
(that is, no non-Standard Model thresholds occur up to 500 GeV), at the minimum
of the $y$-values considered here (i.e., $y=0.001$), one should expect some  
220 events per annum by adopting the D scheme and about 13\% more if one
adopts the C one. However, these rates decrease rapidly as the resolution 
parameter gets larger, by a factor of 50 or so at $y=0.005$ and of 
approximately 800 at $y=0.01$. 
 
\begin{figure}[tbh]
\begin{center}
~\epsfig{file=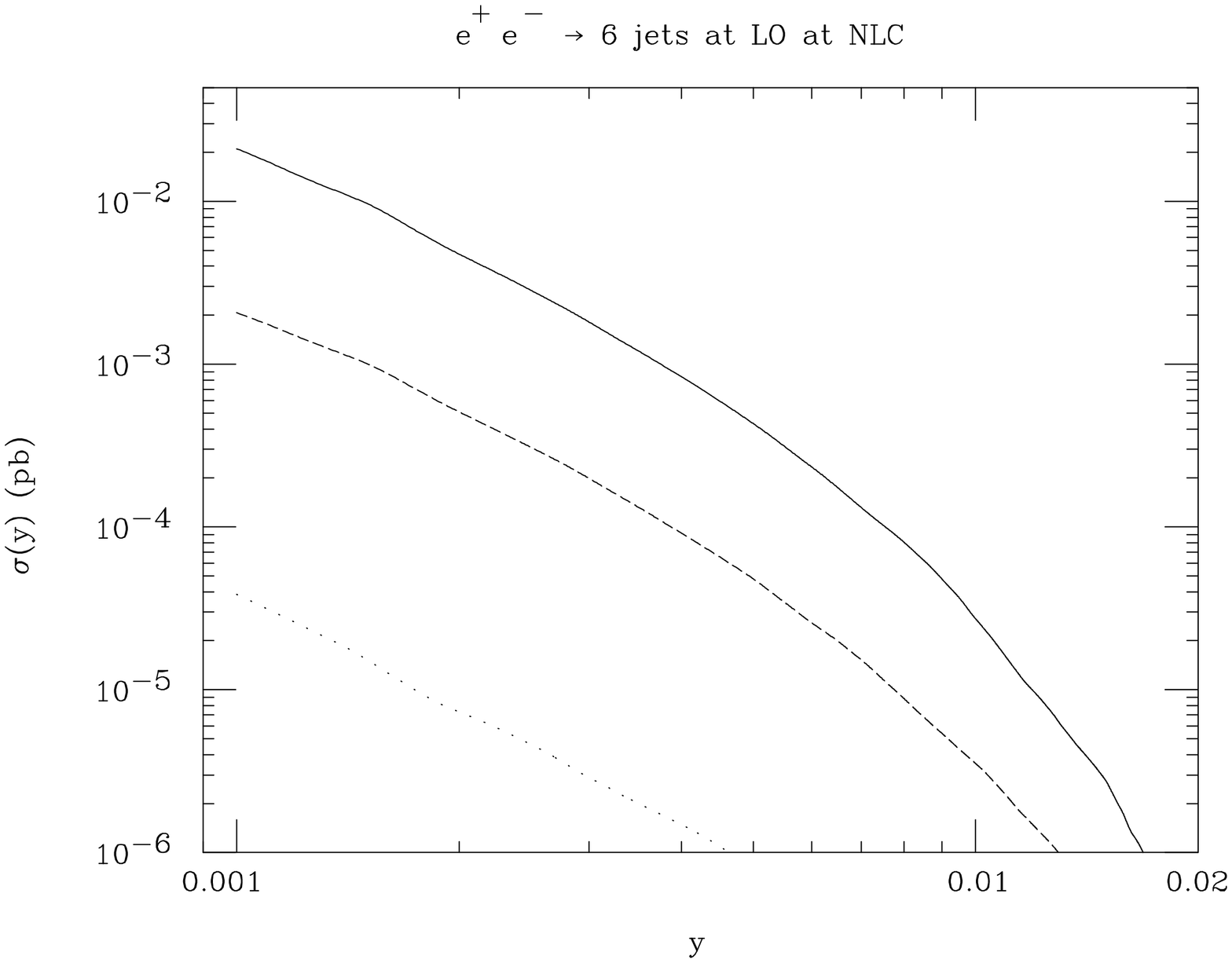,width=12cm,height=16cm,angle=90}
\caption{The total cross section of six-jet events 
at LO in the C scheme, at NLC, decomposed in terms of the three contributions
$e^+e^-\ar q\bar q gggg$ (continuous line), $e^+e^-\ar q\bar q q'\bar q' gg$
(dashed line) and $e^+e^-\ar q\bar q q'\bar q' q''\bar q''$ (dotted line).}
\label{fig_comp6nlc}
\end{center}
\end{figure}

It is also interesting to look at the composition of the 
total rates in terms of the three processes (\ref{qqgggg})--(\ref{qqqqqq}). 
Whereas this is
probably of little concern at LEP1 and LEP2, the capability of the detectors
of distinguishing between jets due to quarks (and among these, bottom flavours
in particular: e.g., in tagging top and/or Higgs decays) 
and gluons, is essential at NLC, in order to
perform dedicated searches for old and new particles. 
The separate behaviours of the 
reactions in (\ref{qqgggg})--(\ref{qqqqqq})
can be appreciated in Fig.~\ref{fig_comp6nlc}, e.g., 
for the case of the C scheme, in terms of total cross sections.
The rates for the D algorithm follow
the same pattern. 

\begin{figure}[tbh]
\begin{center}
~\epsfig{file=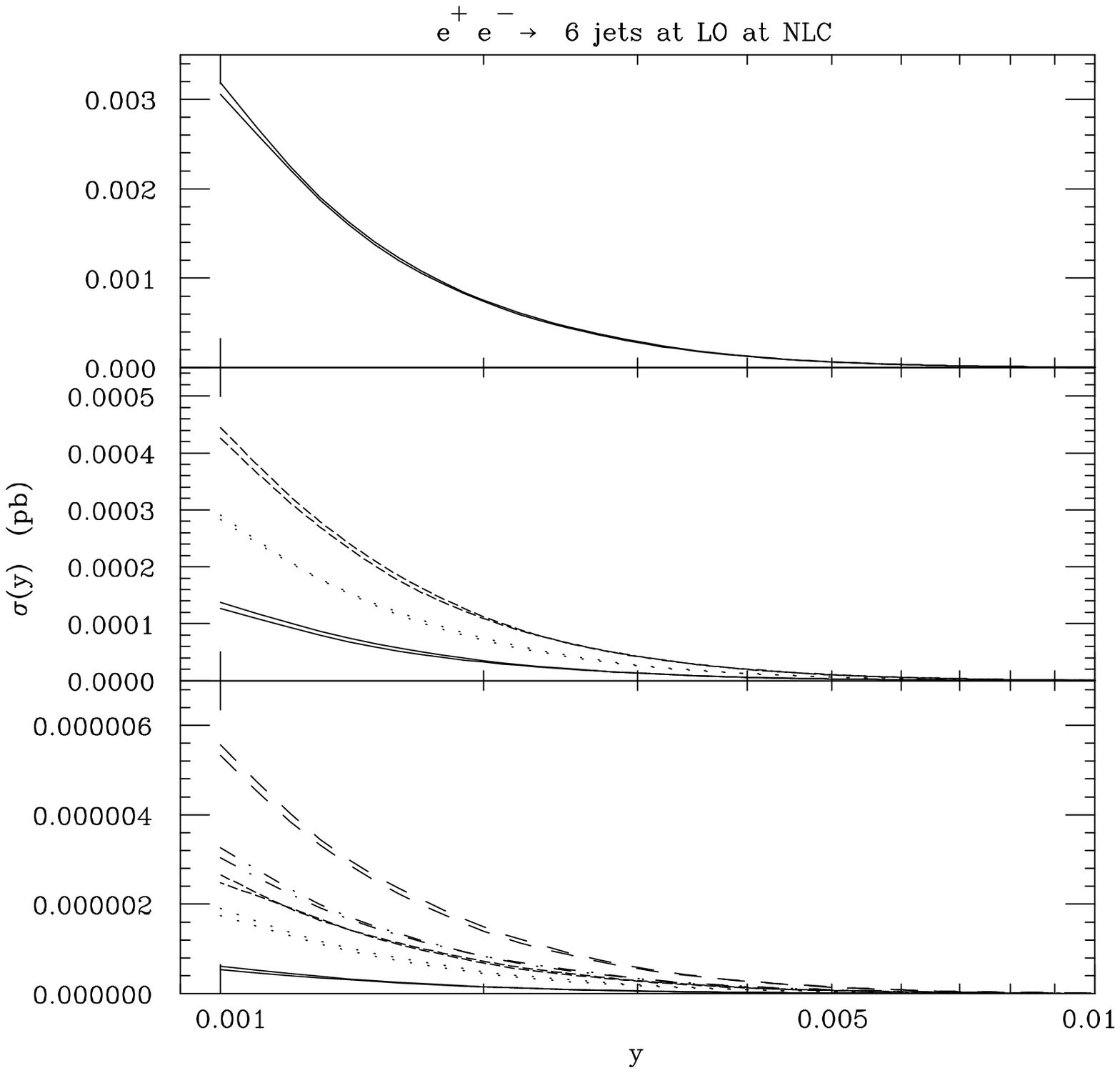,width=16cm,height=18cm,angle=0}
\caption{The total cross section of six-jet events 
at LO in the C scheme, at NLC, decomposed in terms of the contributions
containing $b$-quarks. Upper plot: 
$e^+e^-\ar b\bar b gggg$ (continuous line).
Middle plot: 
$e^+e^-\ar b\bar b b\bar b gg$ (continuous line),
$e^+e^-\ar b\bar b u\bar u gg$ (dashed line) and 
$e^+e^-\ar b\bar b d\bar d gg$ (dotted line).
Lower plot: 
$e^+e^-\ar b\bar b b\bar b b\bar b$ (continuous line),
$e^+e^-\ar b\bar b b\bar b u\bar u$ (short-dashed line),
$e^+e^-\ar b\bar b b\bar b d\bar d$ (dotted line),
$e^+e^-\ar b\bar b u\bar u u\bar u$ (dot-dashed line) and 
$e^+e^-\ar b\bar b u\bar u d\bar d$ (long-dashed line); here,
the case $e^+e^-\ar b\bar b d\bar d d\bar d$ has not been reproduced
as it visually coincides with $e^+e^-\ar b\bar b b\bar b d\bar d$.
In the upper lines $m_b=0$, in the lower lines $m_b=4.25$ GeV.}
\label{fig_b_effect}
\end{center}
\end{figure}

The tagging of $b$-quarks produced, e.g., in the hadronic
top decay chain $t\bar t\ar b\bar bW^+W^-\ar b\bar bjjjj$, is of decisive 
importance, non only to 
suppress the QCD background (mainly proceeding via gluon events,
see Fig.~\ref{fig_comp6nlc}) 
but also to reduce the combinatorics involved
in selecting three jets out of six, in case no flavour 
identification is exploited (twenty combinations in total). If $b$-quark 
jets can be recognised (via their semileptonic decays, 
via $\mu$-vertex devices, etc.), then only six three-jet 
combinations $bjj$ (or $\bar bjj$) need to be selected, one of which 
is such that $M_{bjj(\bar bjj)}\approx m_t$. If $\epsilon_b$ is the efficiency
of tagging one single $b$-jet within a certain fiducial volume,
and two taggings are required, than $\epsilon_b^2$ is the overall rate
for identifying events containing exactly two $b$-quarks
(we assume no correlations between the two $b$-taggings and that the
charge of the heavy quark is recognised). This is the case
which applies to the subprocesses
$e^+e^-\ar b\bar bgggg$ and $e^+e^-\ar b\bar b q\bar q q'\bar q'$ (with 
$q,q'\ne b$)\footnote{For simplicity, we do not consider here the case of
$c$-quark misidentified as $b$-quarks. We can presumably assume that by the
time NLC will be operative the efficiency in distinguishing
between  $c$- and $b$-quark jets will be much higher than at present.
The rejection factor against light quark jets (produced by $u$-, $d$- and
$s$-quarks) as well as against gluon jets is very high already at present,
so that we do not need to worry about their contamination in the present 
analysis.}.
The probability of misidentifying six-jet events produced via four-$b$-quark
reactions
(i.e., $e^+e^-\ar b\bar b b\bar b gg$ or $e^+e^-\ar b\bar b b\bar b 
q\bar q$ (with
$q\ne b$)) as $b\bar bjjjj$ events is then $[2\epsilon_b(1-\epsilon_b)]^2$.
If there are six $b$-quarks in the event, as in
$e^+e^-\ar b\bar b b\bar b b\bar b$, then the corresponding number
is $[3\epsilon_b(1-\epsilon_b)^2]^3$. The total rates as a function
of $y$ in the C scheme for six-parton process involving $b$-quarks are
given in Fig.~\ref{fig_b_effect}, for the case in which only $u$ and $d$
massless contributions are considered (for $c$ and $s$, respectively, one 
gets very similar results, even when the masses, e.g., $m_c=1.35$ GeV 
and $m_s=0.30$ GeV, are retained in the calculation).

If one multiplies the rates given in Fig.~\ref{fig_b_effect} by the
overall $b$-tagging efficiencies mentioned in the previous paragraph, then 
one realises that six-jet events produced via four- and six-$b$-quarks
subprocesses yield negligible rates at NLC
(assuming, e.g., $\epsilon_b=70\%$, the current LEP value). Events 
involving only 
two $b$-quarks 
can instead produce a detectable number of $b\bar bjjjj$ events,
especially at low $y$. For example, for $y=0.001$ in the C scheme one
gets some 15 QCD  background events in $b\bar bjjjj$ samples
(including the $b$-tagging efficiency $\epsilon_b^2$), after a
luminosity  
of ten inverse femtobarns has been collected. Though this is not
a large number per se, considering that $e^+e^-\ar t\bar t\ar 6$ jets
yields about 800 doubly $b$-tagged
events at the same $y$-value in the above scheme, 
one should recall that very high precision measurements of top parameters
(such as $m_t$, $\Gamma_t$, the branching ratios, etc.) 
are expected to be carried out at NLC \cite{tt_theory}. 
Under these circumstances, even small backgrounds need to be quantified
accurately.

Whereas mass effects (markedly, of $b$-quarks) are always negligible
(at LEP1, LEP2 and also NLC) if the sum over all flavour combinations
is performed (that is, when no heavy flavour tagging is exploited),
they can in principle be important in $b$-jet samples, as in those selected
for top studies at NLC. As a matter of fact, in the 
individual contributions to
the six-jet rate mass corrections
can be quite sizable at $\sqrt s=500$ GeV, as shown in 
Fig.~\ref{fig_b_effect}, at the level of 
 ten percent or more. These corrections are more visible
when more $b$-quarks are produced in the final state. However, as 
previously mentioned, these multi-$b$-quark subprocesses
are unlikely to be important in top analyses. In practise, the largest part of
the $b\bar bjjjj$ sample produced from QCD comes from two-quark-four-gluon
events, where $b$-mass effects are naturally smaller because only two massive
particles are involved, over a six-particle phase space and at rather high
energy ($\sqrt s\gg2m_b$). In the end, $b$-mass effects on the complete  
$b\bar bjjjj$ produced by QCD amount to only $4\%$ at the NLC, for typical
values of the resolution parameter in the six-jet region, $y\approx0.001$. 
We have also
verified that the differential distributions (e.g., energy and angles) suffer
imperceptibly from mass corrections, so is the case of the shape variable
spectra. Therefore, in first instance, it is safe to
assume the value $m_b=0$ in numerical simulations of QCD six-jet events.
We will do so in the remainder of the paper.

The concern about possible background effects at the NLC 
due to six-jet events from pure QCD comes about if one further 
considers that they may naturally survive some of the top signal selection 
criteria. These can in fact rely on the analysis of shape variable spectra
of six-jet samples, in particular, the thrust and sphericity
distributions, and since the latter (as already mentioned while commenting 
Figs.~\ref{fig_shape1_6j_lep1}--\ref{fig_shape2_6j_lep1}) might well be
poorly described by the parton shower
approach exploited in the phenomenological MC programs, it
is of decisive importance that also the
non-infrared dynamics of events of the type
(\ref{qqgggg})--(\ref{qqqqqq}) is carefully studied.

It is well known that the
large value of the top mass 
%(here, $m_t=175$ GeV and $m_b=0$) 
leads to rather spherical 
signal events. Therefore, shape variables such us the two 
mentioned above 
represent useful means to disentangle $e^+e^-\ar t\bar t$ events.
For example, a selection strategy that does not exploit
neither lepton identification nor the tagging of $b$-jets was outlined 
in Ref.~\cite{tt}. This approach has a twofold attractiveness:
firstly, its simplicity; secondly, it is a rather general 
procedure, that can be equally 
applied both at and above the threshold $\sqrt s\approx 2m_t$.
The requirements are a large particle multiplicity, a high number of jets
(at least five, eventually forced to six by a clustering scheme) and a rather 
low  thrust (typically, 
below 0.85). Jets are selected according to a jet clustering algorithm (in our
case, the Cambridge one with $y=0.001$, for sake of illustration).

\begin{figure}[tbh]
\begin{center}
~\epsfig{file=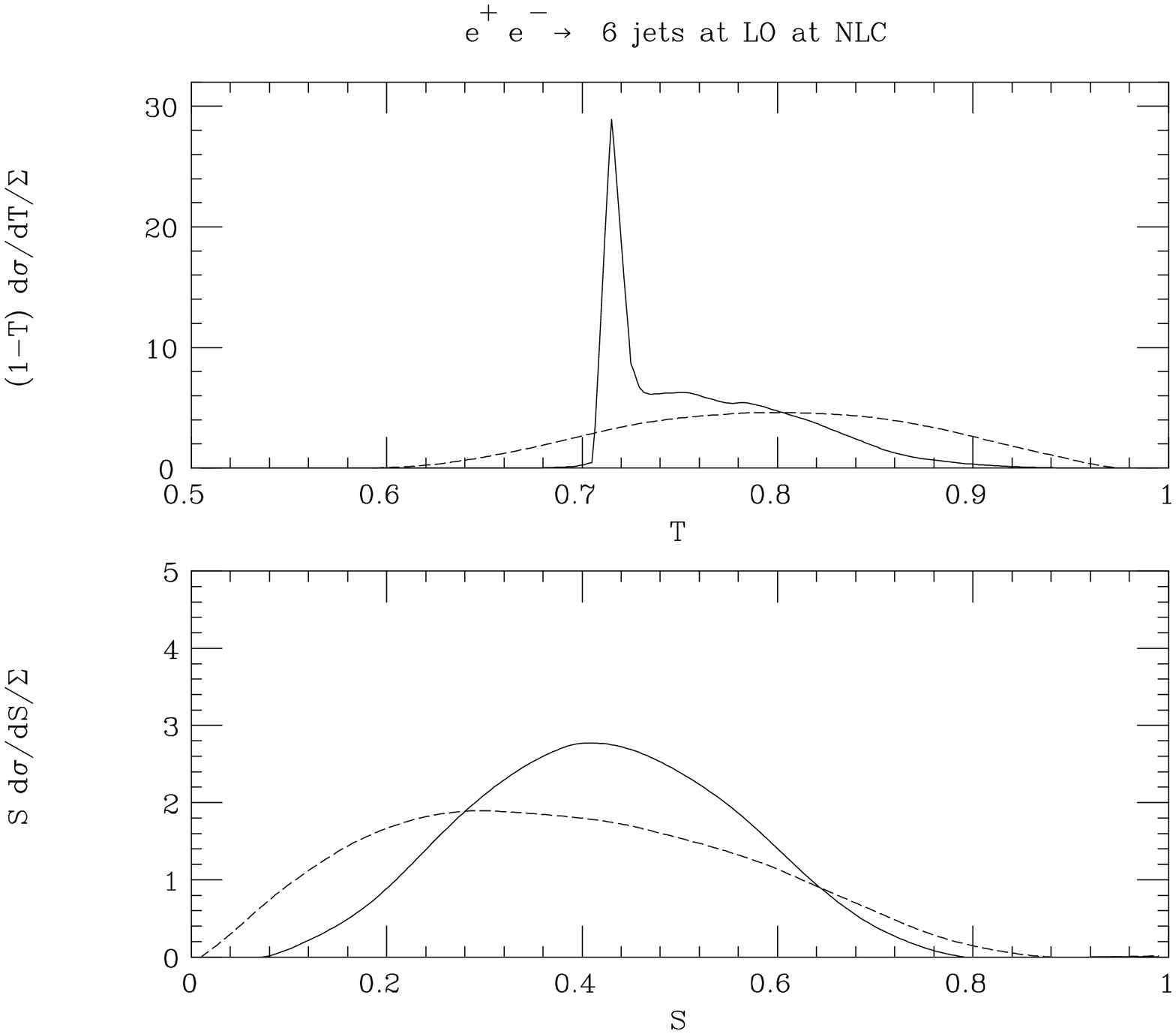,width=12cm,height=16cm,angle=90}
\caption{The distributions in thrust (upper plot) and
sphericity (lower plot) for $e^+e^-\ar t\bar t\ar6$-jet
events (continuous line) 
and for those of the  type (\ref{qqgggg})--(\ref{qqqqqq}) (dashed line) at LO 
in the C scheme with $y=0.001$, at NLC. 
Notice that the distributions have been normalised to a common
factor (one) for readability.}
\label{fig_shape_tt}
\end{center}
\end{figure}

Clearly, the  six-jet events meet the two first criteria. As for the
thrust and sphericity distributions, these are
 shown in Fig.~\ref{fig_shape_tt}.
From there it is evident the overlapping of the two distributions
($t\bar t$ and QCD events)
in the region where the signal events are mostly located, for both
thrust and sphericity.
A cut such as that advocated in Ref.~\cite{tt}, 
i.e., $T<0.85$, would then be rather ineffective
against six-jet events from QCD. Once again, we believe that a correct modelling
of the
phenomenology of the latter based on a ME description will be essential to
quantify correctly the effects of the QCD background, especially in view
of the mentioned 
high precision measurements of top parameters. 

\begin{figure}[tbh]
\begin{center}
~\epsfig{file=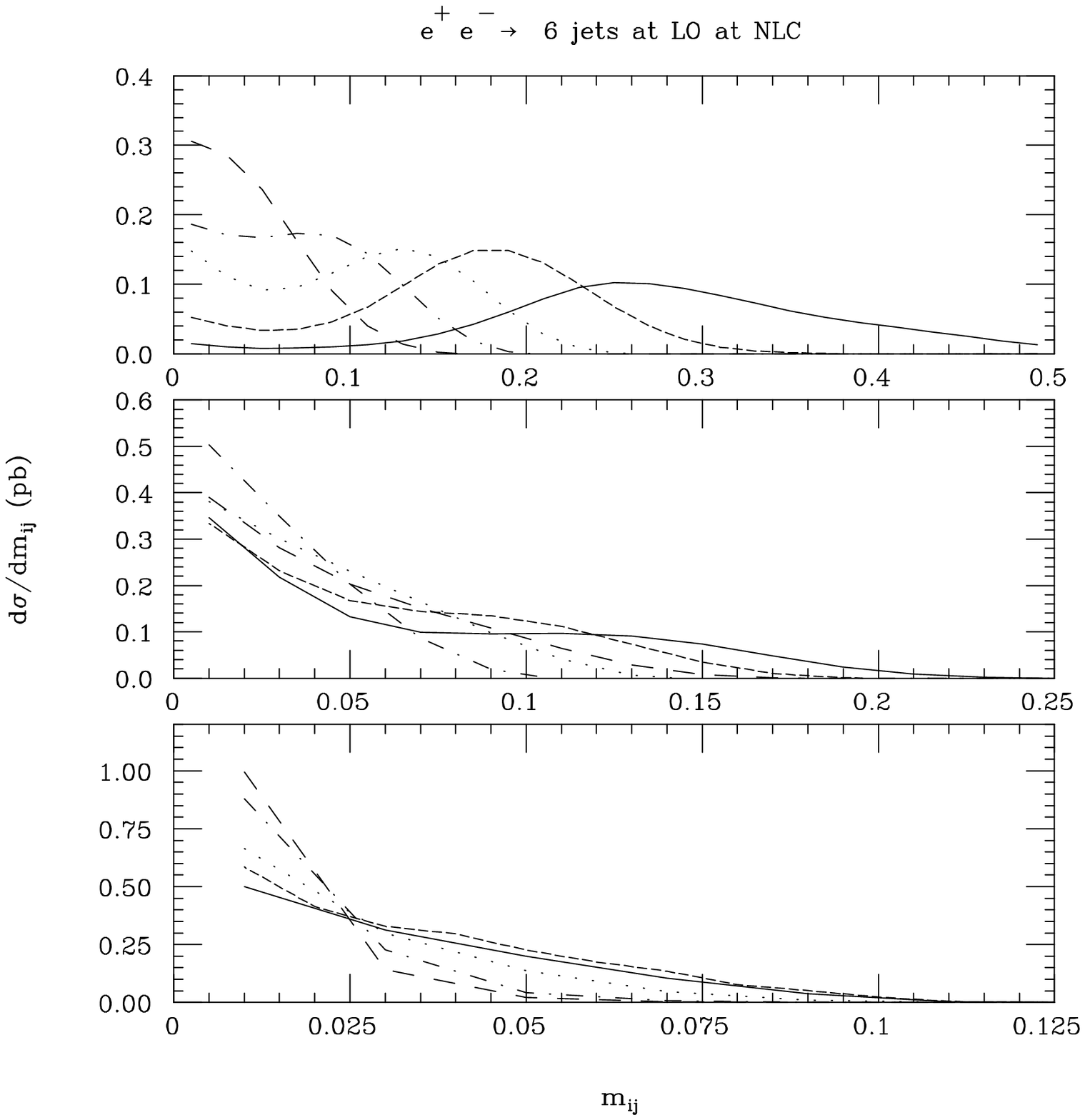,width=16cm,height=18cm,angle=0}
\caption{The distributions in the reduced invariant mass $m_{ij}$ (\ref{mij}) 
for events of the type (\ref{qqgggg})--(\ref{qqqqqq}) at LO in the C scheme
with $y=0.001$, at NLC, for the following combinations of parton pairs $ij$: 
$({12})[{23}]\{{35}\}$ (continuous lines),  
$({13})[{24}]\{{36}\}$ (short-dashed lines),  
$({14})[{25}]\{{45}\}$ (dotted lines),  
$({15})[{26}]\{{46}\}$ (dot-dashed lines) and  
$({16})[{34}]\{{56}\}$ (long-dashed lines),  
in the (upper)[central]\{lower\} frame.}
\label{fig_newmasses_6j_nlc}
\end{center}
\end{figure}

Tests of multiple electroweak self-couplings of  gauge-bosons  
(as well as searches for new resonances) at NLC will often need to 
rely on the mass reconstruction of multi-jet systems (particularly
di-jet ones). Therefore, it is 
instructive to look at the invariant mass distributions
which will be produced  
by all the possible two-parton combinations $ij$ in six-jet events
from QCD at ${\cal O}(\alpha_s^4)$
(with $i=1,...5$ and $j=i+1,...6$). As usual in multi-jet analyses,
we first order the jets in energy, so that $E_1>E_2>...>E_5>E_6$. Then,
we construct the quantities (the equality  strictly holds only for
massless partons $ij$, as is the case here) 
\be
\label{mij}
m_{ij}\equiv\frac{M_{ij}^2}{s}=\frac{2E_iE_j(1-\cos\theta_{ij})}{s},
\ee
where $M_{ij}^2$ represents the Lorentz invariant (squared) mass. 
These fifteen quantities 
are shown in Fig.~\ref{fig_newmasses_6j_nlc} for the C scheme at $y=0.001$,  
their shape being similar for the D algorithm.
We found it convenient to plot the `reduced' invariant masses $m_{ij}$ rather
than the actual ones $M_{ij}$, as energies and angles `scale' with the CM 
energy in such a way that the shape of the distributions is largely unaffected
by changes of the value of $\sqrt s$ in the energy range relevant to NLC.
Therefore, from Fig.~\ref{fig_newmasses_6j_nlc} one should then be able to 
reconstruct
rather accurately the Lorentz invariant mass distributions for a given
CM energy $\sqrt s$ by exploiting eq.~(\ref{mij}). 
Note that the reduced mass spectra are similar in all three reactions 
(\ref{qqgggg})--(\ref{qqqqqq}), their 
integral being however rescaled according to the numbers
given in Fig.~\ref{fig_comp6nlc}. To allow for an easy conversion
of the differential cross sections 
into numbers of events in a certain mass range, 
the spectra in Fig.~\ref{fig_newmasses_6j_nlc} sum to the total cross section of
processes (\ref{qqgggg})--(\ref{qqqqqq}) at NLC.

It is interesting to notice in Fig.~\ref{fig_newmasses_6j_nlc} the `resonant'
behaviour of some of the distributions. This is particularly true
for those involving the most energetic of all the partons.
Using the definition (\ref{mij}) they translate into peak-like structures at 
the true invariant mass values $M_{ij}\approx(250)[212]\{177\}$ GeV,
for the combinations $ij=(12)[13]\{14\}$, 
and, possibly, $M_{15}\approx150$ GeV as well. In all the other cases the
spectra are generally softer and do not show any distinctive kinematic 
feature. Once again, it is important that 
all such behaviours are correctly implemented
in the simulation programs that will be adopted by the NLC experiments, so 
that it will be possible to recognise and eventually subtract the six-jet QCD 
background.

A further general comment that is in order
for six-parton QCD events produced at NLC
is that the much lower value
of $\alpha_s$ at those high energies 
(as compared to that at LEP1 and/or LEP2) in principle
implies a reduced importance of the uncalculated HO strong corrections, so that
the theoretical uncertainty on the rates of 
processes (\ref{qqgggg})--(\ref{qqqqqq}) will be
 more under control at the future collider
than it was/is at the energy scales of  the two LEP phases.

\begin{figure}[tbh]
\begin{center}
~\epsfig{file=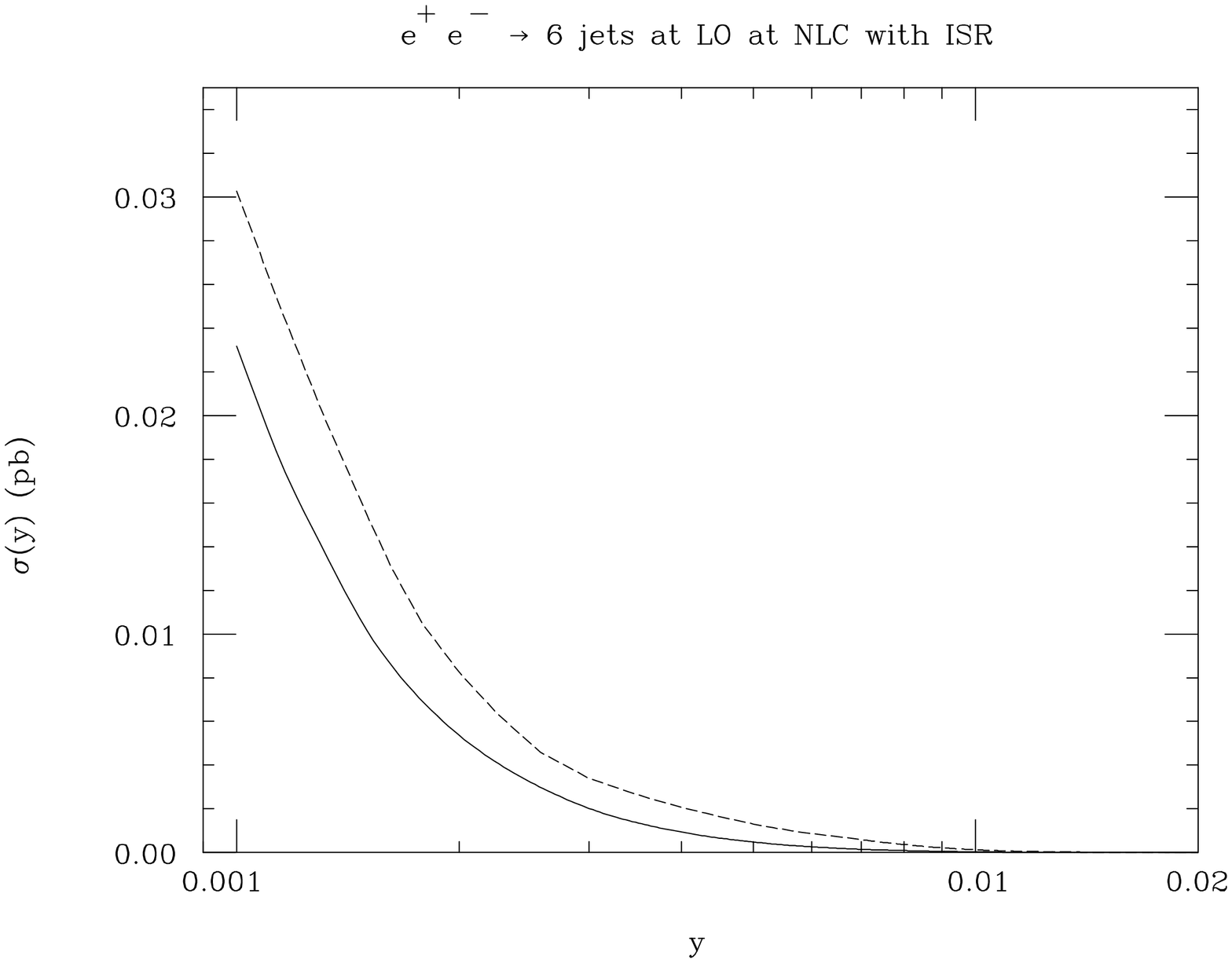,width=12cm,height=16cm,angle=90}
\caption{The total cross section of six-jet events
at LO in the D scheme, at NLC, without (solid line) and with (dashed line)
ISR.}
\label{fig_ISR}
\end{center}
\end{figure}

So far we not included effects of ISR
in any of our calculations for six-jet production.
However, it is well known \cite{ee500} that photon bremsstrahlung generated
by the incoming electron and positron beams
can be quite sizable at  NLC. In contrast, at LEP1 and LEP2 such radiation
is suppressed by the width of the $Z$ and $W^+W^-$ resonances,
respectively\footnote{In addition, peculiar features at  NLC is the presence
of Linac energy spread and beamsstrahlung \cite{bark}. However, their effects
are negligible, as compared to those produced by ISR, at least for the `narrow'
D-D and TESLA collider designs \cite{ISR}. For this reason, we do not consider
them in our analysis. They would however be straightforward to insert.}.
In order to implement ISR we  resort to the mentioned ESF approach. In
particular, we use the expressions given in
Ref.~\cite{Nicro}, through the order $\alpha_{em}^2$. We plot the production
rates for the sum of the processes (\ref{qqgggg})--(\ref{qqqqqq}) in presence
of ISR in Fig.~\ref{fig_ISR}, as a function of the resolution
parameter $y$ in the C scheme. (Effects turn out to be similar in the D 
scheme.) They are compared to those obtained
without ISR (that is, the sum of the rates in Fig.~\ref{fig_comp6nlc}).
No cut is here applied to the soft and collinear 
photons from the initial state.
We see that the curve corresponding to the processes 
(\ref{qqgggg})--(\ref{qqqqqq}) convoluted with ISR lies above the
lowest-order one. This is rather intuitive,
as it is well known that the radiation of photons from the
incoming electron and positrons tends to lower the `effective' CM energy
of the collision \cite{ISR}. Since processes 
(\ref{qqgggg})--(\ref{qqqqqq}) proceed
via $s$-channel diagrams, the expected overall effect would be an enhancement 
of the total rates. In fact, this is what can be seen in Fig.~\ref{fig_ISR}. 
At $y=0.001$, the difference
is around $25\%$ and this tends to increase as $y$ gets larger. However, we
 have verified that the shape of the 
differential distributions (such as those 
in shape variables, see Fig.~\ref{fig_shape_tt}, and in
invariant mass, see Fig.~\ref{fig_newmasses_6j_nlc}) suffer little from
ISR effects.

As announced in the Introduction, we also devote some space to analyse
photoproduction in 
$5\mathrm{jet}+1\gamma$,
$4\mathrm{jet}+2\gamma$,
$3\mathrm{jet}+3\gamma$ and
$2\mathrm{jet}+4\gamma$ events. For this study, we reconsider the case of
LEP1 energies, where the impact of ISR is much less relevant than
at LEP2 and NLC (recall that
we did not implement the resummed logarithmic part of it in photonic events).
In the first two cases we consider both the 
dominant contributions involving only two quarks and the suppressed
ones involving four, whereas in the last two cases only two quarks can
participate in the event. 

Final states including photons are
actively studied experimentally in order to determine the electroweak
couplings of the quarks and as a mean to search for new phenomena
\cite{NP35_36}. 
Furthermore, such particles are the only ones which can be directly revealed
and couple to the quarks at each stage of the perturbative 
parton shower (see Fig.~\ref{fig_cascade}), so that the detection of hard 
isolated photons (or `direct photons') is extremely useful for a 
detailed understanding of
the underlying partonic picture as well as of the 
QCD evolution from large to small energy scales. 
(Conversely, soft photons lying close to
hadronic tracks can give information about the parton 
fragmentation mechanisms, 
though the corresponding rates are not calculable perturbatively.)
In particular, the more 
jets are separated in association with photons, the more likely
will be that photons were radiated by quarks in later stages of such evolution.
In $e^+e^-\ar \mathrm{m~jet~+~n}$ photon events
(with $m+n=6$ and $m\ge2$), $\gamma$-radiation always takes place from the
primary quark-antiquark pair (i.e., that produced in the 
$\gamma,Z$-splitting) in the case of the dominant two-quark events. 
Thus, for the cases $m=4(5)$ and $n=2(1)$, such contributions
should be regarded as a source of background in the mentioned analyses,
with the four-quark contributions being the interesting ones, as in these cases
photons can also be radiated by secondary branching products. 

\begin{figure}[tbh]
\begin{center}
~\epsfig{file=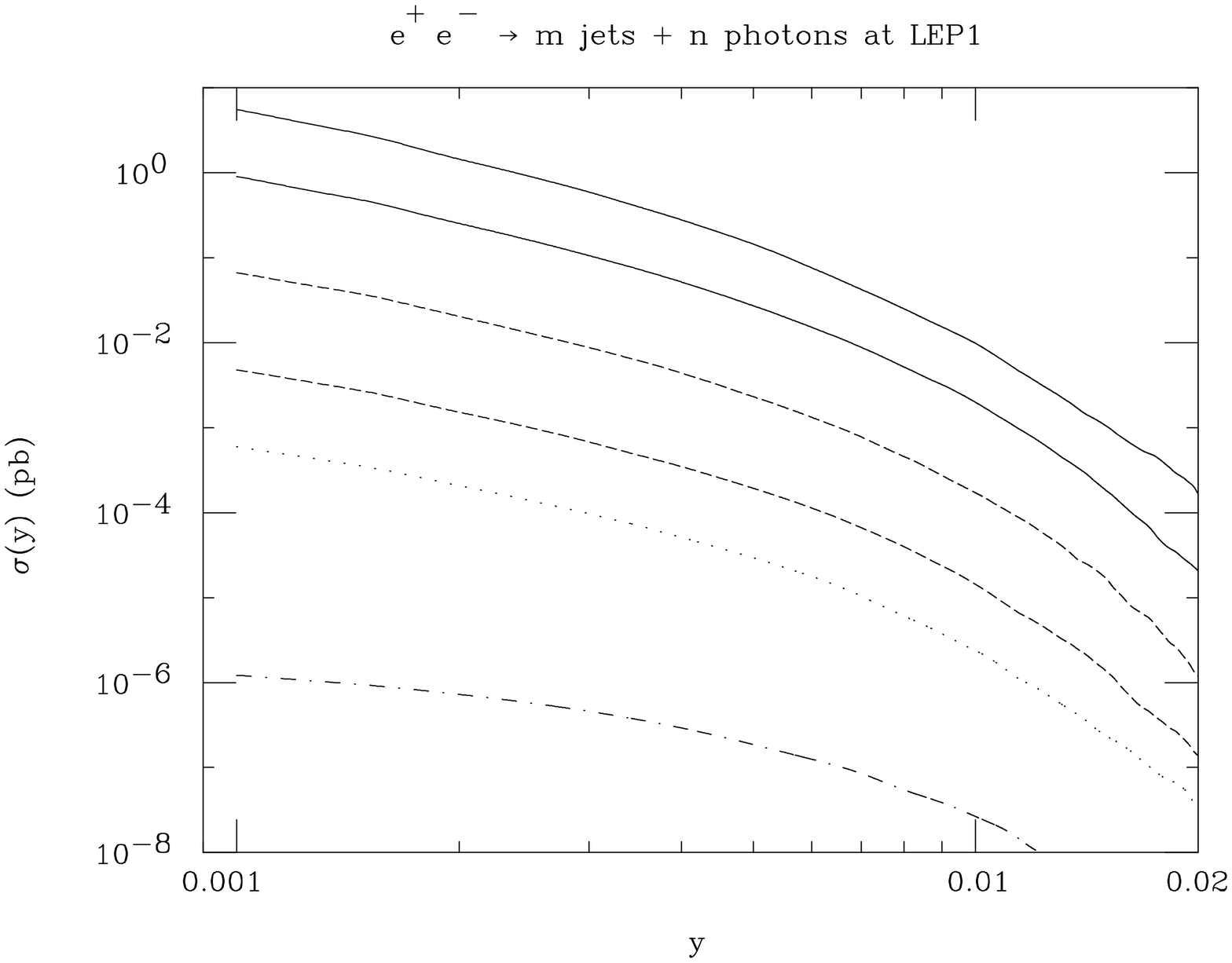,width=12cm,height=16cm,angle=90}
\caption{The total cross section for photoproduction 
at LO in the C scheme, at LEP1, for the processes:
$e^+e^-\ar q\bar q ggg\gamma$ and
$e^+e^-\ar q\bar q q'\bar q'g\gamma$  (m=5, n=1: continuous lines), 
$e^+e^-\ar q\bar q gg\gamma\gamma$ and
$e^+e^-\ar q\bar q q'\bar q'\gamma\gamma$  (m=4, n=2: dashed lines), 
$e^+e^-\ar q\bar q g\gamma\gamma\gamma$ (m=3, n=3: dotted line) and
$e^+e^-\ar q\bar q \gamma\gamma\gamma\gamma$ (m=2, n=4: dot-dashed line).
In the cases m=5, n=1 and m=4, n=2 the upper lines represent the two-quark
contributions whereas the lower ones refer to four-quark processes. 
The following 
additional cuts have introduced on the photons: $p_T^\gamma>5$ GeV
and $|\cos\theta_\gamma|<0.85$.}
\label{fig_photons}
\end{center}
\end{figure}

In this study, we limit ourselves to giving
some total rates. The cross sections for isolated photon events as a function
of the $y$ resolution parameter, applied to both quarks/gluons and photons
on the same footing (in the spirit of the so-called `democratic approach'
\cite{democratic}), in the C scheme, are presented in Fig.~\ref{fig_photons}, 
at LEP1. The mentioned cuts in transverse momentum and polar angle of the
photons have been enforced, in order to screen the singularities
due to infrared (i.e., soft and collinear) EM emission from electrons and
positrons. The hierarchy in the cross sections of
Fig.~\ref{fig_photons} is dictated mainly by the perturbative order
${\cal O}(\alpha_{em}^{m+2} \alpha_{s}^{n})$ for the hard production of
$m$ jets and $n$ photons. 
At LEP1, given, e.g., 
some 400 inverse picobarn of total accumulated luminosity, one
should expect that only the $5\mathrm{jet}+1\gamma$ and 
$4\mathrm{jet}+2\gamma$ events can yield an observable rate, with the 
four-quark contributions being roughly one order of magnitude smaller than
the two-quark ones in both cases.

\section{Summary}
\label{Sect_summa}

In summary, we have studied at the partonic level the tree processes
$e^+e^-\ar q\bar q gggg$, $e^+e^-\ar q\bar q q'\bar q' gg$ and  
$e^+e^-\ar q\bar q q'\bar q' q''\bar q''$ (for massless and massive quarks)
at leading-order in perturbative QCD, by computing their 
exact matrix elements. The latter task has been accomplished thanks to
the use of spinor techniques and colour decomposition methods. Both the
formulae for the Feynman amplitudes and the relevant QCD (i.e., for
$N_C=3$) colour factors have been given explicitly. Several phenomenological
applications at present and future colliders have been discussed. Two 
{\tt FORTRAN} codes have been produced and cross checked with each other
on several different platforms: VMS, OSF, HP.
The fastest of the two requires about $3\times10^{-3}$ CPU seconds 
to evaluate a single event on an alpha-station DEC 3000 - M300X, which makes
it suitable for high statistics Monte Carlo simulations. Further optimisations
are, however, in progress. 

\section{Acknowledgements}

We acknowledge the UK PPARC support. We thank
the Theoretical Physics Groups in Fermilab and Lund
for their kind hospitality while part of this work was carried out.
This research work was supported in part by the Italian Institute of 
Culture `C.M. Lerici' under the grant Prot. I/B1 690, 1997.
Finally, we are grateful to David J. Miller for carefully reading the
manuscript and for several useful discussions and comments.

\vfill\newpage

\section{Appendix I}
\label{Sect_appe1}

In this Appendix we present the analytical expressions for the
helicity amplitudes relevant to the three processes 
(\ref{qqgggg})--(\ref{qqqqqq}). As already mentioned, we make use of the
formalism introduced through Refs.~\cite{KS,mana,method,ioPRD}.
Each amplitude will be expressed in terms of a suitable combination of the
$X$ and $Z$ functions defined in Subsect.~\ref{Subsect_helicity} (and
of scalar products involving the external four-momenta). The directions of
the latter are such that the both the initial state 
electrons and positrons and
all the partons $q,q',q'',\bar q,\bar q',\bar q''g$ 
in the final states are considered outgoing.

In order to minimise the lenght of the formulae, we adopt the following
short-hand notations.
The index $(a)$ (i.e., in round brackets) stands for the momentum
$p_{a}$ and the corresponding polarization $\lambda_a$ of any real particle
$a=1,...8$. 
In the case where external gluons are involved, additional indices
will appear, such as 
$[a]$ (i.e., in squared brackets), to which are associated the 
auxiliary momentum $q_a$ of the gluon $a$ 
and its helicity $\lambda_a$ (see eq.~(\ref{polar})). The notation
$\{k\}$ (i.e., in curly brackets)
refers to both the momentum $p_k$ and the internal helicity $\lambda$
associated to a propagator. We further make the 
convention that all indices $\{k\}$ are summed over, and that the 
sum extends to all possible values of all internal 
helicities, $\lambda,\lambda',...= \pm$, as well as to the momenta 
 appearing in the last lines of the forthcoming expressions for the
amplitudes of processes (\ref{qqgggg})--(\ref{qqqqqq}) (in the spirit
of eq.~(\ref{dirac}))\footnote{Note the order of the indices in the sequence
$\{a,b,c,...\}$ has no importance.}. 

To each of the basic amplitudes are associated propagators functions.
In the case of off-shell fermions, they have the form
\be\label{fdenom}
D_f(\{i\})=\frac{1}{(\sum_{i} p_i)^2-m_f^2},
\ee
whereas in the case of bosons one gets
\be\label{bdenom}
D_V(\{i\})=\frac{1}{(\sum_i p_i)^2-M_V^2+iM_V\Gamma_V}.
\ee 
In eqs.~(\ref{fdenom})--(\ref{bdenom}),
$f$ is the flavour $q,q'$ or $q''$ of a virtual fermion line, whereas 
$V=\gamma,Z$ or $g$, being   $M_{V}=\Gamma_{V}\equiv0$ if $V=\gamma$ or $g$.
In order to assign the correct fermionic propagator functions to the various
amplitudes, one will have to multiply the latter for as many terms
of the form (\ref{fdenom}) as the number of indices $\{i\},...$ 
appearing in the last line of the amplitude equations, with the sums 
$\sum_i$ extending to all the values assumed by $i$. A subscript of the
form $\{...\}_f$ will indicate the flavour 
$f=q,q',q''$ to be assigned to the off-shell fermion line. 
As for the bosonic propagators in the case $V=g$ (i.e., for gluons), 
these will only appear in the functions
$A$, $B$ or $C$ (see below).
Their appropriate expressions will be given while defining the latter. 
Finally, as the incoming $V=\gamma,Z$ currents are a common feature to all
diagrams, we will factorise them in the expressions of the amplitudes, an
operation which will be recalled by the use of the subscript $V$.
Given these simple rules, we believe it trivial to derive the 
correct expressions for all propagators, so these  will not 
appear explicitly in our formulae for the amplitudes. 
In the latter, we will present
the expressions for the topologies of Fig.~\ref{fig_6partons}, in the form
$T_V^i(\Pi,\{\lambda\})$, with $i=1,...8$ for process (\ref{qqgggg}),
              $i=1,...10$ for process (\ref{qqqqgg}), and
              $i=1,...3$ for process (\ref{qqqqqq}), 
and where $\Pi$ represents (a product of) permutations $\pi$ 
of gluon colour indices and 
$\{\lambda\}$ refers to the helicities of the external
particles. 

As for the couplings, the notation $fV$ refers to the pair of
chiral indices $(c_R,c_L)$ entering in the expressions given in
Tabs.~\ref{tab_X}--\ref{tab_Z} and associated with the vertex involving 
a fermion $f=e,q,q',q''$ and a gauge vector $V=\gamma,Z,g$, according
to Tab.~\ref{tab_cRcL}. In particular, 
the  notation $qg$ is meant to describe both a true and a 
fictitious quark-gluon vertex, the latter being used to define the polarization
of the gauge particle (for which  
$(c_R,c_L)\equiv(1,1)$), according to eq.~(\ref{polar}).

Graphically, a $Z$ function is represented as (hereafter, one
symbolically has that $\l\r=(),[]$ or $\{\}$)
\\
\vskip-8.0cm
\begin{center}
\begin{picture}(250,250)
\SetScale{0.5}
\SetWidth{1.2}
\SetOffset(0,0)

\Line(350,350)(450,350)
\Line(350,250)(450,250)
\Photon(400,350)(400,250){5}{5}

\Text(40,152)[]{$Z(\l a\r;\l b\r;\l c\r;\l d\r;fV;f'V) =$}

\Text(233,180)[]{$a$}
\Text(170,181)[]{$b$}

\Text(170,122)[]{$c$}
\Text(233,123)[]{$d$}

\Text(210,153)[]{$V$}
\Text(200,185)[]{$f$}
\Text(200,113.5)[]{$f'$}

\end{picture}
\end{center}
\vskip-5.cm
\begin{equation}\label{Z}
\;\;\;\;\;\;
\end{equation}

\vskip2.0cm
On the same footing as was done in Ref.~\cite{voi}, 
we found it convenient to introduce the function describing a three-gluon
vertex, here generalised to the case of off-shell gauge vectors, each
connected to a fermion line\footnote{Here and in the following, 
we write the momentum of the gluon attached to the fermion line
labelled by $ab$ in terms of those of the others involved. 
This convention will be helpful both in writing the Feynman
diagrams in a more compact analytic form as well as in calculating them 
numerically in a more efficient manner, as this way the labels $a$ and
$b$ only appear as arguments of the $A$, $B$ and $C$ functions, and not in the
internal momentum
summations. In fact, the line $ab$ will eventually be identified 
with the off-shell one connected to the incoming $\gamma,Z$ current in 
Fig.~\ref{fig_6partons}a--c.}, 
\vskip-10.0cm
\begin{center}
\begin{picture}(250,250)
\SetScale{0.5}
\SetWidth{1.2}
\SetOffset(0,0)

\Gluon(400,250)(400,350){5}{5}
\Gluon(400,250)(313.3975,200){5}{5}
\Gluon(400,250)(486.6025,200){5}{5}

\Line(350,350)(450,350)
\Line(511.6025,243.3013)(461.6025,156.6987)
\Line(288.3975,243.3013)(338.3975,156.6987)

\Text(0,130)[]{$A(\l a\r;\l b\r;\l c\r;\l d\r;\l e\r;\l f\r)=$}

\Text(233,180)[]{$a$}
\Text(170,181)[]{$b$}

\Text(139,125)[]{$c$}
\Text(175,76)[]{$d$}

\Text(227,74)[]{$e$}
\Text(264,125)[]{$f$}

\end{picture}
\end{center}
\vskip-2.25cm
\begin{eqnarray}\label{A}                                     \nonumber
\qquad\qquad\qquad\qquad &=& 
           Z(\l a\r;\l b\r;\l c\r;\l d\r;qg;qg)            \\ \nonumber
&  &\times    ~2 X(\l e\r;\{i\};\l f\r;qg)   \\ \nonumber
&- & Z(\l c\r;\l d\r;\l e\r;\l f\r;qg;qg)            \\ \nonumber
&  &\times          ~X(\l a\r;\{i\};\l b\r;qg)          \\   
&-      & {\mbox{same}}[(c,d) \leftrightarrow (e,f)].  
\end{eqnarray}
\vskip0.5in\noindent
To each $A$ term is always associated the propagator
$D_g(\{i\}+\{j\})\equiv 1/(\sum_i p_i+\sum_j p_j)^2$. 
Note that the summation $\{i\}$ extends to
$i=c,d$ if $\l d\r\ne[d]$, or to $i=c$ otherwise and similarly for $\{j\}$
with respect to the indices $e,f$. In addition, if 
$\l d\r\ne[d]$($\l f\r\ne[f]$), the additional term $D_g(\{i\})$($D_g(\{j\})$)
appears.

In addition to eq.~(\ref{A}), we also need the expression 
describing two connected three-gluon vertices and the associated four-gluon
vertex with same colour structure. Assuming again that the gluon propagator
connected to $ab$ is expressed in terms of the momenta associated
to the legs $c, ... , h$, one has 
\vskip1.0cm
$$
B(\l a\r;\l b\r;\l c\r;\l d\r;\l e\r;\l f\r;\l g\r;\l h\r) =
\qquad\qquad\qquad\qquad\qquad\qquad\qquad\qquad\qquad\qquad\qquad
$$
\vskip-20.0cm
\begin{center}
\begin{picture}(250,250)
\SetScale{0.5}
\SetWidth{1.2}
\SetOffset(0,0)

\Gluon(100,250)(170.7107,320.7107){5}{5}
\Gluon(0,250)(-70.71070,320.7107){5}{5}
\Gluon(0,250)(-70.71070,179.2893){5}{5}
\Gluon(100,250)(170.7107,179.2893){5}{5}
\Gluon(0,250)(100,250){5}{5}

\Line(206.0660,285.3553)(135.3553,356.0660)
\Line(-35.35530,356.0660)(-106.06600,285.3553)
\Line(-106.06600,214.6447)(-35.35530,143.9340)
\Line(135.3553,143.9340)(206.0660,214.6447)

\Text(310,140)[]{$a$}
\Text(262,185)[]{$b$}

\Text(237,183)[]{$c$}
\Text(191,140)[]{$d$}

\Text(191,111)[]{$e$}
\Text(239,70)[]{$f$}

\Text(262,70)[]{$g$}
\Text(310,115)[]{$h$}

\Text(149,125)[]{$+$}
\Text(-100,125)[]{$=$}

\Gluon(500,250)(570.7107,320.7107){5}{5}
\Gluon(500,250)(429.2893,320.7107){5}{5}
\Gluon(500,250)(429.2893,179.2893){5}{5}
\Gluon(500,250)(570.7107,179.2893){5}{5}

\Line(606.0660,285.3553)(535.3553,356.0660)
\Line(464.6447,356.0660)(393.9340,285.3553)
\Line(393.9340,214.6447)(464.6447,143.9340)
\Line(535.3553,143.9340)(606.0660,214.6447)

\Text(109,140)[]{$a$}
\Text(61,185)[]{$b$}

\Text(-14,183)[]{$c$}
\Text(-60,140)[]{$d$}

\Text(-60,111)[]{$e$}
\Text(-12,70)[]{$f$}

\Text(61,70)[]{$g$}
\Text(109,115)[]{$h$}

\end{picture}
\end{center}

\begin{eqnarray}\label{B}                                   \nonumber
& =      & Z(\l a\r;\l b\r;\l c\r;\l d\r; qg; qg) \\ \nonumber
& \times & \large[2\large(X(\l g\r;\{i\};\l h\r;qg)
                         +X(\l g\r;\{j\};\l h\r;qg))\times  \\ \nonumber
&        &\phantom{\large[}
                  2X(\l e\r;\{i\};\l f\r;qg)\large]\\ \nonumber
& +      & Z(\l a\r;\l b\r;\l g\r;\l h\r;qg;qg) 
          ~Z(\l c\r;\l d\r;\l e\r;\l f\r;qg;qg)\times \\ \nonumber
&        &        ({2\sum_k p_k}+{\sum_i p_i}+{\sum_j p_j})
                  \cdot({\sum_i p_i}) \\ \nonumber
& +      & Z(\l a\r;\l b\r;\l e\r;\l f\r;qg;qg) 
          ~Z(\l c\r;\l d\r;\l g\r;\l h\r;qg;qg)\times \\ \nonumber
&        &    (\sum_i p_i+\sum_j p_j)^2\\ \nonumber
& -      & Z(\l c\r;\l d\r;\l e\r;\l f\r; qg; qg) \\ \nonumber
& \times & \large[2\large(X(\l g\r;\{i\};\l h\r;qg)
                         +X(\l g\r;\{j\};\l h\r;qg)\large)
                          X(\l a\r;\{i\};\l b\r;qg)+ \\ \nonumber
&        &\phantom{\large[}
                  \large(X(\l a\r;\{k\};\l b\r;qg)
                        -X(\l a\r;\{j\};\l b\r;qg)
                        -X(\l a\r;\{i\};\l b\r;qg)\large)
                          X(\l g\r;\{i\};\l h\r;qg)\large]\\ \nonumber
& +      & 2X(\l e\r;\{i\};\l f\r;qg)\\ \nonumber
& \times &\large[-Z(\l a\r;\l b\r;\l g\r;\l h\r; qg; qg) \\ \nonumber
& &\phantom{\large[}
         \times\large(2X(\l c\r;\{k\};\l d\r;qg)
                      +X(\l c\r;\{j\};\l d\r;qg)\large)\\ \nonumber
&        &\phantom{\large[}
           +Z(\l c\r;\l d\r;\l g\r;\l h\r; qg; qg) \\ \nonumber
& &\phantom{\large[}
         \times\large( X(\l a\r;\{k\};\l b\r;qg)
                      -X(\l a\r;\{j\};\l b\r;qg) 
                      -X(\l a\r;\{i\};\l b\r;qg)\large)\large]\\ 
& -      & {\mbox{same}}[(c,d) \leftrightarrow (e,f)]                  
\end{eqnarray}
The propagators associated to $B$ functions assume the 
form $D_g(\{i\}+\{j\}+\{k\})D_g(\{i\}+\{j\})\equiv
1/(\sum_i p_i + \sum_j p_j + \sum_k p_k)^2
       /(\sum_i p_i + \sum_j p_j)^2$. The summations $\{i\},\{j\}$ and $\{k\}$
are over the indices $c,d$, $e,f$ and $g,h$ if $\l d\r\ne [d]$, 
$\l f\r\ne [f]$ and $\l h\r\ne [h]$, or $c$, $e$ and $g$
otherwise,  respectively. Moreover, as previously explained, if 
$\l d\r\ne[d]$($\l f\r\ne[f]$)[$\l h\r\ne[h]$], 
the additional term $D_g(\{i\})$($D_g(\{j\})$)[$D_g(\{k\})$] has to
be inserted.

Finally, we further need the function 
describing three connected three-gluon vertices and the two
associated graphs with one four-gluon and one three-gluon 
vertex with identical colour matrices,

\begin{center}
\begin{picture}(250,250)
\SetScale{0.5}
\SetWidth{1.2}
\SetOffset(0,0)

\Gluon(550,250)(620.5342,347.0820){5}{7}
\Gluon(450,250)(379.4658,347.0820){5}{7}
\Gluon(450,250)(335.8732,212.9180){5}{7}
\Gluon(500,250)(500,130){5}{7}
\Gluon(550,250)(664.1268,212.9180){5}{7}

\Gluon(450,250)(500,250){5}{3}
\Gluon(500,250)(550,250){5}{3}

\Line(660.9850,317.6927)(580.0833,376.4713)
\Line(419.9167,376.4713)(339.0150,317.6927)
\Line(320.4224,260.4708)(351.3240,165.3652)
\Line(440,130)(560,130)
\Line(648.6759,165.3652)(679.5776,260.4708)

\Text(337,155)[]{$a$}
\Text(285,195)[]{$b$}

\Text(217,193)[]{$c$}
\Text(165,150)[]{$d$}

\Text(155,135)[]{$e$}
\Text(180,76)[]{$f$}

\Text(214,60)[]{$g$}
\Text(287,60)[]{$h$}

\Text(321,76)[]{$i$}
\Text(347,135)[]{$j$}

\Text(20,125)[]{
$C(\l a\r;\l b\r;\l c\r;\l d\r;\l e\r;\l f\r;\l g\r;\l h\r;\l i\r;\l j\r)$ =
}

\end{picture}
\end{center}
\vskip-3.0cm
\begin{center}
\begin{picture}(250,250)
\SetScale{0.5}
\SetWidth{1.2}
\SetOffset(0,0)

\Text(-105,125)[]{$+$}

\Gluon(60,250)(130.5342,347.0820){5}{7}
\Gluon(-40,250)(-110.53421,347.0820){5}{7}
\Gluon(-40,250)(-154.1268,212.9180){5}{7}
\Gluon(-40,250)(10,130){5}{7}
\Gluon(60,250)(174.1268,212.9180){5}{7}

\Gluon(-40,250)(60,250){5}{5}

\Line(170.9850,317.6927)(90.0833,376.4713)
\Line(-70.08331,376.4713)(-150.9850,317.6927)
\Line(-169.5776,260.4708)(-138.67599,165.3652)
\Line(-50,130)(70,130)
\Line(158.6759,165.3652)(189.5776,260.4708)

\Text(92,155)[]{$a$}
\Text(40,195)[]{$b$}

\Text(-28,193)[]{$c$}
\Text(-80,150)[]{$d$}

\Text(-90,135)[]{$e$}
\Text(-65,76)[]{$f$}

\Text(-32,60)[]{$g$}
\Text(41,60)[]{$h$}

\Text(75,76)[]{$i$}
\Text(101,135)[]{$j$}

\Text(130,125)[]{$+$}

\Gluon(550,250)(620.5342,347.0820){5}{7}
\Gluon(450,250)(379.4658,347.0820){5}{7}
\Gluon(450,250)(335.8732,212.9180){5}{7}
\Gluon(550,250)(500,130){5}{7}
\Gluon(550,250)(664.1268,212.9180){5}{7}

\Gluon(450,250)(550,250){5}{5}

\Line(660.9850,317.6927)(580.0833,376.4713)
\Line(419.9167,376.4713)(339.0150,317.6927)
\Line(320.4224,260.4708)(351.3240,165.3652)
\Line(440,130)(560,130)
\Line(648.6759,165.3652)(679.5776,260.4708)

\Text(337,155)[]{$a$}
\Text(285,195)[]{$b$}

\Text(217,193)[]{$c$}
\Text(165,150)[]{$d$}

\Text(155,135)[]{$e$}
\Text(180,76)[]{$f$}

\Text(214,60)[]{$g$}
\Text(287,60)[]{$h$}

\Text(321,76)[]{$i$}
\Text(347,135)[]{$j$}

\end{picture}
\end{center}

\begin{eqnarray}                              \nonumber
& =      & \large\{\large(2X(\l i\r;\{k\};\l j\r;qg)
                  +2X(\l i\r;\{l\};\l j\r;qg) 
                  +2X(\l i\r;\{m\};\l j\r;qg)
                   +X(\l i\r;\{n\};\l j\r;qg)\large)\\ \nonumber
&        &\times\large[2\large(X(\l g\r;\{k\};\l h \r;qg)
                        +X(\l g\r;\{l\};\l h \r;qg)\large)\times\\ \nonumber
&        &\phantom{\times\large[}
                         \large(2X(\l e\r;\{k\};\l f\r;qg)
                                Z(\l a\r;\l b\r;\l c\r;\l d\r;qg;qg)-
           \\ \nonumber
&        &\phantom{\times\large[\large(}
                                X(\l a\r;\{k\};\l b\r;qg)
                               ~Z(\l c\r;\l d\r;\l e\r;\l f\r;qg;qg)\large)
           \\ \nonumber
&        &\phantom{\times}
                        +\large(X(\l a\r;\{m\};\l b\r;qg)
 -X(\l a\r;\{l\};\l b\r;qg)-X(\l a\r;\{k\};\l b\r;qg)\large)\times
           \\ \nonumber
&        &\phantom{\times\large[}
                         \large(2X(\l e\r;\{k\};\l f\r;qg)
                                ~Z(\l c\r;\l d\r;\l g\r;\l h\r;qg;qg)-
           \\ \nonumber
&        &\phantom{\times\large[\large(}
                                X(\l g\r;\{k\};\l h\r;qg)
                               ~Z(\l c\r;\l d\r;\l e\r;\l f\r;qg;qg)\large)
           \\ \nonumber
&        &\phantom{\times}
             -\large(2X(\l c\r;\{m\};\l d\r;qg)
                     +X(\l c\r;\{l\};\l d\r;qg)\large)\times
           \\ \nonumber
&        &\phantom{\times\large[}
               2X(\l e\r;\{k\};\l f\r;qg)
               ~Z(\l a\r;\l b\r;\l g\r;\l h\r;qg;qg) \\ \nonumber
&        &\phantom{\times}
                       +\large(2\sum_m p_m+\sum_l p_l+\sum_k p_k\large)
                        \cdot
                        \large(\sum_k p_k\large)\times
           \\ \nonumber
&        &\phantom{\times\large[}
                                Z(\l a\r;\l b\r;\l g\r;\l h\r;qg;qg)
                               ~Z(\l c\r;\l d\r;\l e\r;\l f\r;qg;qg)\large]
\\ \nonumber
& -      & {\mbox{same}}[(a,b) \leftrightarrow (i,j);
      \{k\}+\{l\}+\{m\}+\{n\}\leftrightarrow -\{n\}] \large\}
\\ \nonumber
& +      & Z(\l a\r;\l b\r;\l i\r;\l j\r;qg;qg)
          ~Z(\l c\r;\l d\r;\l e\r;\l f\r;qg;qg)\\ \nonumber
&        &\times\large[2\large(X(\l g\r;\{k\};\l h\r;qg)
                              +X(\l g\r;\{l\};\l h\r;qg)\large)\times
          \\ \nonumber
&        &\phantom{\times\large[}
          \large(\sum_k p_k+\sum_l p_l+\sum_m p_m+2\sum_n p_n\large)
          \cdot  
          \large(\sum_k p_k\large)
          \\ \nonumber
&        &\phantom{\times}
                +X(\l g\r;\{k\};\l h\r;qg)\times
          \\ \nonumber
&        &\phantom{\times\large[}
          \large(\sum_k p_k+\sum_l p_l+\sum_m p_m+2\sum_n p_n\large)
          \cdot  
          \large(\sum_m p_m -\sum_l p_l -\sum_k p_k\large)
          \\ \nonumber
&        &\phantom{\times}
                -\large(X(\l g\r;\{k\};\l h\r;qg)
                       +X(\l g\r;\{l\};\l h\r;qg)
                      +2X(\l g\r;\{n\};\l h\r;qg)\large)\times
          \\ \nonumber
&        &\phantom{\times\large[}
          \large(\sum_k p_k+\sum_l p_l+2\sum_m p_m\large)
          \cdot  
          \large(\sum_k p_k)\large]
          \\ \nonumber
& +      & 2~Z(\l a\r;\l b\r;\l i\r;\l j\r;qg;qg)
            ~X(\l e\r;\{k\};\l f\r;qg) \\ \nonumber
&        &\times\large[Z(\l c\r;\l d\r;\l g\r;\l h\r;qg;qg) \\ \nonumber
&        &\phantom{\times\large[}
          \large(\sum_k p_k+\sum_l p_l+\sum_m p_m+2\sum_n p_n\large)
          \cdot  
          \large(\sum_k p_k+\sum_l p_l-\sum_m p_m\large)+
          \\ \nonumber    
&        &\phantom{\large[}
          \large(2X(\l c\r;\{m\};\l d\r;qg)
                 +X(\l c\r;\{l\};\l d\r;qg)\large)\times 
          \\ \nonumber
&        &\phantom{\large[}
          \large(X(\l g\r;\{k\};\l h\r;qg)
                +X(\l g\r;\{l\};\l h\r;qg) 
               +2X(\l g\r;\{n\};\l h\r;qg)\large)-
          \\ \nonumber
&        &\phantom{\large[}
         2\large(2X(\l c\r;\{n\};\l d\r;qg)
                 +X(\l c\r;\{m\};\l d\r;qg) 
                 +X(\l c\r;\{l\};\l d\r;qg))\large)\times 
          \\ \nonumber
&        &\phantom{\large[}
          \large(X(\l g\r;\{k\};\l h\r;qg)
                +X(\l g\r;\{l\};\l h\r;qg)\large)\large] \\ \nonumber
& +      &\large(\sum_k p_k+\sum_l p_l+\sum_m p_m\large)^2 \times \\ \nonumber
&        &\{Z(\l a\r;\l b\r;\l g\r;\l h\r;qg;qg)\times     
         \\ \nonumber
&        &\phantom{\large\{}
\large[2Z(\l c\r;\l d\r;\l i\r;\l j\r;qg;qg)
~X(\l e\r;\{k\};\l f\r;qg)- \\ \nonumber
&        &\phantom{\large\{\large[}
 Z(\l c\r;\l d\r;\l e\r;\l f\r;qg;qg)
~X(\l i\r;\{k\};\l j\r;qg)\large] \\ \nonumber
&        &-Z(\l g\r;\l h\r;\l i\r;\l j\r;qg;qg)\times     
         \\ \nonumber
&        &\phantom{\large\{}
\large[2Z(\l a\r;\l b\r;\l c\r;\l d\r;qg;qg)
~X(\l e\r;\{k\};\l f\r;qg)- \\ \nonumber
&        &\phantom{\large\{\large[}
 Z(\l c\r;\l d\r;\l e\r;\l f\r;qg;qg)
~X(\l a\r;\{k\};\l b\r;qg)\large]\} \\ \nonumber
&+       &\large(\sum_k p_k+\sum_l p_l\large)^2
           Z(\l c\r;\l d\r;\l g\r;\l h\r;qg;qg)\times     
         \\ \nonumber
&        &\phantom{\large\{}
\large[2Z(\l a\r;\l b\r;\l e\r;\l f\r;qg;qg)\times \\ \nonumber
&        &\phantom{\large\{}
 (X(\l i\r;\{k\};\l j\r;qg)
 +X(\l i\r;\{l\};\l j\r;qg)
 +X(\l i\r;\{m\};\l j\r;qg))+ \\ \nonumber
&        &\phantom{\large\{\large[}
 Z(\l e\r;\l f\r;\l i\r;\l j\r;qg;qg)\times \\ \nonumber
&        &\phantom{\large\{}
 (X(\l a\r;\{n\};\l b\r;qg)
 -X(\l a\r;\{m\};\l b\r;qg)-\\ \nonumber
&        &\phantom{\large\{\large[}
  X(\l a\r;\{l\};\l b\r;qg)
 -X(\l a\r;\{k\};\l b\r;qg))- \\ \nonumber
&        &\phantom{\large\{\large[}
 Z(\l a\r;\l b\r;\l i\r;\l j\r;qg;qg)\times \\ \nonumber
&        &\phantom{\large\{}
 (2X(\l e\r;\{n\};\l f\r;qg)
  +X(\l e\r;\{m\};\l f\r;qg)
  +X(\l e\r;\{l\};\l f\r;qg))\large] 
\end{eqnarray}
\begin{equation}\label{C}                            
\hskip-11.0truecm{-~{\mbox{same}}[(c,d) \leftrightarrow (e,f)].}
\end{equation}
The propagator functions associated to these terms are of the 
form $D_g(\{k\}+\{l\}+\{m\}+\{n\})D_g(\{k\}+\{l\}+\{m\})D_g(\{k\}+\{l\})
\equiv
1/(\sum_k p_k+\sum_l p_l+\sum_m p_m+\sum_n p_n)^2
       /(\sum_k p_k+\sum_l p_l+\sum_m p_m)^2
       /(\sum_k p_k+\sum_l p_l)^2$.
The summations $\{k\},\{l\}$, $\{m\}$ and $\{n\}$
are here over $c,d$, $e,f$, $g,h$ and $i,j$ if $\l d\r\ne [d]$, 
$\l f\r\ne [f]$, $\l h\r\ne [h]$ and $\l j\r\ne [j]$, or $c$, $e$, $g$ and $i$
otherwise, in the same order. Also in this case, terms of the forms
$D_g(\{k\})$, $D_g(\{l\})$, $D_g(\{m\})$ and $D_g(\{n\})$
will appear if $\l d\r\ne[d]$, $\l f\r\ne[f]$, $\l h\r\ne[h]$ and/or
$\l j\r\ne[j]$, respectively. 

Note that further manipulations would be possible, though in the form
given in eqs.~(\ref{A}), (\ref{B}) and (\ref{C}), the functions $A$, $B$
and $C$ are suitable for immediate implementation in a numerical program, as we
have done in our calculations.
The expressions 
(\ref{Z}), (\ref{A}), (\ref{B}) and (\ref{C}) are now the basic 
ingredients needed to describe the spinor part of all the topologies appearing
in Fig.~\ref{fig_6partons}a--c.

In the remainder of this Appendix, we present the helicity amplitudes
associated to the typical topologies in Fig.~\ref{fig_6partons}a--c, 
stripped off
their colour structure (operation that we recall by means of the introduction
of the argument $\Pi$). For sake of illustration, we assume that the 
labels $a$ and $b$ always refer to the quarks and antiquarks of flavour
$q$ in all reactions (\ref{qqgggg})--(\ref{qqqqqq}), that $c$ and $d$
refer to gluons in reaction (\ref{qqgggg}) and to the quarks and 
antiquarks of flavour $q'$ in reactions (\ref{qqqqgg})--(\ref{qqqqqq})
and that $e$ and $f$ refer to gluons in reactions 
(\ref{qqgggg})--(\ref{qqqqgg}) and to the quarks and 
antiquarks of flavour $q''$ in reaction (\ref{qqqqqq}). The permutations chosen
as example for the gluons labelled $cdef$ in reaction (\ref{qqgggg}) and
$ef$ in reaction (\ref{qqqqgg})  as well those
for the internal coloured gauge vectors in processes 
(\ref{qqqqgg})--(\ref{qqqqqq}) ought to be self-evident in the forthcoming
helicity amplitudes.

\vskip2.0cm
\hskip1.truecm\framebox{$e^+(p_1,\lambda_1)+e^-(p_2,\lambda_2) 
                    \rightarrow
                    q(p_a,\lambda_a)+\bar q(p_b,\lambda_b)
                   +g(p_c,\lambda_c)+     g(p_d,\lambda_d)
                   +g(p_e,\lambda_e)+     g(p_f,\lambda_f)$}

\begin{eqnarray}\label{a1} \nonumber
T^V_1(\Pi,\{\lambda\})
&      = & Z((a);\{i\};(c);[c];qg;qg)   \\ \nonumber
& \times & Z(\{i\};\{j\};(d);[d];qg;qg) \\ \nonumber
& \times & Z(\{j\};\{k\};(1);(2);qV;eV) \\ \nonumber
& \times & Z(\{k\};\{l\};(e);[e];qg;qg) \\ \nonumber
& \times & Z(\{l\};(b);(f);[f];qg;qg),  \\ \nonumber
&        & \{i\}=\{a,c\}_q,
  \quad    \{j\}=\{a,c,d\}_q,       
  \quad    \{k\}=\{1,2,a,c,d\}_q,               \\
&        & \{l\}=\{1,2,a,c,d,e\}_q.       
\end{eqnarray}

\begin{eqnarray}\label{a2} \nonumber
T^V_2(\Pi,\{\lambda\})
&      = & A((a);\{i\};(c);[c];(d);[d])     \\ \nonumber
& \times & Z(\{i\};\{j\};(1);(2);qV;eV) \\ \nonumber
& \times & Z(\{j\};\{k\};(e);[e];qg;qg) \\ \nonumber
& \times & Z(\{k\};(b);(f);[f];qg;qg),  \\
&        & \{i\}=\{a,c,d\}_q,
  \quad    \{j\}=\{1,2,a,c,d\}_q,       
  \quad    \{k\}=\{1,2,a,c,d,e\}_q.       
\end{eqnarray}

\begin{eqnarray}\label{a3-a4} \nonumber
T^V_3(\Pi,\{\lambda\})+T^V_4(\Pi,\{\lambda\})
&      = & B((a);\{i\};(c);[c];(d);[d];(e);[e]) \\ \nonumber
& \times & Z(\{i\};\{j\};(1);(2);qV;eV)     \\ \nonumber
& \times & Z(\{j\};(b);(f);[f];qg;qg),      \\
&        & \{i\}=\{a,c,d,e\}_q,
  \quad    \{j\}=\{1,2,a,c,d,e\}_q.       
\end{eqnarray}

\begin{eqnarray}\label{a5} \nonumber
T^V_5(\Pi,\{\lambda\})
&      = & A((a);\{i\};(c);[c];(d);[d])         \\ \nonumber
& \times & Z(\{i\};\{j\};(1);(2);qV;eV)     \\ \nonumber
& \times & A(\{j\};(b);(e);[e];(f);[f]),        \\ 
&        & \{i\}=\{a,c,d\}_q,
  \quad    \{j\}=\{1,2,a,c,d\}_q.       
\end{eqnarray}

\begin{eqnarray}\label{a6-a7-a8} \nonumber
T^V_6(\Pi,\{\lambda\})+T^V_7(\Pi,\{\lambda\})+T^V_8(\Pi,\{\lambda\})
&      = & T((a);\{i\};(1);(2);qV;eV)                \\ \nonumber
& \times & C(\{i\};(b);(c);[c];(d);[d];(e);[e];(f);[f]), \\ 
&        & \{i\}=\{1,2,a\}_q.       
\end{eqnarray}

\vskip2.0cm
\hskip1.truecm\framebox{$e^+(p_1,\lambda_1) + e^-(p_2,\lambda_2) 
                    \rightarrow
                    q (p_a,\lambda_a)+ \bar q (p_b,\lambda_b)
                   +q'(p_c,\lambda_c)+ \bar q'(p_d,\lambda_d)
                   +g (p_e,\lambda_e)+      g (p_f,\lambda_f)$}

\begin{eqnarray}\label{b1} \nonumber
T^V_1(\Pi,\{\lambda\})
&      = & Z((a);\{i\};(c);(d);qg;qg)   \\ \nonumber
& \times & Z(\{i\};\{j\};(1);(2);qV;eV) \\ \nonumber
& \times & A(\{j\};(b);(e);[e];(f);[f]),    \\ 
&        & \{i\}=\{a,c,d\}_q,
  \quad    \{j\}=\{1,2,a,c,d\}_q.       
\end{eqnarray}

\begin{eqnarray}\label{b2} \nonumber
T^V_2(\Pi,\{\lambda\})
&      = & Z((a);\{i\};(c);(d);qg;qg)    \\ \nonumber
& \times & Z(\{i\};\{j\};(1);(2);qV;eV)  \\ \nonumber
& \times & Z(\{j\};\{k\};(e);[e];qg;qg)  \\ \nonumber
& \times & Z(\{k\};(b);(f);[f];qg;qg),   \\ 
&        & \{i\}=\{a,c,d\}_q,
  \quad    \{j\}=\{1,2,a,c,d\}_q.       
  \quad    \{j\}=\{1,2,a,c,d,e\}_q.       
\end{eqnarray}

\begin{eqnarray}\label{b3} \nonumber
T^V_3(\Pi,\{\lambda\})
&      = & Z((a);\{i\};\{k\};(d);qg;qg)    \\ \nonumber
& \times & Z((c);\{k\};(e);[e];qg;qg)      \\ \nonumber
& \times & Z(\{i\};\{j\};(1);(2);qV;eV)    \\ \nonumber
& \times & Z(\{j\};(b);(f);[f];qg;qg),     \\ 
&        & \{i\}=\{a,c,d,e\}_q,       
  \quad    \{j\}=\{1,2,a,c,d,e\}_q,
  \quad    \{k\}=\{c,e\}_{q'}.       
\end{eqnarray}

\begin{eqnarray}\label{b4} \nonumber
T^V_4(\Pi,\{\lambda\})
&      = & A((a);\{i\};(c);(d);(e);[e])         \\ \nonumber
& \times & Z(\{i\};\{j\};(1);(2);qV;eV)       \\ \nonumber
& \times & Z(\{j\};(b);(f);[f];qg;qg),        \\ 
&        & \{i\}=\{a,c,d,e\}_q,
  \quad    \{j\}=\{1,2,a,c,d,e\}_q.       
\end{eqnarray}

\begin{eqnarray}\label{b5} \nonumber
T^V_5(\Pi,\{\lambda\})
&      = & Z((a);\{i\};(1);(2);qV;eV)         \\ \nonumber
& \times & Z(\{i\};(b);\{k\};(d);qg;qg)     \\ \nonumber
& \times & Z((c);\{j\};(e);[e];qg;qg)         \\ \nonumber
& \times & Z(\{j\};\{k\};(f);[f];qg;qg),        \\ 
&        & \{i\}=\{1,2,a\}_q,
  \quad    \{j\}=\{c,e\}_{q'},
  \quad    \{k\}=\{c,e,f\}_{q'}.       
\end{eqnarray}

\begin{eqnarray}\label{b6} \nonumber
T^V_6(\Pi,\{\lambda\})
&      = & Z((a);\{i\};(1);(2);qV;eV)          \\ \nonumber
& \times & Z(\{i\};(b);\{j\};(d);qg;qg)        \\ \nonumber
& \times & A((c);\{j\};(e);[e];(f);[f]), \\ 
&        & \{i\}=\{1,2,a\}_q,
  \quad    \{j\}=\{c,e,f\}_{q'}.       
\end{eqnarray}

\begin{eqnarray}\label{b7-b8-b9} \nonumber
T^V_7(\Pi,\{\lambda\})+T^V_8(\Pi,\{\lambda\})+T^V_9(\Pi,\{\lambda\})
&      = & Z((a);\{i\};(1);(2);qV;eV)         \\ \nonumber
& \times & B(\{i\};(b);(c);(d);(e);[e];(f);[f]),  \\ 
&        & \{i\}=\{1,2,a\}_q.       
\end{eqnarray}

\begin{eqnarray}\label{b10} \nonumber
T^V_{10}(\Pi,\{\lambda\})
&      = & Z((a);\{i\};(1);(2);qV;eV)        \\ \nonumber
& \times & A(\{i\};(b);\{j\};(d);(f);[f])        \\ \nonumber
& \times & Z((c);\{j\};(e);[e];qg;qg),       \\ 
&        & \{i\}=\{1,2,a\}_q,
  \quad    \{j\}=\{c,e\}_{q'}.       
\end{eqnarray}

\vskip2.0cm
\hskip1.truecm\framebox{$e^+(p_1,\lambda_1) + e^-(p_2,\lambda_2) 
                    \rightarrow
                    q  (p_a,\lambda_a) + \bar q  (p_b,\lambda_b)
                   +q' (p_c,\lambda_c) + \bar q' (p_d,\lambda_d)
                   +q''(p_e,\lambda_e) + \bar q''(p_f,\lambda_f)$}

\begin{eqnarray}\label{c1} \nonumber
T^V_1(\Pi,\{\lambda\})
&      = & Z((a);\{i\};(1);(2);qV;eV)           \\ \nonumber
& \times & Z(\{i\};(b);\{j\};(d);qg;qg)         \\ \nonumber
& \times & Z((c);\{j\};(e);(f);qg;qg),          \\ 
&        & \{i\}=\{1,2,a\}_q,
  \quad    \{j\}=\{c,e,f\}_{q'}.       
\end{eqnarray}

\begin{eqnarray}\label{c2} \nonumber
T^V_2(\Pi,\{\lambda\})
&      = & Z((a);\{i\};(c);(d);qg;qg)           \\ \nonumber
& \times & Z(\{i\};\{j\};(1);(2);qV;eV)         \\ \nonumber
& \times & Z(\{j\};(b);(e);(f);qg;qg),          \\ 
&        & \{i\}=\{a,c,d\}_q,
  \quad    \{j\}=\{1,2,a,c,d\}_q.       
\end{eqnarray}

\begin{eqnarray}\label{c3} \nonumber
T^V_3(\Pi,\{\lambda\})
&      = & Z((a);\{i\};(1);(2);qV;eV)         \\ \nonumber
& \times & A(\{i\};(b);(c);(d);(e);(f)),          \\ 
&        & \{i\}=\{1,2,a\}_q.       
\end{eqnarray}

\section{Appendix II}
\label{Sect_appe2}

In this Appendix we describe how we have proceeded to obtain the QCD colour
factors\footnote{In order to check their correctness 
we have calculated the traces of the QCD matrices both using an analytical
procedure as well as the  computer program {\tt COLOUR} \cite{jari}.}
relevant to the three reactions (\ref{qqgggg})--(\ref{qqqqqq}). In what
follows, we assume that all internal and external colours are always summed
over.

The amplitudes squared for processes (\ref{qqgggg})--(\ref{qqqqqq}), 
summed/averaged over the final/initial colours and spins, can be written
in the general form
\begin{equation}\label{msquare}
{\left|{\overline M}\right|}^2=
  {1\over 4} g_s^4 e^2 \left(\prod_{i} N_i\right) 
                       \left(\prod_p {1\over{n_p!}}\right)
\sum_{\{\lambda\}}
\sum_{\Pi, \Pi^\prime}{M}(\Pi,\{\lambda\})
M^*({\Pi^\prime},\{\lambda\}) C^{\Pi,\Pi^\prime}.
\ee
The factor $1/4$ accounts for the average over the
initial spins whereas
 the EM and QCD couplings are obtained via the relations 
$e=\sqrt{4\pi\alpha_{em}}$ and $g_s=\sqrt{4\pi\alpha_{s}}$ (in natural units)
from the constants introduced in Subsect.~\ref{Subsect_numerics}. 
In the above equation 
 $\Pi^{(')}$ represents (products of) permutations of the gluon indices,
$\{\lambda\}$ indicates the helicities of all external particles
$\lambda_i$ (with $i=1,...8$),
$M({\Pi^{(')}},\{\lambda\})$ is the coefficient of the colour matrix
$T^{\Pi^{(')}}_{f}$ (where $f$ represents a sequence of flavours $ij...$),
that is, the full spinor amplitude (here, $\{i\}=\{1,2\}$)
\be
M(\Pi^{(')},{\{\lambda\}})=\sum_{\pi(qq'(q''))}\sum_{i=1}^{n}
\left(\sum_{V=\gamma,Z} T_V^i(\Pi^{(')},\{\lambda\})D_V(\{i\})\right),
\ee
with $n=8$, 10 and 3 for process (\ref{qqgggg}), (\ref{qqqqgg}) and
 (\ref{qqqqqq}), respectively,
and $C^{\Pi,\Pi^\prime}$ is the appropriate
colour factor, $C^{\Pi,\Pi^\prime} = {\mbox{Tr}}(T^\Pi_f 
T^{\Pi^\prime\dagger}_f)$.  Here, $\sum_{\pi(qq'(q''))}$ refers to
the sum over all possible flavour permutations in events with at least
four quarks.
Furthermore, $n_p!$ is a statistical factor corresponding to 
 each $n_p$-tuple of identical final state particles $p$
appearing in reactions
(\ref{qqgggg})--(\ref{qqqqqq})\footnote{Therefore, our
multidimensional integrations must cover the whole six-particle
phase space.} whereas $N_i$ is the normalisation of the 
fermion current associated to the gluon $i$, see eq.~(\ref{gluon_norm}).
In reaction (\ref{qqgggg}) $p=g$ and $n_g=4$, for 
(\ref{qqqqgg}) one has $p=q,\bar q,g$ with $n_q\equiv n_{\bar q}=1(2)$ if
$q\ne q'(q=q')$ and $n_g=2$, whereas for (\ref{qqqqqq}) one gets
$p=q, \bar q$ with 
$n_q\equiv n_{\bar q}=1(2)[3]$
if $q\ne q'\ne q''(q=q'\ne q'')[q=q'=q'']$. As for the gluon normalisations,
the product $\prod_i$ extends to $i=5,6,7,8$ in 
(\ref{qqgggg}), to $i=7,8$ in (\ref{qqqqgg}) whereas it does not
appear for (\ref{qqqqqq}). 

In order to obtain the colour factors $C^{\Pi,\Pi^\prime}$ we have proceeded
as follows. First, we have decomposed 
a four-gluon vertex into three three-gluon vertices,
each with a different colour and kinematic structure in {\sl factorised} form.
One such decompositions, though not unique, that we have adopted here, 
is the one pictured in Fig.~\ref{fig_4-gluon} (from Ref.~\cite{jari}).
\begin{figure}[h]
  \hbox{\vbox{
    \begin{center}
    \mbox{\psfig{figure=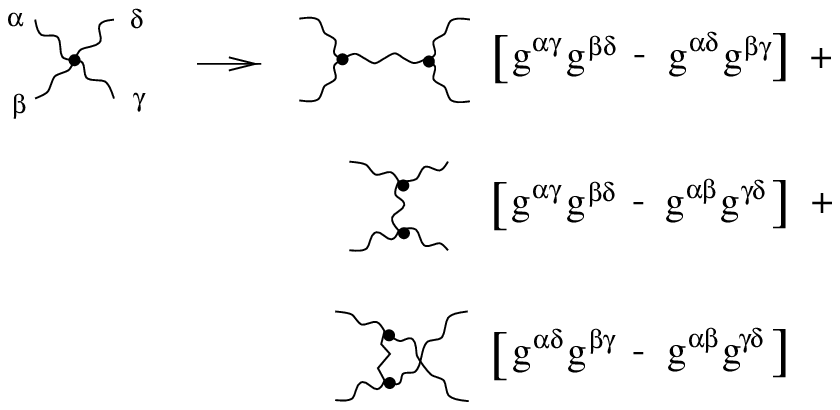}}
    \end{center}
  }}
  \caption{A symbolic representation of the decomposition
of a quadruple-gluon vertex into three triple-gluon vertices.}
  \label{fig_4-gluon}
\end{figure}
Then, we have repeatedly exploited the equation
\be\label{algebra}
[t^A,t^B]_{ij}\equiv t^A_{ik}t^B_{kj}-t^B_{ik}t^A_{kj}={\mathrm{i}}
f^{ABC}t^C_{ij},
\ee
relating the matrices $t^A_{ij}$ (with $A=1,...8$ gluon and $i,j=1,...3$
quark colours) 
of the fundamental representation to the 
structure constants $f^{ABC}$ of the SU(3) gauge group\footnote{As a possible
representation of the generators $t^A_{ij}$,  we have used the Gell-Mann
one.}, in order to rewrite the colour parts of diagrams involving triple and 
quadruple gluon vertices as a combination of the basic colour structures 
typical of the `abelian graphs', i.e., those non involving gluon 
self-couplings. If one does so, than the colour piece of
each diagram can be expressed as an appropriate combination of different
permutations of products involving $t^A_{ij}$ 
matrices only\footnote{Recall that if 
more than one coloured fermion line is present, such a combination is not 
unique, though the final answer obviously is.}. 

In case of 
process (\ref{qqgggg}), all colour structures in Fig.~\ref{fig_6partons}a
are linear combinations of terms of the form (here, $\Pi=\pi$ and $f=ij$)
\be\label{colour_qqgggg}
T^{\pi}_{ij}(q\bar qgggg)=(t^{A}t^{B}t^{C}t^{D})^{\pi(5678)}_{ij},
\ee
where $A,B,C,D(i,j)$ are the gluon(quark) colour and $\pi(5678)$ is
one the possible permutations of the gluon indices $A,...D=5,...8$. 
For four external gluons, one has
$4!$ of these, that is 24, twelve of which will be cyclical and twelve
anticyclical,
so that the number of fundamental colour factors
$C^{\pi,\pi'}$ 
needed to describe the colour content of process (\ref{qqgggg}) 
is 576 in total. 
These are given in Tab.~\ref{tab_qqgggg},
for the cyclical $\pi'$-permutations only, as the remaining colour factors
can be obtained by symmetry. 

\begin{table}[!t]
\begin{center}
\begin{tabular}{|c||c|c|c|c|c|c|c|c|c|c|c|c|}
\hline
\multicolumn{13}{|c|}
{Tr$[T_{ij}^{\pi}(q\bar qgggg)(T_{ij}^{\pi'}(q\bar qgggg))^\dagger]$} 
\\ \hline\hline
$C^{\pi,\pi'}$
     & 5678 & 5786 & 5867 & 6758 & 6875 & 6587 & 7856 & 7568 
     & 7685 & 8576 & 8657 & 8765 
\\ \hline \hline
5678 & 256 & 4 & 4 & 4 & -5 & 4 & 71/2 & 4 & -5 & -5 & -5 & -14
\\ \hline
5786 & 4 & 256 & 4 & -5 & -14 & -5 & 4 & 4 & -5 & 4 & 71/2 & -5
\\ \hline
5867 & 4 & 4 & 256 & 71/2 & -5 & 4 & -5 & -5 & -14 & 4 & 4 & -5
\\ \hline
6758 & 4 & -5 & 71/2 & 256 & 4 & 4 & -5 & 4 & 4 & -14 & -5 & -5
\\ \hline
6875 & -5 & -14 & -5 & 4 & 256 & 4 & -5 & 71/2 & 4 & -5 & 4 & 4
\\ \hline
6587 & 4 & -5 & 4 & 4 & 4 & 256 & -14 & -5 & -5 & -5 & 4 & 71/2
\\ \hline
7856 & 71/2 & 4 & -5 & -5 & -5 & -14 & 256 & 4 & 4 & 4 & -5 & 4
\\ \hline
7568 & 4 & 4 & -5 & 4 & 71/2 & -5 & 4 & 256 & 4 & -5 & -14 & -5
\\ \hline
7685 & -5 & -5 & -14 & 4 & 4 & -5 & 4 & 4 & 256 & 71/2 & -5 & 4
\\ \hline
8576 & -5 & 4 & 4 & -14 & -5 & -5 & 4 & -5 & 71/2 & 256 & 4 & 4
\\ \hline
8657 & -5 & 71/2 & 4 & -5 & 4 & 4 & -5 & -14 & -5 & 4 & 256 & 4
\\ \hline
8765 & -14 & -5 & -5 & -5 & 4 & 71/2 & 4 & -5 & 4 & 4 & 4 & 256
\\ \hline
5687 & -32 & 40 & -32 & -1/2 & -1/2 & -32 & 31 & -1/2 & 31 & -1/2 & 40 & 31
\\ \hline
5768 & -32 & -32 & 40 & 40 & 31 & -1/2 & -1/2 & -32 & -1/2 & -1/2 & 31 & 31
\\ \hline
5876 & 40 & -32 & -32 & 31 & 31 & -1/2 & 40 & -1/2 & 31 & -32 & -1/2 & -1/2
\\ \hline
6785 & -1/2 & 31 & 31 & -32 & -32 & 40 & -1/2 & -1/2 & -32 & 31 & -1/2 & 40
\\ \hline
6857 & -1/2 & 31 & 40 & 40 & -32 & -32 & 31 & 31 & -1/2 & -1/2 & -32 & -1/2
\\ \hline
6578 & -32 & -1/2 & -1/2 & -32 & 40 & -32 & 31 & 40 & -1/2 & 31 & -1/2 & 31
\\ \hline
7865 & 31 & -1/2 & 31 & -1/2 & 40 & 31 & -32 & 40 & -32 & -1/2 & -1/2 & -32
\\ \hline
7586 & -1/2 & -32 & -1/2 & -1/2 & 31 & 31 & -32 & -32 & 40 & 40 & 31 & -1/2
\\ \hline
7658 & 40 & -1/2 & 31 & -32 & -1/2 & -1/2 & 40 & -32 & -32 & 31 & 31 & -1/2
\\ \hline
8567 & -1/2 & -1/2 & -32 & 31 & -1/2 & 40 & -1/2 & 31 & 31 & -32 & -32 & 40
\\ \hline
8675 & 31 & 31 & -1/2 & -1/2 & -32 & -1/2 & -1/2 & 31 & 40 & 40 & -32 & -32
\\ \hline
8756 & 31 & 40 & -1/2 & 31 & -1/2 & 31 & -32 & -1/2 & -1/2 & -32 & 40 & -32
\\ \hline
\end{tabular}
\caption{The colour factors $C^{\pi,\pi^\prime}$, multiplied by 27, 
for $e^+e^-\rightarrow q\bar q gggg$. Permutations $\pi$ are along
columns, permutations $\pi'$ are along rows. Only half of the 24
$\pi'$-permutations are given (cyclic ones), as the remaining colour factors 
can be 
obtained by symmetry: the $(1-12)\times(1-12)$ submatrix is identical
to the $(13-24)\times(13-24)$ one (transposed), 
whereas the $(13-24)\times(1-12)$ 
submatrix is identical to the $(1-12)\times(13-24)$ one transposed.}
\label{tab_qqgggg}
\end{center}
\end{table}

In case of 
process (\ref{qqqqgg}), the colour structures appearing 
in Fig.~\ref{fig_6partons}b can be rewritten as
linear combinations of terms of the four forms (here, 
$\Pi=\pi(\pi\pi')$ in the first(last) two cases and $f=ijkl$)

\be\label{colour_qqqqgg1}
{T^{\pi}_1}_{ijkl}(q\bar q q'\bar q'gg)=(t^{X}t^{A}t^{B})^{\pi(X78)}_{ij}
                                        (t^{X})_{kl},
\ee

\be\label{colour_qqqqgg2}
{T^{\pi}_2}_{ijkl}(q\bar q q'\bar q'gg)=(t^{X})_{ij}
                                        (t^{X}t^{A}t^{B})^{\pi(X78)}_{kl}
\ee

\be\label{colour_qqqqgg3}
{T^{\pi\pi'}_3}_{ijkl}(q\bar q q'\bar q'gg)=(t^{X}t^{A})^{\pi (X7)}_{ij}
                                            (t^{X}t^{B})^{\pi'(X8)}_{kl},
\ee

\be\label{colour_qqqqgg4}
{T^{\pi\pi'}_4}_{ijkl}(q\bar q q'\bar q'gg)=(t^{X}t^{B})^{\pi (X8)}_{ij}
                                            (t^{X}t^{A})^{\pi'(X7)}_{kl},
\ee
where $A,B,X(i,j,k,l)$ are the gluon(quark) colour indices and 
$\pi^{(')}(X78)$, $\pi^{(')}(X7)$ and $\pi^{(')}(X8)$ 
are again gluon index permutations. The
symbol $X$ refers to any internal gluon $X=1,...8$ connecting the two 
quark-antiquark lines of colour $ij$ and $kl$ (and flavours $q$ and $q'$,
respectively).
For two external gluons, one has $3!\times1!$ permutations of the former 
two kinds
and $2!\times2!$ of the latter, for a total of 400 basic colour factors
$C_{mn}^{\Pi,\Pi'}$, with $m,n=1,2,3,4$. 
We present these in Tabs.~\ref{tab_qqqqgg11}--\ref{tab_qqqqgg34},
for the combinations $mn=11,12,13,33$ and 34.
The colour factors 
$C_{22(44)}^{{\pi,\pi^\prime}({\pi\pi^\prime,\pi''\pi'''})}$ are identical 
to the $C_{11(33)}^{{\pi,\pi^\prime}({\pi\pi^\prime,\pi''\pi'''})}$ ones
and the corresponding tables coincide with those numbered
\ref{tab_qqqqgg11} and \ref{tab_qqqqgg33} here, respectively
(tacitly assuming that the same gluon permutations are now made along different
quark colour lines).
Further notice that the colour factors 
$C_{23}^{\pi,\pi'\pi''}$ can be obtained from 
Tab.~\ref{tab_qqqqgg13} by exchanging $7$ with $8$ in the $\pi'$ and $\pi''$ 
permutations, as can easily be understood by 
comparing the corresponding colour diagrams.
Finally, the cases $C_{14}^{\pi,\pi^\prime\pi''}$ 
are identical to the $C_{23}^{\pi,\pi^\prime\pi''}$ ones whereas
the $C_{24}^{\pi,\pi^\prime\pi''}$'s coincide with 
the $C_{13}^{\pi,\pi^\prime\pi''}$'s, so that the combinations
$mn=14$ and $24$ can both be
deduced from Tab.~\ref{tab_qqqqgg13}, after appropriately assigning the quark
colours $ij$ and $kl$ to the gluon permutations $\pi,\pi'$ and $\pi''$. 

Decomposing the colour structure of the Feynman diagrams in 
Fig.~\ref{fig_6partons}b in terms of 
(\ref{colour_qqqqgg1})--(\ref{colour_qqqqgg4}) is very convenient when 
dealing with diagrams which only differ in the exchange of  
the flavours, $q\leftrightarrow q'$.
In fact, in this  case, in order to obtain 
the colourless part of new diagrams one flips
between themselves both the quark and antiquark
labels in the expressions of the amplitudes (\ref{b1})--(\ref{b10}):
 i.e., $a\leftrightarrow c$ and
       $b\leftrightarrow d$ 
(this way, both the momenta and helicities get interchanged). 
As for the colour factors,
these can easily be obtained from Tabs.
\ref{tab_qqqqgg11}--\ref{tab_qqqqgg34}, if one notices that
the colour matrices in (\ref{colour_qqqqgg2}) are identical to those
of eq.~(\ref{colour_qqqqgg1}) when one interchanges the (anti)quark colour
labels, i.e., $i\leftrightarrow k$ and $j\leftrightarrow l$,
and so is the case for those in (\ref{colour_qqqqgg4}) with respect
to eq.~(\ref{colour_qqqqgg3}).   

\begin{table}[!t]
\begin{center}
\begin{tabular}{|c||c|c|c|c|c|c|}
\hline
\multicolumn{7}{|c|}
{Tr$[{T_1}_{ijkl}^{\pi}(q\bar qq'\bar q'gg)
    ({T_1}_{ijkl}^{\pi'}(q\bar qq'\bar q'gg))^\dagger]$} 
\\ \hline\hline
$C_{11}^{\pi,\pi'}$
     & X78 & 78X & 8X7 & 87X & X87 & 7X8 
\\ \hline \hline
X78      & 32(-32) &  1/2(-1/2) &  1/2(-1/2) & 5(-5) & -4(4) & -4(4) 
\\ \hline
78X      &  1/2(-1/2) & 32(-32) &  1/2(-1/2) & -4(4) & 5(-5) & -4(4)  
\\ \hline
8X7      &  1/2(-1/2) &  1/2(-1/2) & 32(-32) & -4(4) & -4(4) & 5(-5)  
\\ \hline
87X      & 5(-5) & -4(4) & -4(4) & 32(-32) &  1/2(-1/2) &  1/2(-1/2)  
\\ \hline
X87      & -4(4) & 5(-5) & -4(4) &  1/2(-1/2) & 32(-32) &  1/2(-1/2)  
\\ \hline
7X8      & -4(4) & -4(4) & 5(-5) &  1/2(-1/2) &  1/2(-1/2) & 32(-32)  
\\ \hline
\end{tabular}
\caption{The colour factors $C_{11}^{\pi,\pi^\prime}$, multiplied by 9, 
for $e^+e^-\rightarrow q\bar q q'\bar q'gg$ ($q\ne q'$). 
In round brackets are given the additional ones needed in the case of
identical flavours $q=q'$ (for which the exchange of indices
$i\leftrightarrow k$
is understood in one of the two interfering colour structures), 
multiplied by 27. Permutations $\pi$ are along
columns, permutations $\pi'$ are along rows. The first three are cyclical
permutations, whereas the latter three are anticyclical.
Note the symmetry in the
colour factors: the $(1-3)\times(1-3)$ submatrix is identical
to the $(4-6)\times(4-6)$ one (transposed), whereas the $(1-3)\times(4-6)$
submatrix is identical to the $(4-6)\times(1-3)$ one (transposed).}
\label{tab_qqqqgg11}
\end{center}
\end{table}

\begin{table}[!t]
\begin{center}
\begin{tabular}{|c||c|c|c|c|c|c|}
\hline
\multicolumn{7}{|c|}
{Tr$[{T_1}_{ijkl}^{\pi}(q\bar qq'\bar q'gg)
    ({T_2}_{ijkl}^{\pi'}(q\bar qq'\bar q'gg))^\dagger]$} 
\\ \hline\hline
$C_{12}^{\pi,\pi'}$
     & X78 & 78X & 8X7 & 87X & X87 & 7X8 
\\ \hline \hline
X78  & -3(31) & 57/2(-1/2) & -3(31) & -3(-5) & 6(-14) & -3(-5) 
\\ \hline
78X  & -3(31) & -3(31) & 57/2(-1/2) & -3(-5) & -3(-5) & 6(-14) 
\\ \hline
8X7  & 57/2(-1/2) & -3(31) & -3(31) & 6(-14) & -3(-5) & -3(-5)  
\\ \hline 
87X  & -3(-5) & -3(-5) & 6(-14) & -3(31) & -3(31) & 57/2(-1/2)   
\\ \hline
X87  & 6(-14) & -3(-5) & -3(-5) & 57/2(-1/2) & -3(31) & -3(31) 
\\ \hline
7X8  & -3(-5) & 6(-14) & -3(-5) & -3(31) & 57/2(-1/2) & -3(31)  
\\ \hline
\end{tabular}
\caption{The colour factors $C_{12}^{\pi,\pi^\prime}$, multiplied by 9, 
for $e^+e^-\rightarrow q\bar q q'\bar q'gg$ ($q\ne q'$). 
In round brackets are given the additional ones needed in the case of
identical flavours $q=q'$ (for which the exchange of indices
$i\leftrightarrow k$
is understood in one of the two interfering colour structures), 
multiplied by 27. Permutations $\pi$ are along
columns, permutations $\pi'$ are along rows. The first three are cyclical
permutations, whereas the latter three are anticyclical.
Note the symmetry in the
colour factors: the $(1-3)\times(1-3)$ submatrix is identical
to the $(4-6)\times(4-6)$ one transposed, whereas the $(1-3)\times(4-6)$
submatrix is identical to the $(4-6)\times(1-3)$ one transposed.}
\label{tab_qqqqgg12}
\end{center}
\end{table}

\begin{table}[!t]
\begin{center}
\begin{tabular}{|c||c|c|c|c|}
\hline
\multicolumn{5}{|c|}
{Tr$[{T_1}_{ijkl}^{\pi}(q\bar qq'\bar q'gg)
    ({T_3}_{ijkl}^{\pi'\pi''}(q\bar qq'\bar q'gg))^\dagger]$} 
\\ \hline\hline
$C_{13}^{\pi,\pi'\pi''}$
     & (X7)(X8) & (X7)(8X) & (7X)(X8) & (7X)(8X) 
\\ \hline \hline
X78  &      1(-5) & -7/2(-1/2) &    1(31)   & -7/2(71/2)  
\\ \hline
78X  & -7/2(71/2) &    1(-1/2) &    28(31)  &     -8(-5)  
\\ \hline
8X7  &      28(4) &    -8(-32) & -7/2(-1/2) &       1(4)   
\\ \hline 
87X  & -7/2(-1/2) &       1(4) &   -7/2(-5) &      1(40)    
\\ \hline
X87  &     -8(40) &      28(4) &     1(-5)  & -7/2(-1/2)  
\\ \hline
7X8  &      1(31) &   -7/2(-5) &    -8(-14) &     28(31)    
\\ \hline
\end{tabular}
\caption{The colour factors $C_{13}^{\pi,\pi^\prime\pi''}$, multiplied by 9, 
for $e^+e^-\rightarrow q\bar q q'\bar q'gg$ ($q\ne q'$). 
In round brackets are given the additional ones needed in the case of
identical flavours $q=q'$  (for which the exchange of indices
$i\leftrightarrow k$
is understood in one of the two interfering colour structures), 
multiplied by 27. Permutations $\pi$ are along
columns, permutations $\pi'$ and $\pi''$ 
are along rows.}
\label{tab_qqqqgg13}
\end{center}
\end{table}

\begin{table}[!t]
\begin{center}
\begin{tabular}{|c||c|c|c|c|}
\hline
\multicolumn{5}{|c|}
{Tr$[{T_3}_{ijkl}^{\pi\pi'}(q\bar qq'\bar q'gg)
    ({T_3}_{ijkl}^{\pi''\pi'''}(q\bar qq'\bar q'gg))^\dagger]$} 
\\ \hline\hline
$C_{33}^{\pi\pi',\pi''\pi'''}$
     & (X7)(X8) & (X7)(8X) & (7X)(X8) & (7X)(8X) 
\\ \hline \hline
(X7)(X8)  &   32(-32) &     -4(4) &     -4(4) & 1/2(-1/2)     
\\ \hline
(X7)(8X)  &     -4(4) &    32(40) & 1/2(-1/2) &    -4(-5)    
\\ \hline
(7X)(X8)  &     -4(4) & 1/2(-1/2) &    32(40) &    -4(-5)     
\\ \hline 
(7X)(8X)  & 1/2(-1/2) &    -4(-5) &    -4(-5) &    32(31) 
\\ \hline
\end{tabular}
\caption{The colour factors $C_{33}^{\pi\pi',\pi''\pi'''}$, multiplied by 9, 
for $e^+e^-\rightarrow q\bar q q'\bar q'gg$ ($q\ne q'$). 
In round brackets are given the additional ones needed in the case of
identical flavours $q=q'$  (for which the exchange of indices
$i\leftrightarrow k$
is understood in one of the two interfering colour structures), 
multiplied by 27. Permutations $\pi$ and $\pi'$
are along columns, permutations $\pi''$ and $\pi'''$ 
are along rows.}
\label{tab_qqqqgg33}
\end{center}
\end{table}

\begin{table}[!t]
\begin{center}
\begin{tabular}{|c||c|c|c|c|}
\hline
\multicolumn{5}{|c|}
{Tr$[{T_3}_{ijkl}^{\pi\pi'}(q\bar qq'\bar q'gg)
    ({T_4}_{ijkl}^{\pi''\pi'''}(q\bar qq'\bar q'gg))^\dagger]$} 
\\ \hline\hline
$C_{34}^{\pi\pi',\pi''\pi'''}$
     & (X8)(X7) & (X8)(7X) & (8X)(X7)   & (8X)(7X) 
\\ \hline \hline
(X7)(X8)  &     -3(31) &   -3(-5) &   -3(-5) & 57/2(-1/2)     
\\ \hline
(X7)(8X)  &     -3(-5) & -3(-1/2) &    6(40) &      -3(4)     
\\ \hline
(7X)(X8)  &     -3(-5) &    6(40) & -3(-1/2) &      -3(4)      
\\ \hline 
(7X)(8X)  & 57/2(-1/2) &    -3(4) &    -3(4) &    -3(-32) 
\\ \hline
\end{tabular}
\caption{The colour factors $C_{34}^{\pi\pi',\pi''\pi'''}$, multiplied by 9, 
for $e^+e^-\rightarrow q\bar q q'\bar q'gg$ ($q\ne q'$). 
In round brackets are given the additional ones needed in the case of
identical flavours $q=q'$ (for which the exchange of indices
$i\leftrightarrow k$
is understood in one of the two interfering colour structures), 
multiplied by 27. Permutations $\pi$ and $\pi'$
are along columns, permutations $\pi''$ and $\pi'''$ 
are along rows.}
\label{tab_qqqqgg34}
\end{center}
\end{table}

In case of identical flavours in process
(\ref{qqqqgg}), $q= q'$, one further needs to calculate the diagrams which  
differ (in the spinor part) from the original ones of the case $q\ne q'$  
in the exchange of the four-momentum and helicity of {\sl one}
 of the, say, quarks with 
those of the other, $a\leftrightarrow c$. Because of Fermi
statistics, a minus sign factorises too. However, (anti)quark colour quantum
numbers have to be exchanged as well. This introduces 
new colour structures, identical to those in 
eqs.~(\ref{colour_qqqqgg1})--(\ref{colour_qqqqgg4}) except for the exchange
 $i\leftrightarrow k$. Clearly, the  colour factors building up the
squared amplitude of the new set of graphs are identical to those already
given in Tabs.~\ref{tab_qqqqgg11}--\ref{tab_qqqqgg34}. In contrast, new ones
are needed for the interference terms. These appear in the above tables in
round brackets (note their different normalisation).

In case of 
process (\ref{qqqqqq}), we recognise the following three  basic
colour structures, as the fundamental ones to calculate all the colour
factors pertinent to the
graphs in Fig.~\ref{fig_6partons}c (here, $\Pi=\pi$ and $f={ijklmn}$):

\be\label{colour_qqqqqq1}
{T_1}_{ijklmn}^{\pi}
(q\bar q q'\bar q' q''\bar q'')=(t^{X}t^{Y})^{\pi(XY)}_{ij}
                                (t^{X})_{kl}
                                (t^{Y})_{mn},
\ee

\be\label{colour_qqqqqq2}
{T_2}_{ijklmn}^{\pi}
(q\bar q q'\bar q' q''\bar q'')=(t^{X})_{ij}
                                (t^{X}t^{Y})^{\pi(XY)}_{kl}
                                (t^{Y})_{mn},
\ee

\be\label{colour_qqqqqq3}
{T_3}_{ijklmn}^{\pi}
(q\bar q q'\bar q' q''\bar q'')=(t^{X})_{ij}
                                (t^{Y})_{kl}
                                (t^{X}t^{Y})^{\pi(XY)}_{mn},
\ee
where $X,Y(i,j,k,l,m,n)$ are the gluon(quark) colour indices and 
$\pi(XY)$ represents gluon index permutations. The
symbols $X$ and $Y$ refers to any combination of
internal gluons $X,Y=1,...8$ interconnecting the three 
quark-antiquark lines of colour $ij$, $kl$ and $mn$ (and flavours $q$,
$q'$ and $q''$, respectively). In this case, 
the calculation of the colour factors is much easier. In fact, it is 
trivial to see that there are only four
fundamental ones, of the form $C_{op}^{\pi,\pi'}$, with $o,p=1,2,3$.
Two correspond to the case $o=p$, $4/3$ and $-1/6$, and two to the opposite
condition, $o\ne p$, $-1/3$ and $7/6$. 
These values are obtained depending on the 
way  three coloured  fermion loops (of flavour $q\ne q'\ne q''$) are connected 
to each other via four gluons. If $o=p$, one of the loops has four attached
vectors and the others have two.
When $o\ne p$, two loops have three connections and the remaining one only
one. (Note that in case of the colour diagrams of process (\ref{qqqqqq})
gluon permutations along quark 
lines are constrained, as two (and only two) 
gluon connections from each of the interfering diagrams must always be 
adjacent in the full colour graph.)
Once again, it is rather straightforward to 
calculate all diagrams, and corresponding colour 
structures, differing only by flavour exchanges,
given the explicit expressions 
of the formulae (\ref{c1})--(\ref{c3}) and of the colour matrices in 
eqs.~(\ref{colour_qqqqqq1})--(\ref{colour_qqqqqq3}), in terms of 
external momenta/helicities and flavours, respectively.

Also in case of process (\ref{qqqqqq}), if two or more flavours in the final 
state are identical, more (spinor and colour) diagrams are needed. 
For the spinor part, the number of graphs doubles(triples) depending on 
whether two(three) flavours coincide and the additional ones 
(with respect to the case of all different flavours) can 
again be obtained by simple relabelling
of the relevant  momenta and helicities in the formulae (\ref{c1})--(\ref{c3}),
as previously explained. As for colour diagrams, it can immediately
be seen that for
the cases $q=q'\ne q''$, $q'=q''\ne q$ and $q''=q\ne q'$ one has to compute
graphs in which two bubbles are connected by 2, 3 or 4 gluons with
2, 1 and 0 additional gluons coupled to only one of the loops, respectively.
In correspondence of the three sets of colour graphs, one gets 3, 2 and
1 independent colour factors: i.e., $(1/18, 5/9,-4/9)$, $(-7/18,1/9)$
and $(-1/3)$. Finally, 
in the case $q=q'=q''$, the additional colour factors needed for the 
calculation of the complete ME for process (\ref{qqqqqq}) have already been 
derived, as they appear among the numbers given in Tab.~\ref{tab_qqgggg}. 
Indeed, one only needs five of those: i.e., 31, $-5$, $-1/2$, $71/2$ and
4 (all to be further divided by 27). Also in case of identical flavours
the same restrictions as above on the possible gluon permutations in the
colour diagrams squared apply.

To appropriately combine the spinor and colour parts of the amplitudes
into the matrix elements of the form (\ref{msquare}) for all
 processes (\ref{qqgggg})--(\ref{qqqqqq})
 is a tedious but trivial labour that we
leave as an exercise for the reader, 
as it would only results here in cumbersome
expressions carrying 
no additional information with respect to that already given 
in the previous formulae and tables. 
This procedure has been implemented
in our {\tt FORTRAN} programs, that we make available to the public 
upon request.

\end{document}